%% file: fibers.tex
\documentclass[11pt]{article}
\pdfoutput=1

\usepackage[no-natbib-sort]{jheppub}
\usepackage[T1]{fontenc} 

\usepackage{amssymb}
\usepackage{amsfonts}
\usepackage{amsbsy}
\usepackage{amsmath}
\usepackage{amsthm}
\usepackage{graphicx}
\usepackage{epstopdf}
\usepackage[vcentermath]{youngtab}
\usepackage{multirow}
\usepackage{latexsym} 
\usepackage{array}
\usepackage{color}
\usepackage{verbatim}
\usepackage{comment}
\usepackage[utf8]{inputenc}
\usepackage{soul}
\usepackage{subcaption}

\usepackage{tikz}

\input cyracc.def 
\newfont{\twelvecyr}{wncyr10 at 12pt}

\def\Adj{{\rm Adj }}
\def\trace{{\mathrm Tr }}

\def\Z{\mathbb{Z}}
\def\F{\mathbb{F}}

\def\C{\mathbb{C}}

\def\P{\mathbb{P}}

\def\bbQ{\mathbb{Q}}
\def\bbF{\mathbb{F}}

\def\n3a{t}

\def\ord{\mathrm{ord}}

\def\bbP{\mathbb{P}}
\newcommand{\rk}[0]{\operatorname{rk}}

\newcommand{\SU}[0]{\mathrm{SU}}

\newcommand{\ignore}[1]{}

\def\agsu{{\mathfrak{su}}}

\newcommand{\wati}[1]{\footnote{\textcolor{blue}{\textbf{WT:\ #1}}}}
\newcommand{\fatima}[1]{\footnote{\textcolor{red}{\textbf{FA:\ #1}}}}
\newcommand{\richard}[1]{\footnote{\textcolor{magenta}{\textbf{RN:\ #1}}}}

\newcommand{\clean}{
\renewcommand{\fatima}[1]{}
\renewcommand{\richard}[1]{}
\renewcommand{\wati}[1]{}
}
\clean

\newcommand{\eqq}[1]{\begin{align}#1\end{align}}

\title{Classifying Fibers and Bases in Toric Hypersurface Calabi-Yau
  Threefolds}

\author{Fatima Abbasi,}
\author{Richard Nally,}
\author{and Washington Taylor}

\affiliation{Center for Theoretical Physics\\
Department of Physics\\
Massachusetts Institute of Technology\\
77 Massachusetts Avenue\\
Cambridge, MA 02139, USA}

\emailAdd{fabbasi at mit.edu}
\emailAdd{rnally at mit.edu}
\emailAdd{wati at mit.edu}

\preprint{MIT-CTP/5958}

\abstract{We carry out a complete analysis of the toric elliptic and
  genus-one fibrations of all 474 million reflexive polytopes in the
  Kreuzer-Skarke database.  Earlier work with Huang showed that all
  but 29,223 of these polytopes have such a fibration.  We identify
  2,264,992,252 distinct fibrations, and determine the fiber and base
  structure in each case; after accounting for
  automorphisms of the ambient polytope, these fibrations furnish
  2,250,744,657 equivalence classes.  We summarize generic features
  and identify exotic special cases among these fibrations.
These fibrations
illustrate many features  that have been explored in the context of 6D F-theory,
 including
  gauge groups hosted on non-toric divisors, 
automatic enhancement of gauge groups,
and implicit non-toric
  bases 
  and high-rank 6D SCFTs
  associated with non-flat fibers,
  as well as novel geometric features such as singular
  bases for genus-one fibrations with multisections. 
  This analysis illustrates the power
  of elliptic and genus-one fibrations, and the 
 geometro-physical language of F-theory as a tool for understanding the
  structure of Calabi-Yau threefolds.  }
  
\begin{document}
\maketitle
\flushbottom

\section{Introduction}
\label{sec:Intro}

Calabi-Yau manifolds have been a central subject of study in
theoretical physics and in mathematics since the realization over 40
years ago that these spaces can be used to construct a variety of
compactifications of string theory \cite{Candelas:1985en}.  Despite a
tremendous amount of work in this area, it remains unknown whether the
number of distinct topological types of Calabi-Yau threefolds is
finite or infinite.
Nonetheless, 
an important class of
Calabi-Yau threefolds are those that admit an elliptic or genus-one
fibration, and  it is known that there exist a finite number of
topological types of such geometries \cite{Grassi, Gross}.
Elliptic Calabi-Yau
manifolds play a central role in the approach to string theory known
as F-theory \cite{Vafa-F-theory, Morrison-Vafa-I, Morrison-Vafa-II},
which connects  the geometry of an elliptic Calabi-Yau to a physical
theory of gravity, gauge fields, and matter.
 A systematic approach can in principle be taken to constructing
and classifying all elliptic Calabi-Yau threefolds, by first classifying allowed bases
and then classifying possible ``tunings'' of Weierstrass models over
each base.  While there are a number of technical complications and
subtleties in realizing a complete enumerated classification using
this approach, it is clear that combining physical insights from
F-theory with the mathematical structure of elliptic fibrations gives
a powerful set of tools and perspectives with which to analyze
Calabi-Yau threefolds (and fourfolds).
Moreover, in recent years a
growing body of evidence has shown that most known Calabi-Yau manifolds
have the form of an elliptic or
genus-one fibration (or at least lie in an extended K\"ahler cone that has a chamber with such a fibration) \cite{Candelas:2012uu, Gray:2014fla,
  Anderson:2016cdu,Anderson:2017aux,Anderson:2018kwv,
  Huang-WT-2018-Long},
and there are general arguments for
why this should be the case, particularly for Calabi-Yau threefolds with large Picard number \cite{Huang-WT_2018-Short}, as is also expected in the mathematics literature \cite{Gross, wilson1998existence}.
%


The largest known set of Calabi-Yau threefolds are the toric
hypersurfaces in reflexive 4D polytopes \cite{Batyrev:1993oya}.  A complete classification of the 473,800,776 reflexive 4D polytopes was made some time ago by
Kreuzer and Skarke (KS) \cite{Kreuzer:2000xy}.
This represents the richest database of examples of Calabi-Yau threefolds available;
it remains unclear, however, precisely how many distinct Calabi-Yau threefolds can be
built from these polytopes
\cite{Demirtas:2020dbm,Gendler:2023ujl,Chandra:2023afu,MacFadden:2023cyf}.

Considering toric hypersurface Calabi-Yau threefolds
from the perspective of elliptic fibrations can give a useful
framework for systematically studying this large set of geometries.
In previous work \cite{Huang-WT-fibers}, Huang and one of the authors of this work
showed
that, of the 474 million reflexive 4D polytopes, all but 29,223
contain at least one reflexive 2D subpolytope, indicating that at
least one flop phase of each of the associated Calabi-Yau threefolds
has an elliptic
(or genus-one) fibration structure.
Almost 15 years ago,
Braun \cite{Braun:2011ux}
identified and classified all  such toric elliptic fibrations over
the simplest 2D base $\bbP^2$.
In this work, we extend  these earlier analyses
by performing a complete classification of all 2D reflexive
subpolytopes of 4D polytopes in the KS database, along with the 2D
toric base polytopes to which the associated elliptic Calabi-Yau
threefolds project along the fiber. 
In this paper, we summarize the
results of this classification, including a statistical analysis of typical
fibers and bases, along with an examination of some more exotic
structures that are found in these fibrations. Data files containing the full results of the analysis, as well as the software used to do the analysis, are included in an
associated Zenodo archive \cite{archive}.
\wati{ put in a placeholder for now in this reference}

The structure of this paper is as follows: 
in \S\ref{sec:fiber-base}, we review some known facts about toric
fibers and bases for elliptic Calabi-Yau threefolds.
In \S\ref{sec:algorithm},
we describe the algorithm and approach used to perform the complete
analysis of fibrations in the KS database.
In \S\ref{sec:statistics}, we describe general features of the
statistics of the 2.25 billion fiber/base pairs found in the
analysis.  In \S\ref{sec:examples}, we go into more detail on some
interesting examples and features found in the analysis.
\S\ref{sec:conclusions} contains some concluding remarks.

\section{Toric Bases, Fibers, and Fibrations}
\label{sec:fiber-base}

In this section, we recall some key background for our work.
We begin with some general comments to provide perspective on how the
analysis of this paper fits into the
framework of a broader research agenda.

Two of the most powerful approaches known for constructing large
classes of Calabi-Yau threefolds are: (a) as hypersurfaces (or, more
generally, complete intersections) in toric varieties, (b) through
elliptic (or, more generally, genus-one) fibrations.  For (a), the KS
database of 474 million polytopes provides the richest set of concrete
examples of Calabi-Yau threefolds available, which
has been used in many contexts
in the physics and mathematics literature.  For (b), the finite nature
of the set of fibered CY threefolds,
the fact that almost all known Calabi-Yau threefolds have an elliptic
or genus-one-fibered phase,
the possibility of direct
construction through tuned Weierstrass models, and the geometric
interpretation from F-theory make this a very powerful framework for
understanding the structure of CY threefolds.  The goal of this paper
is to combine these perspectives and show that almost
all of the many Calabi-Yau threefolds that can be constructed from
polytopes in the Kreuzer-Skarke database live in families (defined by
the extended K\"ahler cone) that contain an elliptic or genus-one
fibered representative, and that this allows us to classify and understand the
structure of these spaces.

One motivation for this work is to explore the range of possible fibration
structures that arise in this large class of Calabi-Yau threefolds,
and how they fit into the full set of possible F-theory models.  A
second motivation is to illustrate how the geometric/physical language
of F-theory provides a powerful tool for understanding and identifying
most known Calabi-Yau threefolds.  In general, a toric hypersurface
Calabi-Yau is defined by a suitable triangulation of one of the 474
million reflexive polytopes in the Kreuzer-Skarke list; as reviewed in
e.g. \cite{Gendler:2023ujl,Chandra:2023afu}, such a geometry is
characterized topologically as a real manifold by its ``Wall data,''
consisting of its Hodge numbers, triple intersection numbers, and
second Chern class, and as a complex manifold by its Mori cone.
These data are somewhat opaque, in the
sense that they do not convey much intuitive understanding of the
structure of a given Calabi-Yau, or enable us to easily classify
structures or connect related Calabi-Yau threefolds.  On the other hand, the F-theory language for Calabi-Yau
threefolds provides a rich set of insights into the structure of these
varieties.  In particular, from this perspective, rather than
identifying an elliptic Calabi-Yau in terms of its Wall data and Mori cone,
we can characterize it by features with physical and geometric
significance; in this language, a given polytope may
correspond to a family of Calabi-Yau threefolds organized into a toric extended K\"ahler cone that
have several different fibered phases, each of which can be described
in F-theory language
as, e.g., an elliptic fibration over the base Hirzebruch
$\F_3$, with a gauge factor SU(3) over the base curve of
self-intersection -3 and a gauge factor $E_6$ tuned over the base
curve of self-intersection +3.  The latter geometric language
provides a powerful framework for understanding not only the structure
of individual Calabi-Yau threefolds, as fibration data reflects key information about
the intersection numbers, Hodge numbers, and Mori cone of the Calabi-Yau,
but also, because all elliptic Calabi-Yau threefolds are
connected by a combination  of geometric transitions with physical
interpretation, provides a way of understanding how different Calabi-Yau
threefolds are connected.
These geometric transitions include tensor transitions corresponding to blowing
up points in the base, Higgs transitions in which the Hodge number
$h^{1,1} (X)$ decreases representing a reduction in the physical gauge
group, and more obscure matter transitions \cite{Anderson:2015cqy},
in which the gauge group
and base structure remain invariant.

In the remainder of this section, we will introduce the background necessary to pursue those goals.
Heuristically, an elliptic threefold is one
that can be viewed as a torus varying over the points of a
two-dimensional base; we begin
in \S\ref{sec:elliptic}
with basic definitions of elliptic
curves and elliptic fibrations to make this intuition precise before
describing the toric technology we use throughout the paper.
In \S\ref{sec:toric-bases} and \S\ref{sec:toric-fibers}, we describe
the set of 61,539 toric bases and 16 toric fibers that serve as the
building blocks for toric elliptic fibrations, which are described in
\S\ref{sec:toric-fibrations}.  In \S\ref{sec:genus-1}, we clarify
the distinction between genus-one fibered and elliptically fibered
Calabi-Yau threefolds, and
 in \S\ref{sec:Nagell}, we describe Nagell's algorithm, a unified framework for
 analyzing the geometry of all elliptic and genus-one fibrations.
 We conclude in \S\ref{sec:F-theory} with an overview of 6D F-theory,
 which essentially provides a dictionary between structures in 6D
 supergravity theories and the geometry of elliptic and genus-one fibrations.

\subsection{Elliptic curves and elliptic fibrations}
\label{sec:elliptic}

Fix a field $F$; we call a genus-one curve $E$ defined over $F$ an \textit{elliptic curve} if it admits at least one point over $F$. This is because the points of $E$ furnish a group under a geometrically defined addition law; any group must have an identity element, so to be an elliptic curve $E$ must have at least one point. For instance, because by the adjunction formula quartic hypersurfaces in $\bbP^{1,1,2}$ are Calabi-Yau onefolds, i.e. complex tori,  the curve \eqq{x^4+y^4=z^2} has genus one, but over the rational numbers $\bbQ$, this polynomial has no nonzero solutions, so its graph does not furnish an elliptic curve over $\bbQ$.  It is a general fact that the geometry of an elliptic curve can be summarized in terms of two invariants, $j$ and $\Delta$. The distinguishing characteristic of the discriminant $\Delta$ is that $E$ is smooth unless $\Delta=0$, in which case we also necessarily have $j\to i\infty$.

Our primary interest is not in elliptic curves themselves, but rather in elliptic fibrations. We call a complex manifold $X$ torus-fibered over a $d$-dimensional base $B_d$ if there exists a map $\pi:X\to B_d$ such that the preimage of a generic point $b\in B_d$ is a torus. In analogy to the requirement that an elliptic curve has a point over its field of definition, we say that a torus-fibered manifold $X$ is \textit{elliptically fibered} if this fibration admits a section, and \textit{genus-one fibered} if it does not. In fact, this is more than an analogy; an elliptic fibration can equivalently be viewed as an elliptic curve whose field of definition is the function field of $B_d$, with the group of points being realized by the group of sections of the fibration. 

The geometry of elliptically fibered Calabi-Yau threefolds (CY3s) is strongly constrained. As summarized in the introduction, elliptic CY3s can
in principle be systematically classified by first classifying all
compact complex K\"ahler surfaces $B_2$ that can support an elliptic or genus-one fibered
Calabi-Yau threefold, and then classifying all possible fibrations over each
base.  We review here briefly some aspects of this classification in
the context of toric bases and toric hypersurface fibers.  We focus first on elliptic fibrations; genus-one fibrations are defined and discussed in more detail in \S\ref{sec:genus-1}.

It is a classical fact, which we review in \S\ref{sec:Nagell}, that an elliptic fibration over a base $B_2$ can be described as a
Weierstrass model
\begin{equation}
 y^2 = x^3 + f x + g \,,
\label{eq:Weierstrass}
\end{equation}
where $f, g$ are sections of the line bundles ${\cal O} (-4 K), {\cal
  O} (-6 K)$, where $-K$ is the anticanonical class of the base. In terms of $f$ and $g$, the discriminant $\Delta$ is given by \eqq{\Delta = -16\left(4f^3+27g^2\right).}
Tuning $f, g$ to vanish to various orders over  curves (divisors) and
points in the base thus gives singularities in the fibration.  The
codimension one
singularities were classified by Kodaira \cite{Kodaira}; when the
orders of vanishing of $(f, g)$ do not exceed $(4, 6)$, the
singularity is associated with a Dynkin diagram and
can be resolved to give a smooth elliptic Calabi-Yau threefold.
A key insight of F-theory is that the Dynkin diagram associated with
the singularity corresponds to the Lie algebra of the gauge group in the
associated 6-dimensional supergravity theory \cite{Vafa-F-theory, Morrison-Vafa-I,
Morrison-Vafa-II, BershadskyEtAlSingularities}. 

Over any base $B_2$, there are a finite number of distinct topological
types of elliptic Calabi--Yau threefolds that can be ``tuned''
by choosing special values of the Weierstrass coefficients $f, g$
\cite{Gross,Kumar:2010ru}.
Roughly, each such tuning corresponds to an algebraic subvariety of
the full moduli space of Weierstrass models, and by the Hilbert Basis
Theorem, there are a finite number of distinct such strata.
In the language of F-theory, which we describe in further detail in
\S\ref{sec:F-theory},
and as reviewed in e.g. \cite{Weigand:2018rez},
various
physical features correspond to increasingly subtle aspects of the
fibration.
Nonabelian gauge groups are associated
with codimension one singularities; to understand these gauge groups
for 6D theories,
one must generalize  the Kodaira classification to include nontrivial monodromy for non-simply
laced groups \cite{BershadskyEtAlSingularities}.
The vanishings that can appear in F-theory, i.e. those
for which the orders of vanishing of $(f, g)$ do not exceed $(4, 6)$,
and their associated gauge algebras are listed in Table
\ref{tab:kodaira}.
Note that for any base that is not weak Fano, where there are curves
of self-intersection $C \cdot C \leq -3$, the curvature of the normal
bundle over such curves forces an automatic Kodaira singularity
associated
in the F-theory picture
with a
geometrically rigid
non-Higgsable gauge group, or ``non-Higgsable cluster''  (NHC)
\cite{Morrison:2012np}.
Codimension
two singularities encode local matter fields, and are relatively well
understood, at least for matter charged under nonabelian gauge fields
(see, e.g., \cite{BershadskyEtAlSingularities,Katz:1996xe,Grassi:2011hq,Klevers:2017aku}).
Global aspects of the fibration, in particular the Mordell-Weil group
and Tate-Shafarevich/Weil-Chatalet groups, correspond to abelian U(1)
and discrete gauge factors, respectively, and are not understood in a
completely systematic way for  either U(1)$^k$
beyond $k = 2$ \cite{Cvetic:2015ioa}, or $\Z_m$ structures (see, e.g.,
\cite{Pioline:2025uov} for recent progress in this direction). 

\begin{table}[]
    \centering
    \begin{tabular}{|c|c|c|c|c|c|} \hline
    Type & Singularity & $\operatorname{ord}(f)$ & $\operatorname{ord}(g)$ & $\operatorname{ord}(\Delta)$ & Gauge algebra \\\hline\hline
         $I_0$ & $\emptyset$ & $\geq0$ & $\geq0$ & 0 & $\emptyset$ \\\hline
         $I_0^*$ &  $D_4$ & $\geq$2 & $\geq$ 3 & 6 & $\mathfrak{so}(8)$, $\mathfrak{so}(7)$, or $\mathfrak{g}_2$ \\\hline
         $I_n$ & $A_{n-1}$ & 0 & 0 & $n\geq2$ & $\mathfrak{su}(n)$ or $\mathfrak{sp}(\lfloor n/2 \rfloor)$ \\\hline
         $I_n^*$ & $D_{n-2}$ & 2 & 3 & $n\geq7$ & $\mathfrak{so}(2n-4)$ or $\mathfrak{so}(2n-5)$ \\\hline
         $II$ & $\emptyset$ & $\geq1$ & 1 & 2 & $\emptyset$\\\hline
         $II^*$ & $E_8$ & $\geq4$ & 5 & 10 & $\mathfrak{e}_8$\\\hline
         $III$ & $A_1$ & 1 & $\geq2$ & 3 & $\mathfrak{su}(2)$ \\\hline
         $III^*$ & $E_7$ & 3 & $\geq5$ & 9 & $\mathfrak{e}_7$ \\\hline
         $IV$ & $A_2$ & $\geq2$ & 2 & 4 & $\mathfrak{su}(2)$ or $\mathfrak{su}(3)$\\\hline
         $IV^*$ & $E_6$ & $\geq3$ & 4 & 8 & $\mathfrak{e}_6$ or $\mathfrak{f}_4$\\\hline

    \end{tabular}
    \caption{The Kodaira classification of singularities in elliptic
      fibrations. For each Kodaira class, we give the ADE type of the
      singular fiber, the orders to which $f$, $g$, and $\Delta$
      vanish, and the possible nonabelian gauge algebras that the
      singularity can give rise to in F-theory. For the classes with
      more than one gauge algebra, additional monodromy information
      is
      needed to specify a gauge group.}
    \label{tab:kodaira}
\end{table}

\subsection{Toric bases for elliptic Calabi-Yau threefolds}
\label{sec:toric-bases}

We  next discuss the possible bases for toric elliptic fibrations.
From \cite{Grassi, Gross}, it is known that all bases $B_2$ that
support elliptic or genus-one fibered Calabi-Yau threefolds can be
described as
complex 2D projective space $\P^2$,  Hirzebruch surfaces $\F_m$
with $m \leq 12$, blowups of one of these surfaces at one or more
points,
or the Enriques surface.  For the Enriques surface,
the anticanonical class of the base $- K$ is nonzero only up to torsion, and is
of limited interest in this context, so we focus on the other cases.

Starting with $\P^2$ and $\F_m$, which are all toric, one can
systematically blow up toric points in all possible ways in an
iterative fashion, constrained by the condition that a curve
$C$ of
self-intersection $C \cdot C \leq -13$ forces a Kodaira type $(4, 6)$ singularity in
the fibration, which does not have a good resolution.  Thus, bases
containing such curves cannot support smooth elliptic Calabi-Yau
threefolds, and the systematic blowup process must stop when such
curves are encountered.  In \cite{MorrisonTaylorToric}, this approach was used
to construct a comprehensive list of 61,539 toric bases that support elliptic
Calabi-Yau threefolds.
It is natural to expect that this set
corresponds to the set of bases found from toric projections of
elliptic Calabi-Yau threefolds in the Kreuzer-Skarke database, and
indeed we find that this is the case; every base in this list appears
as the base of a fibration in the KS database, and all bases found
from explicit elliptic fibrations in the database are either in this list or a blow-down of a base in the toric list. The singular cases with blowdowns are discussed further in \S\ref{sec:singular}.

Note that in constructing this list, any toric points where the
Weierstrass coefficients $f, g$ are forced to vanish to orders (4, 6)
are blown up; such points can be interpreted in F-theory as
corresponding to 
strongly coupled superconformal field theories (SCFTs) coupled to the
6D supergravity theory
\cite{HeckmanMorrisonVafa,Heckman:2015bfa}.  However, some bases,
particularly those containing curves of self-intersection $-9, -10,
-11$ (such as, e.g., $\F_9, \F_{10},\F_{11}$) contain non-toric (4, 6)
points in generic Weierstrass models over these bases.  We will
discuss similar situations further in \S\ref{sec:SCFT}, where the SCFT
appears only for certain fibrations over a base that does not force
(4, 6) points in generic Weierstrass models; this can occur even for
the simplest base $\P^2$.
In all these cases, blowing up the (4, 6) point on the base,
corresponding to going onto the tensor branch of the SCFT, gives a
fibration over a non-toric base without the SCFT, with $h^{1,1} (B)$
increasing by one for each such blowup.
A smooth elliptic Calabi-Yau threefold over a toric base with a (4, 6)
point in the fibration structure generally has a non-flat fiber over
this point
\cite{
Intriligator:1997pq,
Candelas:2000nc,
Braun:2011ux,
Lawrie:2012gg,
Braun:2013nqa,
Cvetic:2013uta,
Dierigl:2018nlv,
Apruzzi:2018nre,
Apruzzi:2019opn}; blowing up the base corresponds to a flop in the Calabi-Yau
threefold that leaves $h^{1,1} (X)$ unchanged.  When the SCFT is forced by the
existence of a $-9, -10,$ or $-11$ curve on the base, we can think of
it as a ``non-Higgsable SCFT,'' or NHSCFT, and 
it will be
helpful in some situations to include the number of
tensors in the SCFT in the effective Hodge number of the base:
\begin{equation}
h^{1,1}_*(B) = h^{1,1} (B)+ {\rm rk(NHSCFT)}\,,
\label{eq:h11s}
\end{equation}
where each $-9, -10,
-11$ curve in the base contributes 3, 2, 1, respectively, to rk(NHSCFT).

While there is not yet a complete list of all possible non-toric bases
$B_2$ that support elliptic or genus-one fibered Calabi-Yau threefolds, some progress has
been made in this direction.  In \cite{MartiniTaylorSemitoric}, a
complete list of 162,404 semi-toric bases, i.e. bases that are invariant under a single $\C^*$ action, was compiled.  Moreover, in
\cite{TaylorWangNon-toric}, Wang and one of the present authors
systematically identified all non-toric bases where either $h^{1,1}
(B_2) <8$ or the base supports an elliptic Calabi-Yau threefold $X$
with $h^{2,1} (X) \geq 150$.  Completing the classification of
non-toric bases presents some technical complications that provide an
interesting challenge for further research.  As discussed below, while our analysis is based on toric structures, resolving singularities in these toric fibrations can provide insight into the structure of some non-toric bases.

\subsection{Toric fibers}
\label{sec:toric-fibers}

Next we proceed to discuss toric fibers. In the fibrations we consider here, the fibers will be hypersurfaces in toric varieties, and so are defined by two-dimensional reflexive polytopes. 
To understand the structure of the fibers more explicitly,
it is helpful
to review the basics of the construction of  Calabi-Yau varieties as
anticanonical
hypersurfaces in toric varieties.
For a more detailed review of toric geometry and this construction,
see \cite{Skarke:1998,Huang-WT-2018-Long}.
Briefly, given a reflexive polytope $\nabla$ in dimension $d +1$, and an associated fan
giving a  compact ambient toric
variety $V$, the adjunction formula guarantees that an
anticanonical hypersurface in $V$ gives a
Calabi-Yau variety of
dimension $d$.  This reduces the construction of a large class of
Calabi-Yau varieties to a combinatorial problem \cite{Batyrev:1993oya,Kreuzer:2000xy}.  Specifically,
the defining polynomial $P_\nabla$ of a Calabi-Yau hypersurface in a
reflexive polytope $\nabla$ is given in terms of its points $p_i$, as
well as the points $q_j$ of the polar dual polytope $\nabla^\circ$, as
\eqq{P_\nabla = \sum_{q_j \in \nabla^\circ} a_j\prod_{p_i}
  x_i^{<p_i,q_j>+1},\label{eq:definingPolynomial}}
 where the coefficients $a_j$ are generically
 complex, the $x_i$ are formal
 (redundant) quasi-homogeneous
 variables whose vanishing loci are the
prime toric divisors, and the product is taken only over points on the
boundary of $\nabla$ not interior to its facets.

Just as four dimensional reflexive polytopes give rise to Calabi-Yau
threefolds, two dimensional reflexive polytopes give rise to
Calabi-Yau one-folds, i.e. complex tori. There are sixteen 2D
reflexive polytopes; we show these 16 polytopes, in the coordinates we
use throughout the paper, in Figure \ref{fig:16fibers}.
Hypersurfaces in these 16 polytopes correspond to
 16 distinct fiber types, which were systematically analyzed in
\cite{Klevers:2014bqa}.

\input{fibers_tikz}

In this paper, we study toric elliptic fibrations, where the fiber is encoded in a 2D reflexive subpolytope of a 4D reflexive polytope. 
Each fiber type can be thought of as a particular class of restricted
Weierstrass models (\ref{eq:Weierstrass}), where $f, g$ are tuned to
take a specific form.\footnote{Technically, for
three of the toric hypersurface fiber types, the fibration does not have a
section and is a genus-one fibration; in these cases there is a
closely related Weierstrass model associated with the Jacobian
fibration, as discussed further in \S\ref{sec:genus-1}, \S\ref{sec:Nagell}.}  Thus, these
16 fiber types can be thought of as describing particular subclasses
of the full finite set of elliptic fibration types over any given base.
  Each of these subclasses generically has certain kinds of
singularities and global structures associated with
  nonabelian, abelian, and discrete gauge groups in the F-theory
  context.
One of the goals of this study is to understand what range of such
structures arise naturally in this simple class of constructions, and
to understand what more exotic features might go outside this class of constructions.

\subsection{Toric elliptic fibrations}
\label{sec:toric-fibrations}

We are interested here in understanding elliptic fibrations that are
explicit in the structure of a toric hypersurface Calabi-Yau
threefold through torically fibered polytopes
\cite{Avram:1996pj,Kreuzer:1997zg,Skarke:1998}.
A 4D reflexive polytope $\nabla$ is fibered by a 2D reflexive
subpolytope $\nabla_2$ when $\nabla_2$ is contained in $\nabla$ along a 2D
sublattice containing the origin.
It was argued in \cite{Huang-WT-fibers}
that any polytope fibration of this kind can be associated with at
least one triangulation giving a  toric cone structure  that is
compatible with the toric morphism that projects out the 2D reflexive
subpolytope, and hence provides an elliptic fibration structure for
the resulting Calabi-Yau threefold.  Note that when the toric base
onto which this fibration is projected contains curves of
self-intersection -3 or below, this triangulation is in general not a
standard (i.e.\ fine, regular, and star) triangulation, but rather in the class
studied in \cite{Berglund:2016nvh},
i.e. of a type that is often referred to as a ``vex'' triangulation.
The standard analysis of Batyrev \cite{Batyrev:1993oya} generalizes straightforwardly to this kind of triangulation \cite{MacFadden:2025ssx}.
Note, however, that in most of the analysis of this paper, we do not
deal with explicit triangulations; rather, we reformulate the
hypersurface defining equation (\ref{eq:definingPolynomial}) in terms
of a Weierstrass model, relying implicitly on the existence of a
triangulation compatible with the toric morphism.

The relationship between toric structure and Weierstrass model is
easiest to illustrate in the case of a ``standard stacking'' with the
fiber $\P^{2, 3, 1}$ (fiber $F_{10}$), associated with the generic
elliptic fibration over a given base $B_2$,  
where $B_2$ is described as
a 2D toric
variety having rays $v_i$.
The details of this correspondence and further references are
spelled out in detail in \cite{Candelas:1997eh,Skarke:1998,Huang-WT-2018-Long}; we summarize the story very briefly here.
In the simplest case, where the base is
weak Fano (and therefore also one of the 16 polytopes in
Figure~\ref{fig:16fibers}), the polytope is defined by the vertices
(1, 0; 0, 0), (0, 1; 0, 0) and $\{ (-2, -3; v_i), v_i \in B_2\}$.
In this situation, the defining equation (\ref{eq:definingPolynomial})
becomes the long Weierstrass form, sometimes also called
\textit{Tate form},
\eqq{y^2 + a_1xyz + a_3yz^3 = x^3 + a_2 x^2z^2 + a_4 xz^4 +
  a_6z^6,\label{eq:longWeierstrass}}
 where $x,y,z$ are  coordinates on $\bbP^{2,3,1}$ and the $a_i$ are
 determined by the sets of monomials $(j, k; m)$ in the dual polytope
 $\nabla^\circ$; for instance, points $(0, 0; m)$
 correspond to monomials in $a_1$, points
 $(1, -1; m)$
 correspond to monomials in $a_2$, and so on.
 Note that in this correspondence, a key insight is that the existence
 of a subpolytope $\nabla_2 \subset \nabla$ corresponds in the dual
 polytope
 $\nabla^\circ$
 to a projection to the 2D polytope that is dual to the fiber, $\nabla^\circ \rightarrow \nabla^\circ_2$.  Building on this basic construction,
related polytopes associated with tuned gauge group factors can be
realized by removing monomials from the $a_i$ so that these line
bundle sections vanish to various degrees over the base toric
divisors (so-called ``Tate tunings''), tables of which with increasing
precision can be found in
\cite{BershadskyEtAlSingularities,Grassi:2011hq,Huang-WT-2018-Long}; this corresponds in the original polytope $\nabla$ to adding
points associated with a ``top'' characterizing the associated gauge
group.  
Similarly, for non-weak Fano bases, there is an analogous procedure but to get
a reflexive polytope one must take the ``dual of the dual'' of this
construction,\footnote{To be precise, stacking $F_{10}$ over a
non-Fano base yields a non-reflexive polytope, whose dual contains
rational (rather than integral) points; taking the dual of only the
integral points yields a reflexive
enhancement
of the initial polytope,
which is the correct analogue of the standard stacking for
non-weak-Fano bases.}
which automatically includes additional points in
$\nabla$ associated with non-Higgsable (rigid) gauge factors \cite{Morrison:2012np} lying
over curves in the base of self-intersection
-3 or below.

Note that the
(when necessary, dual of the dual of the)
``standard stacking'' with fiber $F_{10}$, i.e. a fibration where the base is stacked over the point $(-2,-3)$ of the fiber, in general
gives the generic elliptic fibration over a given base.  A fibration
with a different fiber or different $F_{10}$ structure
will in general involve tuning additional nonabelian or abelian gauge factors.
Thus, for
bases with large $h^{1,1} (B_2)$ and many  toric curves of self-intersection
$C \cdot C \ll -2$ that support large rigid/non-Higgsable gauge factors,
where such tunings are difficult or impossible, we expect standard
stacking $F_{10}$ fibrations to dominate.

Nonabelian gauge factors in standard stackings of $F_{10}$ are well
understood through Tate tunings or equivalently through the study of
polytope tops. However, for other fibers besides $F_{10}$, or indeed
for nonstandard stackings of $F_{10}$, these analyses fall short. To
use tops, for instance, one would need to use distinct tops for each
fiber type; some progress towards such a geometric analysis of tops for
each of the 16 fiber types is given in
\cite{Bouchard:2003bu}.
Instead of working explicitly with the geometry of tops, we find it convenient to use a more unified
framework, in which the different fiber types can all be analyzed with
just the Kodaira classification in Table \ref{tab:kodaira}. We turn to
this more detailed analysis in \S\ref{sec:Nagell} after first briefly
discussing genus-one fibrations.

\subsection{Genus-one, elliptic, and Jacobian fibrations}
\label{sec:genus-1}

The existence of a section in a fibration by fiber $F_n$ can
immediately be seen from the presence of a curve $C$ of
self-intersection $C \cdot C = -1$ in the fiber.\footnote{This follows
from the Riemann-Roch formula for surfaces: $- K \cdot C = C \cdot C
+2 = 1$, so the anticanonical hypersurface contains a point in each fiber.}  Torically, such a curve is realized by a ray that is the sum of the neighboring rays.
From this, it is easy to see that all fibers other than $F_1, F_2,$
and $F_4$ give an elliptic fibration with at least one section.  For
these three fibers, however, there is not a global section, rather
there is a $k$-fold {\it  multi-section}.  For example, following the
same logic as above for $F_{10}$, the defining equation for a
Calabi-Yau threefold arising from a toric variety with a toric fiber
$F_1$ is a general cubic, which is well known to have a 3-section,
associated with 3 rational points in each fiber that may be connected
by nontrivial monodromy around the base.  Such a fibration is called a
{\it  genus-one} fibration, since the fiber is still geometrically a
torus (i.e., a genus-one Riemann surface), but lacks the single
rational section needed to make a consistent fibration by elliptic
curves with related marked points.
The fibers $F_2$ and $F_4$ generically provide 2-sections.

Genus-one fibrations with $k$-sections are related in F-theory to discrete
abelian gauge groups.  In general, genus-one fibrations with
$k$-sections live in Tate-Shafarevich or Weil-Chatalet groups $\Z_k$
that share a common $\tau$ function describing the fibration locally,
but with different monodromy structure; the {\it Jacobian} fibration,
corresponding to the identity element of the group, is a true elliptic
fibration.   F-theory on the Jacobian fibration gives a 6D
supergravity theory with gauge group $\Z_k$; compactification of this
theory on a circle $S^1$ with a nontrivial Wilson line gives a 5D
theory that can be described as M-theory compactified on the other
elements of the Tate-Shafarevich group \cite{Braun:2014oya, Morrison:2014era}.
  To understand general elliptic and genus-one fibered Calabi-Yau
  threefolds
with any fiber $F_n$ in a parallel fashion to the description above of
$F_{10}$ standard stackings  with tops and Tate models, we follow a
standard approach to the construction of Weierstrass models for the
Jacobian fibration of any genus-one or elliptically fibered CY3.

\subsection{Weierstrass models and Nagell's algorithm}
\label{sec:Nagell}

Any elliptic curve $E$ over a field $K$ can be written in \textit{long
  Weierstrass form} (\ref{eq:longWeierstrass}).
If the characteristic of $K$ is not 2 or 3\footnote{We work in characteristic zero throughout. For recent work on F-theory and related string vacua in positive characteristic, see e.g. \cite{Candelas:2019llw,Kachru:2020sio,Kachru:2020abh,Candelas:2023yrg}.}, we can complete the square in $x$ and complete the cube in $y$ to pass to \textit{(short) Weierstrass form}, \eqq{y^2 = x^3 + fxz^4 + gz^6,\label{eq:weierstrass}} where $f,g$ are given in terms of the $a_i$ by \begin{subequations}
\label{eq:shortWeierstrassInTermsOfTate}
\begin{align}
b_2 &= a_1^2+4a_2\\
b_4 &= a_1a_3+2a_4\\
b_6 &= a_2^3+4a_6\\
b_8 &= b_2a_6-a_1a_3a_4+a_2a_3^2-a_4^2\\
f &= -\frac{1}{48}\left(b_2^2-24b_4\right)\\
g &= -\frac{1}{864}\left(-b_2^3+36b_2b_4-216b_6\right).
\end{align}
\end{subequations}
Note that the map from long Weierstrass form to short Weierstrass form
is many-to-one, and hence is generically not invertible. Furthermore,
note that the point $(1 : 1: 0)$, which (largely for historical
reasons) is usually called the point at infinity, satisfies
Eq.~(\ref{eq:weierstrass}) for any $f,g$. Thus, $E$ must be an
elliptic, rather than a genus-one, curve to be put into Weierstrass
form. The complex structure of $E$  
is encoded in its $j$-invariant
and discriminant $\Delta$, which are given in terms of the Weierstrass
coefficients $f,g$ as \begin{subequations}
    \begin{align}
        \Delta &= -16\left(4f^3+27g^2\right) \label{eq:Delta}\\
        j &= 1728 \frac{4f^3}{4f^3+27g^2} \label{eq:j}.
    \end{align}
\end{subequations}
There exists an algorithm, called Nagell's algorithm, for rewriting an
arbitrary elliptic curve $E$, equipped with a zero point $p_0$, in
Weierstrass form; this algorithm is implemented in e.g. Sage
\cite{sagemath}.

The same procedure of going from an arbitrary defining polynomial to short Weierstrass form can be carried out for an elliptic fibration (with section); this is because elliptic fibrations can equivalently be viewed as elliptic curves whose field of definition is the function field of the base $B_2$. 
The basic idea, as described succinctly in \cite{Braun:2014oya}, following Deligne, is that given a rational section $z$ of ${\cal  O}(-
K_{B_2})$, one can use the Riemann-Roch theorem to compute the numbers
of independent sections of multiples of $- K_{B_2}$;  the number of polynomials in the generating sections at 
degrees below 6 exceed the number of sections  at this degree, and a
linear relation follows, which is the Weierstrass equation.
Nagell's algorithm depends crucially on the choice of
$p_0$, and hence does not apply if $E$ does not have a point over $K$,
i.e. if $E$ is a genus-zero curve rather than an elliptic curve;
however, in the genus-one context, it is quite natural to instead
apply Nagell's algorithm to the Jacobian $J(E)$.
This is done explicitly in \cite{an2001jacobians} for  the cases of
2-, 3-, and 4-sections.
To analyze fibrations
of the 16 toric fibers in a uniform way, therefore, one can
use Nagell's algorithm to write the defining polynomial of the
fibration in Weierstrass form, where the zeroes of the discriminant
can be read off straightforwardly from Eq.~\ref{eq:Delta}. This
strategy was used in \cite{Klevers:2014bqa} to give explicit
Weierstrass forms for elliptic and (the Jacobians of) genus-one fibrations
for all 16 of the 2D toric fiber types.

The most relevant results
of this general analysis are summarized in
Appendix~\ref{sec:Weierstrass}.
We give one specific example here and
outline the application of Nagell's algorithm to
the case of general $F_{10}$ fibrations (i.e., including
nonstandard stackings).
This completes the discussion of \S\ref{sec:toric-fibrations},
describing a fully general toric fibration with fiber $F_{10}$, and
will be useful in describing specific fibrations in which F-theory
gauge groups are tuned over non-toric divisors, as will be studied in \S\ref{sec:non-toric-gauge}.
In a general toric fibration, the defining polynomial of
$F_{10}$ takes a more general form than that given in
Eq.~\ref{eq:longWeierstrass}, namely \eqq{\alpha y^2 + a_1xyz +
  a_3yz^2 = \beta x^3 + a_2x^2z^2 + a_4xz^4 +
  a_6z^6.\label{eq:longWeierstrassWithAlphaBeta}} Nagell's algorithm
allows this to be rewritten in a standard short Weierstrass form as in
Eq.~\ref{eq:weierstrass}, where $f,g$ depend on the coefficients
$\alpha,\beta$ as well as the $a_i$, and are given explicitly in
Eq.~\ref{eq:F10weierstrass}. From the form of $f,g$, we find that
\eqq{\Delta \sim \alpha^3\beta^2 \tilde{\Delta},} where
$\tilde{\Delta}$ is a generic polynomial in
$\alpha,\beta,a_i$. Importantly, neither $\tilde{\Delta}$ nor the
Weierstrass coefficients $f,g$ vanish
automatically
when $\alpha$ or $\beta$
do. Thus, on $\alpha=0$ we have $\ord{(f,g,\Delta)} = (0,0,3)$;
comparing to Table \ref{tab:kodaira}, we see that this corresponds to
an $I_3$ singularity, and accordingly we expect an $SU(3)$ gauge group
whenever $\alpha$ can vanish, i.e. when it consists of more than one
monomial. Similarly, on the $\beta=0$ locus we have
$\ord{(f,g,\Delta)} = (0,0,2)$, and we expect an $SU(2)$ gauge
group. This is the origin of the ``generic'' $SU(2)\times SU(3)$ gauge
group listed for $F_{10}$ in Table \ref{tab:16fibers},
as originally identified in \cite{Klevers:2014bqa}.
Note that, as in the discussion of \S\ref{sec:toric-fibrations}, when
the fiber $F_{10}$ is put in standard form in the first two
coordinates of the polytope, the monomials in $\alpha, \beta$
correspond to points in the dual lattice of the form $(-1, 1; m)$ and $(2,
-1; m)$, respectively.

A similar analysis applies to the remaining fiber types. In general, the set of monomials over  the dual 2D polytope to a given fiber type will
automatically give rise to some additional structure that can include
nonabelian gauge factors, U(1) factors, and discrete factors. 
A full list of the properties of the 16 fiber types,
including whether there is a section and the generic gauge group, is
given in Table~\ref{tab:16fibers}.
\footnote{Thanks to Thorsten Schimannek  for a helpful discussion on
the global structure of the group in the case of fiber $F_4$.}
For more details on the specific Weierstrass models for each fiber,
see
 Appendix~\ref{sec:Weierstrass}.

\begin{table}[h]
\begin{centering}
\begin{tabular}{|c||c|c|c||c|c|c|}\hline
Polytope & Dual & $n_{v}$ & $n_p$ & Section? & Weierstrass model & Gauge group \cite{Klevers:2014bqa} \\\hline \hline
$F_1$ & $F_{16}$ & 3 & 4 & No & \ref{eq:F1weierstrass} & $\bbZ_3$ \\\hline
$F_2$ & $F_{15}$ & 4 & 5 & No & \ref{eq:F2weierstrass} & $U(1)\times\bbZ_2$\\\hline
$F_3$ & $F_{14}$ & 4 & 5 & Yes & \ref{eq:F1weierstrass}, $a_{003}=0$ & $U(1)$ \\\hline
$F_4$ & $F_{13}$ & 3 & 5 & No & \ref{eq:F4weierstrass} & $(SU(2)\times\bbZ_4)/\Z_2$ \\\hline
$F_5$ & $F_{12}$ & 5 & 6 & Yes & \ref{eq:F1weierstrass}, $a_{030} = a_{003} = 0$ & $U(1)^2$ \\\hline
$F_6$ & $F_{11}$ & 4 & 6 & Yes & \ref{eq:F1weierstrass}, $a_{012}=a_{003} = 0$ & $U(1)\times SU(2)$ \\\hline
$F_7$ & $F_{7}$ & 6 & 7 & Yes & \ref{eq:F1weierstrass}, $a_{300}=a_{030}=a_{003} = 0$ & $U(1)^3$ \\\hline
$F_8$ & $F_{8}$ & 4 & 7 & Yes & \ref{eq:F1weierstrass}, $a_{030}=a_{012} = a_{003} = 0$ & $SU(2)^2\times U(1)$ \\\hline
$F_9$ & $F_9$ & 5 & 7 & Yes & \ref{eq:F1weierstrass}, $a_{030}=a_{102} = a_{003}=0$ & $SU(2)\times U(1)^2$\\\hline
$F_{10}$ & $F_{10}$ & 3 & 7 & Yes & \ref{eq:F10weierstrass} & $SU(2)\times SU(3)$ \\\hline
$F_{11}$ & $F_{6}$ & 4 & 8 & Yes & \ref{eq:F10weierstrass}, $a_6$ = 0 & $(SU(2)\times SU(3) \times U(1))/\Z_6$ \\\hline
$F_{12}$ & $F_5$ & 5 & 8 & Yes & \ref{eq:F1weierstrass}, $a_{120}=a_{030}=a_{102}=$ & $SU(2)^2\times U(1)^2$ \\ & & & & & $a_{003}=0$ & \\\hline
$F_{13}$ & $F_{4}$ & 3 & 9 & Yes & \ref{eq:F4weierstrass}, $a_{310}=a_{130}=a_{201}=$ & $SU(4)\times SU(2)^2/\bbZ_2$  \\ & & & & & $a_{021}=0$&\\\hline
$F_{14}$ & $F_{3}$ & 4 & 9 & Yes & \ref{eq:F1weierstrass}, $a_{210}=a_{120}=a_{030}=$ & $SU(3)\times SU(2)^2 \times U(1)$  \\ & & & & & $a_{102}=a_{003}=0$& \\\hline
$F_{15}$ & $F_{2}$ & 4 & 9 & Yes & \ref{eq:F1weierstrass}, $a_{300}=a_{120}=a_{030}=$ & $SU(2)^4/\bbZ_2 \times U(1)$ \\ & & & & & $a_{102}=a_{003}=0$&\\\hline
$F_{16}$ & $F_{1}$ & 3 & 10 & Yes & \ref{eq:F1weierstrass}, $a_{210} = a_{120} = a_{201} =$ & $SU(3)^2/\bbZ_3$ \\
& & & & &  $a_{102} = a_{021} = a_{012} = 0$ & \\\hline
\end{tabular}
\caption{Properties of the 16 two-dimensional reflexive polytopes and
  ``generic'' elliptic fibrations with each fiber. For each polytope,
  we give its dual polytope, the numbers $n_v$ and $n_p$ of vertices
  and points, respectively, whether the generic elliptic fibration
  admits a section or only a multi-section, the Weierstrass model of
  the elliptic fibration, and the generic gauge group, as computed in
  \cite{Klevers:2014bqa}.
Note that some of these gauge groups may have a quotient by a discrete
center not indicated here; we have included some discrete quotient
where known, or where they can be inferred from the matter spectrum.
}
\label{tab:16fibers}
\end{centering}
\end{table}


\subsection{6D F-Theory Models}
\label{sec:F-theory}

We have described so far various aspects of the geometry of
elliptically fibered CY3s.
The physics of F-theory essentially provides a dictionary
that allows us to relate geometric features of elliptic CY3s to
aspects of the 6D supergravity theories that one obtains
by compactifying F-theory on these geometric spaces.
In
this subsection, we will describe this relationship in a little more
detail; for a much more detailed
review, see e.g. \cite{Weigand:2018rez}.
The material in this subsection is particularly relevant for some of the specific examples that we explore in \S\ref{sec:examples}.

F-theory can be thought of as a nonperturbative version of type IIB
string theory, in which the axiodilaton encodes a complex structure
for an auxiliary torus that can vary over the base space.  For
compactification to 6D, we must have a compact complex K\"ahler
surface $B$ (4 real dimensions) with an effective anticanonical class;
supersymmetry imposes the condition that the total space forms an
elliptic fibration over the base $B$, such that the total space of  the fibration is
an elliptic Calabi-Yau threefold $X$.

Physically, such a compactification gives rise to
a 6D supergravity theory with $(1,0)$
supersymmetry.
 The basic building blocks for such a theory are supergravity
 multiplets, which include the gravity multiplet, vector
multiplets, tensor multiplets, and hypermultiplets. There is always
one gravity multiplet, and we  write the
numbers of the other multiplets in the effective theory as $V$, $T$, and
$H$, respectively. As usual in string compactifications, these numbers
are determined by the topology of the compactification manifold. For
instance, in the absence of (4, 6) loci giving SCFT's, the number of tensor multiplets is fixed straightforwardly
by the Hodge numbers of the base: \eqq{T = h^{1,1}(B)-1.}

As described above, the gauge group of the compactification is
determined by the details of the fibration;  the nonabelian gauge group is controlled by the
singular fibers by means of the Kodaira classification in Table
\ref{tab:kodaira} and its refinement to include monodromy data,
and the number of $U(1)$ gauge
factors is given by the rank of the Mordell-Weil group of the
fibration. For a given structure of the (continuous) gauge group $G$, the number $V$ of vector
multiplets is given by the dimension of  $G$: \eqq{V =
  \operatorname{dim}(G).} The gauge group dimension is related to the
topological data of the fibration through the Shioda-Tate-Wazir (STW)
formula \cite{wazir2004arithmetic},
\begin{align}
    h^{1,1}(X) = h^{1,1}(B) + 1 + \operatorname{rk}
    \operatorname{MW(X)}
    + \operatorname{rk} G_{\text{nonabelian}} (+ \operatorname{rk} \operatorname{SCFT}),
    \label{eq:STW}
\end{align}
which, geometrically, states that
a basis of divisors for the threefold can be formed from the divisors
on the base, 
the zero-section,  Mordell-Weil sections, and
the components of fibral divisors giving rise to
Cartan elements of
a nonabelian gauge
group.
Note that we have included in the last term also
contributions for singular fibrations with
(4, 6) loci corresponding to the number of tensors in the associated
SCFT, generalizing the usual STW formula for smooth Calabi-Yau
manifolds; in the flop phase where these points in the base are blown up these are
incorporated into $h^{1,1} (B')$ for the blown up base, and otherwise this
contribution corresponds to non-flat fibers.

We next turn to the matter content of the theory, which is controlled
by codimension two singularities in the elliptic fibration, as
mentioned above (and in some cases by the topology of the divisor
supporting the nonabelian gauge group).   We can classify the
hypermultiplets of the theory by whether they are charged or neutral
under the gauge group. The total number $H$ of hypermultiplets and the
numbers $H_{\text{neutral}}$ and $H_{\text{charged}}$ of neutral and
charged hypermultiplets, respectively, are controlled by the
topological data of the fibration and the gauge group by a powerful set
of
gravitational, gauge-gravitational, and pure gauge
anomaly cancellation constraints \cite{GreenSchwarzWest6DAnom}.  The simplest constraints are those
that involve only the numbers of charged and uncharged matter fields: \begin{subequations}
    \begin{align}
        h^{2,1}(X) &= H_{\text{neutral}} - 1 \label{eq:h21=Hneutral-1} \\
        H - V &= 273 - 29T \label{eq:HminusV} \\
         h^{2,1}(X) &= 272 + V - 29T - H_{\text{charged}} \label{eq:272}.
    \end{align}
\end{subequations}
The anomaly constraints on charged matter fields are slightly more
intricate, and play a key role in the F-theory/geometry dictionary.
For a simple
nonabelian gauge group $G$, these can be written as
\cite{Kumar:2010ru}
\begin{subequations}
\begin{align}
a \cdot a &= 9 - T\,, \label{eq:nonabelACa} \\
a \cdot b &= -\frac{1}{6} \lambda
	\left(\sum_{R} x_{R} A_{R}
		- A_{\Adj}\right)\,,
	\label{eq:nonabelACA} \\
0 &= \sum_{R} x_{R} B_{R}- B_{\Adj}\,,
	\label{eq:nonabelACB} \\
b \cdot b &= \frac{1}{3} \lambda^2
	\left(\sum_{R} x_{R} C_{R}
		- C_{\Adj}\right)\,,
	\label{eq:nonabelACC} 
\end{align}
\end{subequations}
where $x_R$ is the number of matter hypermultiplets transforming in
representation $R$ of $G$, $\lambda$ is a constant depending on the
type of gauge factor (e.g. $\lambda = 1$ when $G = SU(N)$), and
 the group theory coefficients $A_R$, $B_R$, and $C_R$ are
defined by
\begin{equation}
\trace_R F^2 = A_R \trace F^2\,,
	\quad \trace_R F^4 = B_R \trace F^4 + C_R \left(\trace F^2\right)^2\,.
\end{equation}
A key aspect of the physics-geometry dictionary of F-theory is that
the anomaly coefficients
$a, b$, which are certain terms appearing in the 6D supergravity
action, are mapped one-to-one to the divisor classes $K, \Sigma$ in
the F-theory base $B$ corresponding to the canonical class and the
divisor class supporting the gauge factor $G$, respectively;
the inner product here
becomes the intersection product in the F-theory base.
For semi-simple groups, similar equations hold, with separate anomaly
coefficients $b_i$ for each simple factor $G_i$ of $ (G =\prod_i
G_i)/\Gamma$.

For a given geometry, these anomaly equations are almost completely
sufficient to determine the matter content of a given theory.  For a
given gauge factor $G$, knowing the anomaly coefficients $a, b$ fixes
a unique set of {\it generic} \cite{BershadskyEtAlSingularities,Taylor:2019ots} matter representations for
hypermultiplet matter fields in the theory.  For example, for SU($N$),
generic matter fields are those in the fundamental, two-index
antisymmetric, and adjoint representations.
Similarly, generic matter charged under a pair of gauge factors
$G_i \times G_j$
arises at the intersection point between the divisors $b_i, b_j$, such
as SU($N$) $\times$ SU($M$) bifundamental fields $(\mathbf{N}, \bar{\mathbf{M}})$.
While more exotic
representations can arise in special contexts (see, e.g.,
\cite{Klevers:2017aku}), in this paper we encounter only constructions
with generic matter content.  Tables of the generic matter fields
associated with specific gauge group factors
(and pairs of factors)
on base curves with given
intersection numbers can be found in, e.g., \cite{Johnson:2016qar}.

For abelian factors, a similar set of anomaly equations to those above
constrains the matter content (see, e.g., \cite{Erler6DAnom,Park:2011wv,ParkAnomalies}); in this case, for a single U(1)
factor, generic matter has charges $\pm 1$ and $\pm 2$.  Abelian
factors have an anomaly coefficient $b$ that plays an analogous role
to that of nonabelian factors; for multiple abelian factors, this
becomes a two-index object $b_{i j}$. In the cases we encounter here, where
the abelian factors arise from sections associated with -1 curves in
the fiber, the anomaly coefficient for U(1) factors is essentially the canonical class
$K$, and we can  think of the matter content as simply that of a
U(1) factor that arises from an SU(2) factor tuned on the genus-one
divisor  $- K$, broken on the single associated adjoint field.

Finally, there can be a discrete gauge group in the theory.  This is
harder to see directly from the structure of an elliptic CY3,
but ties into the role of genus-one fibered Calabi-Yau threefolds.
Physically, as discussed above, genus-one fibered threefolds have a
multisection, and live in a group $\Z_k$ containing the Jacobian
fibration, a genuine
elliptic fibration; they are associated with the geometry of F-theory
models compactified to 5D with a Wilson line that breaks the discrete
gauge group.
We will not go deeply into this structure here, but see
\cite{Braun:2014oya,Morrison:2014era,Anderson:2023wkr,Anderson:2023tfy}
for more on this subject.

\section{Fiber and Base Analysis}
\label{sec:algorithm}

In this section we give an overview of the methods
and classification schemes
used here for analyzing
the toric fibration structures in the KS database.
Subsection~\ref{sec:identifying-fibers} describes the algorithm for
identifying fibers and bases, subsection~\ref{sec:automorphisms} describes the
methodology for identifying fibrations that are equivalent under
automorphisms of the ambient polytope,
subsection~\ref{sec:implementation} gives some details on the
implementation and computation, and in subsection \ref{sec:organization} we explain the format of our ancillary data files.

\subsection{Overview of the algorithm for identifying fibers and bases}
\label{sec:identifying-fibers}

Let $L = \nabla \cap \bbZ^4$ be the set of lattice points of a 4D
reflexive polytope, and $V$ be the set of vertices of the dual polytope $\nabla^\circ$ given by:
\begin{align*}
    \nabla^\circ = \{ w:\langle w, p \rangle \geq -1 \ \forall p \in \nabla\}
\end{align*}
Following the approach developed in
\cite{Huang-WT-fibers},
for each point $p \in L$, we define  the function $\displaystyle s(p) = \text{max}_{v \in V} \langle p, v \rangle$ and define sets $S_i = \{p \in L : s(p) = i \}$. Then,  $\nabla$ contains a 2D reflexive subpolytope
if and only if it satisfies one of the following conditions:
\begin{eqnarray*}
    (1) && \exists x, y \in S_1 : x\neq -y \text{ and}\ -x, -y \in S_1\\
    (2) &&\exists x, y \in S_2 : -(x+y) \in S_1 \text{ or}\ -(x+y) \in S_2 \\
    (3) && \exists x, y \in S_3 : -\frac{(x+y)}{2} \in S_1
\end{eqnarray*}

In order to find subpolytopes, we look for a plane spanned by two rays
$x$ and $y$ which satisfy one of the above conditions.
For any such $x$ and $y$, we can then solve a set
of linear equations to find all points of $\nabla$ that are linear
combinations of $x$ and $y$. This yields all points in the subpolytope. To
identify which specific fiber is realized by each such
subpolytope, we can use the sets
$S_i$. The sizes of the sets $S_1, S_2, S_3$, calculated in Table~\ref{tab:ssets}, distinguish all
2D reflexive polytopes except fibers 6 and 10
from Figure~\ref{fig:16fibers}.
But these polytopes have different numbers of points
 and since we have all the points in the subpolytope,
a quick check on the number of points allows us to completely identify
all the different fiber types.

\begin{table}[]
    \centering
\begin{tabular}{|c|c||@{\hspace{2em}}||c|c|} \hline
Polytope & $(|S_1|, |S_2|, |S_3|)$ & Polytope & $(|S_1|, |S_2|, |S_3|)$ \\
\hline\hline 
$F_1$ & $(0, 3, 0)$ & $F_9$ & $(4, 2, 0)$ \\\cline{1-2}\cline{3-4}
$F_2$ & $(4, 0, 0)$ & $F_{10}$ & $(2, 2, 1)$ \\\cline{1-2}\cline{3-4}
$F_3$ & $(2, 2, 0)$ & $F_{11}$ & $(4, 2, 1)$ \\\cline{1-2}\cline{3-4}
$F_4$ & $(2, 0, 2)$ & $F_{12}$ & $(6, 1, 0)$ \\\cline{1-2}\cline{3-4}
$F_5$ & $(4, 1, 0)$ & $F_{13}$ & $(4, 2, 2)$ \\\cline{1-2}\cline{3-4}
$F_6$ & $(2, 2, 1)$ & $F_{14}$ & $(6, 2, 0)$ \\\cline{1-2}\cline{3-4}
$F_7$ & $(6, 0, 0)$ & $F_{15}$ & $(8, 0, 0)$ \\\cline{1-2}\cline{3-4}
$F_8$ & $(4, 1, 1)$ & $F_{16}$ & $(6, 3, 0)$ \\\cline{1-2}\cline{3-4}
\end{tabular}
    \caption{The sizes of sets $|S_i|$ used to identify each fiber \cite{Huang-WT-fibers}.}
    \label{tab:ssets}
\end{table}

For a given polytope, we can scan over all pairs of rays $x, y$ to
identify all possible fibers.  Each fiber is one of the 16 toric
fibers reviewed in \S\ref{sec:toric-fibers}.  For each distinct fiber,
we then wish to identify the corresponding base.

To find the base, we want to project out the 2D reflexive
subpolytope. Let $x, y$ be two rays that span the fiber. We find
a $GL(4, \bbZ)$ transformation $M$ such that $Mx \to (1, 0, 0,
0)$. Finding this transformation requires repeated applications of the
Euclidean algorithm. In this basis, projecting onto the latter three
coordinates is equivalent to throwing away one dimension of the
subpolytope. We repeat this procedure one more time. Starting with a
3D polytope, we find a $GL(3, \bbZ)$ transformation $N$ such that
$N\tilde{y} \to (1, 0, 0)$ where $\tilde{y}$ is the projection of the
vector $y$ in the 3D polytope. Projecting onto the final two
coordinates returns the base rays.

This projection gives a set of 2D  rays from which we can identify the
base.
First, we reduce any non-primitive rays that are multiples of an
integer 2D vector to the associated primitive ray; this gives a set of
primitive rays $v_i = (x_i, y_i), i = 1, \ldots, k$.  Next, we order the
rays by angle and check to confirm that the base is smooth by
checking whether $ x_i y_{i +1}- x_{i +1} y_i= 1$ (imposing
cyclicity, so  subscripts are taken mod $k$).
When the base  constructed from the projected rays is not smooth, we
identify an associated smooth base by adding all rays in the convex hull.

We identify the resulting smooth base using the set of intersection numbers of the
toric divisors. After ordering the rays in the convex hull
by angle, we calculate the
sequence of intersection numbers $(c_1, \ldots, c_k)$ using the
equation (mod $k$): 
\begin{equation}
    c_i (x_i, y_i) + (x_{i -1}, y_{i -1}) + (x_{i +1}, y_{i +1}) = 0 
\end{equation}
In all cases, the resulting sequence of intersection numbers
corresponds precisely to one of the Morrison-Taylor (MT) bases identified in
\cite{MorrisonTaylorToric}.

This algorithm gives the maximal toric base for an elliptic CY3
associated with a triangulation of  $\nabla$.  In many cases, there
are also triangulations corresponding to flop phases in which
fewer base divisors are compatible with the toric morphism, and there
is thus a non-flat fibration to a base with smaller $h^{1,1}
(B)$.\footnote{For example, the polytope associated with the standard
stacking over Hirzebruch $\F_1$ has a triangulation giving the generic
elliptic CY3 over $\F_1$, but there is also a triangulation giving a
non-flat fibration over $\P^2$ \cite{rw-forthcoming}.}  In
general, we find it useful to organize fibrations using the phase
where the base has maximal $h^{1,1} (B)$; this is also part of the
reason that it is useful to track $h^{1,1}_*(B)$, as defined in
(\ref{eq:h11s}), which extends this consideration to flop phases where
non-flat fibers automatically forced from the structure of the base are traded for base divisors, but
the base is no longer toric.

The result of this analysis for a given polytope in the KS database is a list of
fibrations, where each fibration can be identified by a fiber with
fiber ID 1--16 (as summarized in \S\ref{sec:toric-fibers}), and a base
with base ID 1--61,539 (as summarized in \S\ref{sec:toric-bases}),
where in some cases a subset of the base divisors have been blown
down, indicated by a list of true-false values describing the set of
base divisors that are still present.

\subsection{Identifying equivalence of fibrations under polytope automorphisms}
\label{sec:automorphisms}
In general, a polytope $\nabla$ might enjoy nontrivial
\textit{automorphisms}, i.e. linear transformations
$\Lambda\in\operatorname{SL}(4,\bbZ)$ that map the vertices of
$\nabla$ into each other. The automorphisms of $\nabla$ form a group,
$\operatorname{Aut}(\nabla),$ whose elements can be efficiently
computed in e.g. CYTools \cite{Demirtas:2022hqf}. For instance, it is
straightforward to verify that the polytope whose vertices are given
by the columns of the matrix \eqq{\left(\begin{array}{cccccc}
    1&0&-2&0&0&0\\ 0&1&-3&0&0&0\\ 0&0&0&1&0&-2\\ 0&0&0&0&1&-3
  \end{array}\right),\label{eq:autExample}}
i.e. the combination of two copies of fiber $F_{10}$ on orthogonal two-planes,
has a $\bbZ_2$ group of automorphisms generated by exchanging the axes pairs containing the two
$F_{10}$ fibers.  \ignore{, consisting of the identity matrix and the
  nontrivial automorphism \eqq{\Lambda = \left(\begin{array}{cccc}
      0&0&1&0\\ 0&0&0&1\\ 1&0&0&0
      \\0&1&0&0\end{array}\right),\label{eq:autExample2}} which acts
  by permuting the two copies of $\bbP^2$.}  While the computation of
polytope automorphisms is by now standard, to the best of our
knowledge, no systematic study of their properties has appeared in the
literature. Some statistics of polytope automorphisms which we
collected in the course of our work and which we found helpful to its
completion are presented in Appendix \ref{sec:autostats}.

If $\nabla$ has nontrivial automorphisms and contains multiple 2D reflexive subpolytopes, in general the subpolytopes might be mapped into each other by automorphisms of the ambient polytope. For instance, the polytope given in Eq.~\ref{eq:autExample} contains  
(along with an additional $F_2$ fiber)
two $F_{10}$ subpolytopes, which by construction are mapped into each other by the polytope's unique nontrivial automorphism. We call two subpolytopes \textit{automorphism equivalent} if there exists an automorphism of the ambient polytope that exchanges them, and \textit{automorphism inequivalent} otherwise. 

Not all pairs of two subpolytopes can even potentially be mapped into
each other by automorphisms; clearly, two subpolytopes can only be
automorphism-equivalent if they have the same fiber type. In fact,
more is true; two subpolytopes can only be automorphism equivalent if
their associated fibrations have the same base as well. In what
follows, it will be convenient to consider \textit{fibration
  structures}, groupings of subpolytopes with the same fiber type and
base type that can potentially be automorphism equivalent. Note that
in general, different subpolytopes of a fixed polytope whose fiber
type and base type are the same, i.e. different representatives of the
same fibration structure, need not be automorphism equivalent. We
therefore have the bounds \eqq{\#\left(\text{fibration
    structures}\right) \le \#\left(\text{automorphism inequivalent
    fibrations}\right) \le \#\left(\text{total fibrations}\right).}

Fix a 4D reflexive polytope $\nabla$. For fibrations with a smooth
base, the data defining a fibration structure are simply the fiber
type, in the conventions of Table \ref{tab:16fibers}, and the base
type, i.e. the member of the MT list
\cite{MorrisonTaylorToric} with the same toric intersection
structure as the base
of the fibration. For fibrations with singular bases, in addition to
the topological type of the resolved base, we also include a
classification of the curves which are blown down to yield the
singular base. Thus, in total, a fibration structure is identified by
five pieces of information: the ambient polytope, the fiber type, the
base type, whether the base is smooth, and a (possibly empty) list of
curves that have been blown down.\footnote{An additional complication
arises when the base of the fibration is singular and the resolved
base itself has nontrivial automorphisms, as then the specification of
blowdowns becomes ambiguous. In this rare occurrence, we resolve this
ambiguity by selecting the lexicographically minimal representative of
the specification of blown down curves under the base automorphisms.}

Geometrically, automorphism-equivalent fibrations should be
identified. Furthermore, from the physics perspective, as emphasized
in \cite{Braun:2011ux}, because compactifying F-theory on elliptic
fibrations defined by automorphism equivalent subpolytopes gives
identical physical theories, it is also natural
in this context
to count the number of
automorphism equivalence classes of fibrations, rather than the total
number of fibrations. Thus, given a 4D reflexive subpolytope $\nabla$,
once we have identified the fibers and bases of all of its elliptic
fibrations using the algorithm above, we also compute the
automorphisms of $\nabla$ and check whether any subpolytopes are
mapped into each other by those automorphisms. In practice, we first
sort the subpolytopes into fibration structures, and only compute
automorphisms if at least one fibration structure has more than one
representative.

\subsection{Computational details}
\label{sec:implementation}
In the last two subsections, we have spelled out an algorithm for
completely classifying the automorphism-inequivalent toric elliptic
fibrations of a 4D reflexive polytope $\nabla$ from the Kreuzer-Skarke
list. We have applied this algorithm, which we have implemented in
both CYTools \cite{Demirtas:2022hqf} and Julia
\cite{bezanson2017julia}, to all 473,800,776 4D reflexive polytopes to
obtain a complete classification of all toric elliptic
fibrations. Before presenting the results of our analysis in
\S\ref{sec:statistics}, we briefly outline the technical details of
this computation, which was carried out on MIT's subMIT cluster
\cite{Bendavid:2025arv}.

The KS database \cite{KS} classifies polytopes by the number of
vertices of their dual polytope. Thus, the file v05.gz contains the
1,561 polytopes whose duals have five vertices each.\footnote{One
immediate way to verify that the KS database organizes its files by
the number of vertices of the dual, rather than the number of vertices
of the polytope itself, is to query it for polytopes with
$h^{1,1}=1,h^{2,1}=101$; doing so returns the vertices of the mirror
quintic rather than those of the quintic itself.} We have kept this
organizational scheme throughout our analysis, including in the
database accompanying this publication, but have split the v$n$.gz
files into more manageable files of (up to) one million polytopes each,
labeled v$n$-1.txt and so on, when appropriate.

Our analysis proceeds in three steps. Given a list of input polytopes
as described above, we begin by using the PALP package to preprocess the raw KS 4D polytope files to include some additional information about each polytope.
We then find all 2D reflexive subpolytopes 
and classify them into fibration structures. This is accomplished
by means of a Julia script which implements the algorithm in
\S\ref{sec:identifying-fibers}, and which we have included as an
ancillary file in the associated Zenodo archive. Once the fibration structures and their
representatives have been specified, we select all polytopes with at
least one fibration structure with more than one representative. These
polytopes are input into CYTools, where we compute the number of
automorphism inequivalent representatives of each fibration structure;
the CYTools script used for this calculation is again included as an
ancillary file. Finally, we merge the outputs of the Julia and CYTools
analyses to create our final output files, which contain counts of the
fibration structures, inequivalent fibrations, and total fibrations
for all 473,800,776 4D reflexive polytopes.


\subsection{Organization of results and nomenclature}
\label{sec:organization}
The results of the analysis are organized in a set of data files that
are publicly available at \cite{archive}.  The details of
the formatting of the files are described in more detail in the
README.txt file associated with the archive, but we give a broad
description here.

In the remainder of this paper and the Zenodo database, we refer to a
given polytope using the syntax ``v$xx$-$yyy$,'' where (following the
format used in the KS database \cite{KS}), $xx$ denotes the number of
vertices of the dual polytope (2 digits, ranging from 05--36) and
$yyy$ denotes indexing in the order of the KS database for polytopes
with that number of vertices (varying numbers of digits).  Fibers
1--16 are ordered as in \cite{Klevers:2014bqa} and
Figure~\ref{fig:16fibers}, and bases 1--61,539 are ordered canonically
as in the database associated with \cite{MorrisonTaylorToric}, i.e. they are ordered
first by $h^{1,1} (B_2)$, and entries at each value of $h^{1,1}$ are ordered
canonically according to (the negatives of the) intersection numbers
of the toric base divisors, so e.g. bases $1, 2, \ldots, 13$
correspond to $\P^2,\F_0, \ldots,\F_{12}$).
For convenience, that database is also included in the Zenodo archive.

With these conventions set, we can explain the format of the files in the database. Each 4D reflexive polytope appears in the database as a single line, which begins with some basic topological data specifying the polytope, namely its label (v$xx$-$yyy$, as specified above, which for machine-readability we specify as [$xx$,$yyy$]) and Hodge numbers, in the form $[h^{1,1},h^{2,1}]$. Next we give the total number of 2D subpolytopes, before proceeding to specify how these fibrations are broken up into fibration structures, etc. 

Fibration structures are presented as a list, where each structure is
specified first by base, and then within each
base by fiber types. We specify the base as follows:

\vspace*{0.05in}
\texttt{[base singular?, base ID, $h_*^{1,1}(B)$, [blown down curves],...}
  \vspace*{0.05in}

\noindent
where $h^{1,1}_*(B)$ is given by (\ref{eq:h11s}).

We then specify the different fiber types that appear and the number of subpolytopes of that type as 

\vspace*{0.05in} \texttt{
    [fiber ID, \#(total fibrations in this structure), \#(inequivalent fibrations)]
} \vspace*{0.05in}
Putting it all together, the full database entry for a single polytope
takes the form

\vspace*{0.05in}

\noindent  
\texttt{
\hspace*{0.06in}  [[xx,yyy],[$h^{1,1}(X)$,$h^{2,1}(X)$],\#(total fibrations),\\ 
\hspace*{0.10in} \      [[base 1 singular?, base 1 ID, $h_*^{1,1}$(base 1), [blown down curves in base 1], \\
\hspace*{0.10in} \ \ \        [[fiber 1 ID, \#(total fibrations with base 1-fiber 1), \\
\hspace*{0.10in} \ \  \  \ \  \#(inequivalent fibrations)],\\
\hspace*{0.10in} \ \  \  \       [fiber 2 ID, \#(total fibrations with base 1-fiber 2), \\
\hspace*{0.10in} \ \  \  \ \ \#(inequivalent fibrations)],...]], \\
 \hspace*{0.10in} \ \      [base 2 singular?, base 2 ID, $h_*^{1,1}$(base 2), [blown down curves in base 2], \\
 \hspace*{0.10in} \ \ \  [[fiber 1 ID, \#(total fibrations with base 2-fiber 1), \\
\hspace*{0.10in} \ \  \  \ \ \#(inequivalent fibrations)],\\
\hspace*{0.10in} \ \ \ \   [fiber 2 ID, \#(total fibrations with base 2-fiber 2), \\
\hspace*{0.10in} \ \  \  \ \ \#(inequivalent fibrations)],...]]]
    ,...]
} \vspace*{0.05in}

Let us see how this works for a particular example. Consider the polytope with vertices given by the columns of the matrix \eqq{\left(\begin{array}{ccccc} -1&-1&-1&-1&2\\-1&-1&-1&2&-1\\1&1&3&-1&-1\\0&4&0&-1&0
\end{array}\right),} which has Hodge numbers $h^{1,1}(X)=h^{2,1}(X)=19$. The database entry corresponding to this polytope is 

\begin{verbatim}
    [[5, 724], [19, 19], 3, [[0, 150, 6, [], [[10, 2, 1]]], 
     [1, 4162, 14, [1, 1, 0, 0, 1, 0, 0, 1, 0, 0, 1, 0, 0, 1, 1, 1], [[1, 1, 1]]]]].
\end{verbatim}

From this line, we see that its label (as specifed above) is v05-724, it has Hodge numbers $h^{1,1}(X)=h^{2,1}(X)=19$, as expected, and it has three total elliptic fibrations, coming from two different fibration structures:
\begin{enumerate}
    \item $F_{10}$ fibered over base 150, which has
      $h^{1,1}_*(B)=h^{1,1}(B)=6$, with no blowdowns. There are two
      fibrations with this structure, and they are related by an
      automorphism.
    \item $F_1$ fibered over base 4162, which has $h^{1,1}_*(B)=h^{1,1}(B)=14$, with blowdowns. The  (negative) intersection form of this base is
      \eqq{[6, 1, 2, 2, 2, 2, 2, 2, 2, 2, 2, 2, 2, 1, 6, 0],} and the string specifiying the blowdowns is \eqq{[1, 1, 0, 0, 1, 0, 0, 1, 0, 0, 1, 0, 0, 1, 1, 1].} A zero in this string indicates that the corresponding curve in the intersection sequence has been blown down, and a one indicates that it has not. Thus, to construct the base of this of this fibration, we have blown down eight curves of self-intersection -2. There is a single fibration in this structure.  
\end{enumerate}

\section{Distributions of Fibers and Bases}
\label{sec:statistics}

From the complete analysis of all toric 2D subpolytopes in the KS
database, we identify 2,264,992,252 fibrations, with 2,177,857,036 distinct ``structures,''
i.e. combinations of base, fiber, and ambient polytope as defined above.
After accounting for automorphism equivalence, the number of inequivalent fibrations is 2,250,744,657.

In this section, we
describe some general aspects of the statistics of these fibrations.
In Subsection~\ref{sec:polytopes} we organize top-level statistics in
terms of polytopes, in subsection~\ref{sec:statistics-fibers} we
describe statistics of fiber types, and in
subsection~\ref{sec:statistics-bases} we describe some statistics in
terms of the bases.


\subsection{Fibrations per polytope}
\label{sec:polytopes}

The average number of fibrations per polytope is 4.78, of which 4.75 are automorphism-inequivalent.
It is interesting to compare this with the average number of ``obvious'' fibrations for CICYs, which was found to be 9.85 in \cite{Anderson:2016cdu}.

The polytope with the most fibrations has 362 fibrations.  This is
realized for polytope v08-242 (in the nomenclature of
Section~\ref{sec:organization}), which gives CY3's with Hodge numbers
$(h^{1,1} (X), h^{2,1} (X)) = $ (68, 4).  These fibrations furnish ten fibration structures and, after accounting for polytope automorphism, each fibration structure has a single equivalence class of representatives.

\begin{figure}[h!]
    \centering
\begin{picture}(200,450)(- 100,- 240)
\put(-120, 115){\makebox(0,0){\includegraphics[width=0.55\linewidth]{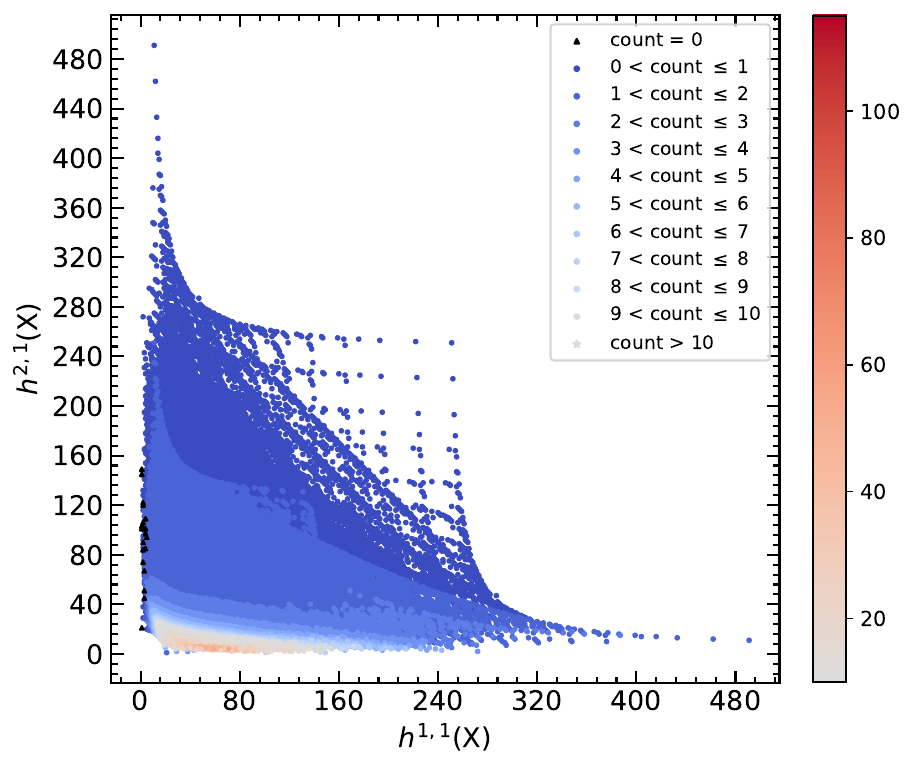}}}
\put(120, 115){\makebox(0,0){\includegraphics[width=0.55\linewidth]{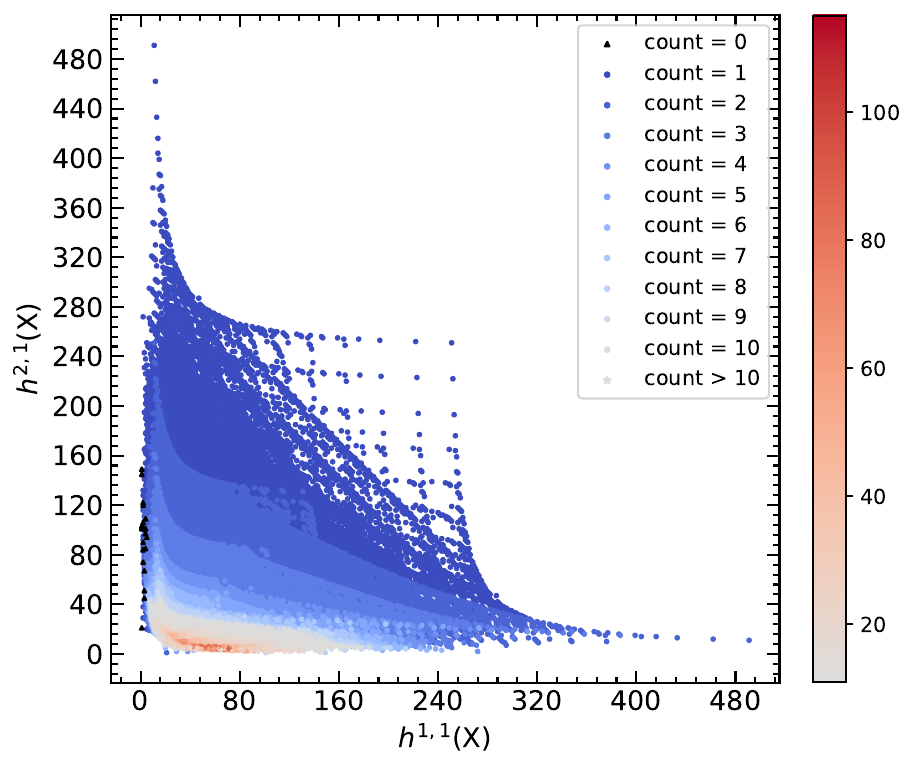}}}
\put(0, -120){\makebox(0,0){\includegraphics[width=0.55\linewidth]{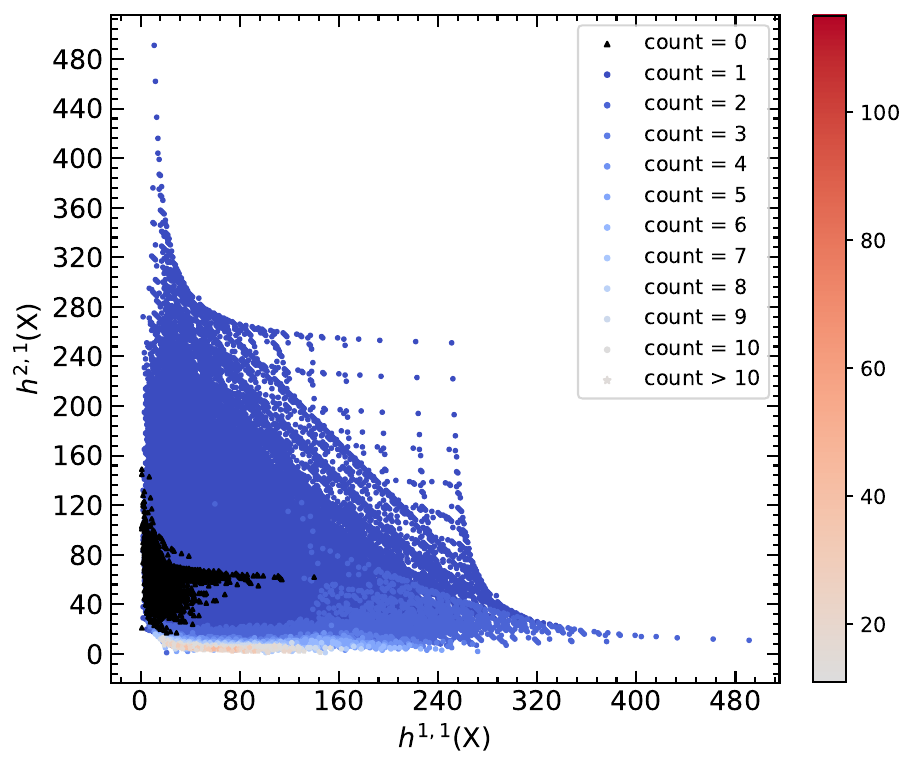}}}
\put(-120, 10){\makebox(0,0){(a) Mean fibrations}}
\put(120, 10){\makebox(0,0){(b) Max fibrations}}
\put(0, -230){\makebox(0,0){(c) Min fibrations}}
\end{picture}
\caption{The average (a), maximum (b), and minimum (c) number of
  automorphism-inequivalent fibrations over a polytope
giving Calabi-Yau threefolds
  $X$ with Hodge
  numbers $h^{1, 1}(X)$ and $h^{2, 1}(X)$ are plotted. The black
  triangles indicate zero structures, the blue colors indicate smaller
  number of structures, up to 10, and the color bar on the right
  indicates the color corresponding to larger number of fibrations
  (more red means larger number of structures).}
    \label{fig:fibrations}
\end{figure}

We plot the average number of total (automorphism inequivalent) fibrations
on a polytope with a given $h^{1,1}$ and $h^{2, 1}$
in Figure~\ref{fig:fibrations}(a); plots of the maximum
and minimum number of (again, automorphism equivalent)
fibration structures that a polytope with a
given Hodge pair can host appear in Figures~\ref{fig:fibrations}(b, c).



In general, we see that the polytopes with many fibrations tend to
have small $h^{2,1} (X)$, and the number of fibrations for polytopes
with small $h^{2,1} (X)$ peaks at intermediate values of $h^{1,1}
(X)$.  The plot of the maximum number of fibrations at fixed Hodge
numbers has an interesting structure; when both Hodge numbers are large,
there is generally only a single fibration.  A second fibration
generally appears below a boundary with a peak around $(140, 140)$ and
a slope that cuts off polytopes with large $h^{2,1} (X)$, but
continues out to large $h^{1,1} (X)$; an example of an interesting
polytope with two fibrations is the polytope with maximum $h^{1,1} (X)
= 491$, discussed in \S\ref{sec:large-h11}.
\footnote{From the structure of this maximal polytope, the analyses of
\cite{Aspinwall:1997ye,Anderson:2016cdu}, and the related
structure of other polytopes with large $h^{1,1} (X)$, it seems
plausible that this second fibration is often associated with
structures that are natural in a dual Spin(32)/$\Z_2$ heterotic
theory, while the first, upper class of fibrations corresponds more
naturally with an $E_8 \times E_8$ heterotic dual.}
A further boundary that more closely approximates a diagonal line
separates out polytopes with at most three fibrations, and so on.
It would be interesting to understand in more detail the structures
responsible for this pattern.

\subsection{Fiber types}
\label{sec:statistics-fibers}

\begin{figure}
    \centering
    \includegraphics[width=0.6\linewidth]{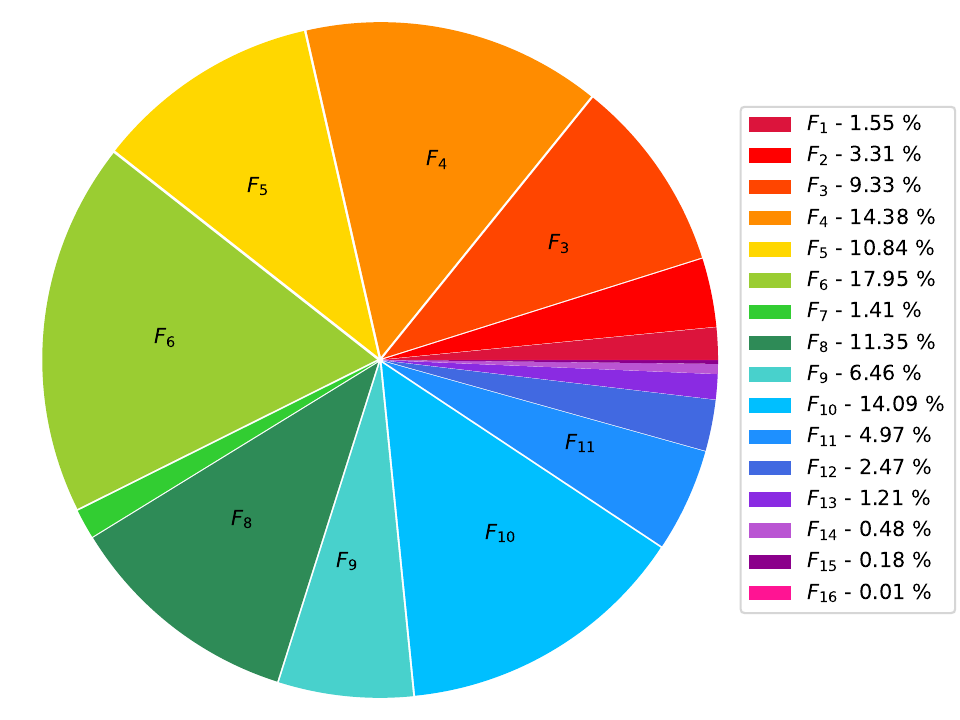}
    \caption{The density of inequivalent fibrations by each of the sixteen fiber types.}
    \label{fig:pie}
\end{figure}

\begin{figure}
    \centering
    \includegraphics[width=.75\linewidth]{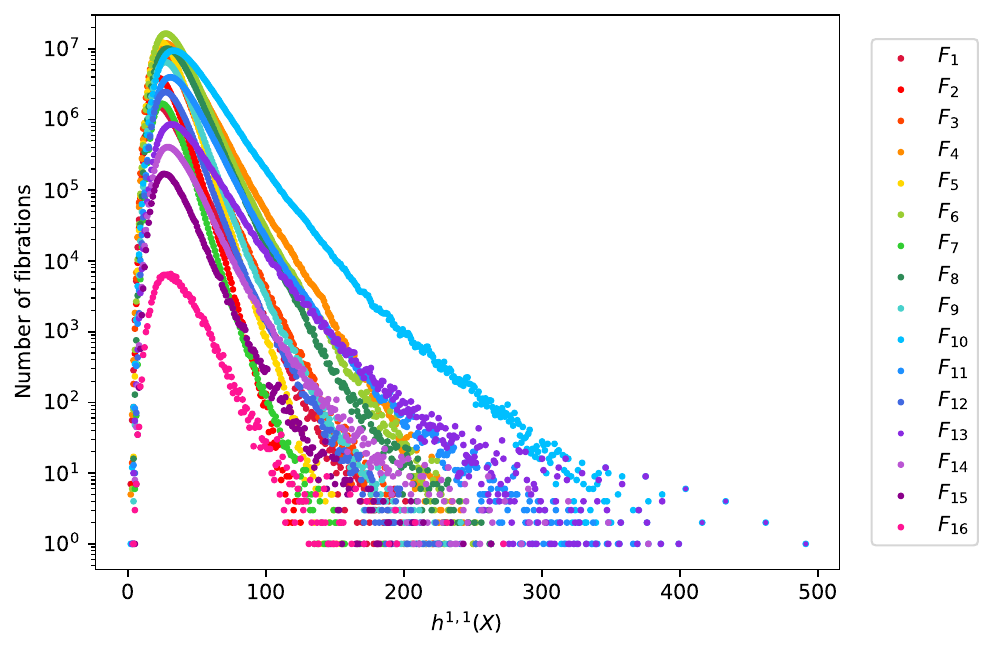}
    \caption{The total number of automorphism-inequivalent fibrations of each fiber type as a function
      of $h^{1,1}(X)$.
Note that at large $h^{1,1} (X)$, above $h^{1,1} (X)\sim 400$, all
polytopes with a fiber $F_{13}$ also have a fiber $F_{10}$.\ignore{, but the
former overlies the latter in the figure, so the $F_{10}$ fibers are not apparent,
though still present.}
    }
    \label{fig:h11x_fibers}
\end{figure}

We find fibrations with each of the sixteen fibers.
This is guaranteed since for any pair of fibers $F, F'$ we can construct a
reflexive 4D polytope as a product (with rays $(v\in F, v'\in F')$)
or a ``sum'' (with rays $(v \in F, 0), (0, v' \in F')$) of these two
fibers.
\footnote{A detailed analysis of these polytopes and their behavior
under mirror symmetry will be given in \cite{ot-forthcoming}.}
The
distribution of fiber types is given in Table \ref{tab:fiberstats} and illustrated graphically in Figure \ref{fig:pie}.
We find that $F_6$ is the most common fiber ($\sim 18\%$), with 403,991,733 automorphism-inequivalent fibrations,  while $F_{16}$ is the least common fiber ($\sim 0.01\%$),
with only 182,915 inequivalent fibrations.  On the other hand, for polytopes with large
$h^{1,1} (X)$, $F_{10}$ dominates, as can be seen in Figure
\ref{fig:h11x_fibers};
the highest $h^{1, 1}$ of a polytope that has no $F_{10}$ fibers is $h^{1, 1} = 227$.
The dominance of fiber $F_6$ was to us somewhat unexpected;
because every base admits at least one fibration with fiber $F_{10}$
(the standard stacking, \S\ref{sec:toric-fibrations}), and this fiber
is closest to the generic Weierstrass model, one might have expected
$F_{10}$ to dominate.
Indeed, much of the previous analysis of fibrations
focused
implicitly on $F_{10}$ fibers.
We explore the geometry of fibrations with fiber
$F_6$ in further detail in \S\ref{sec:common}.

\begin{table}[]
    \centering
    \begin{tabular}{|c||c|c|c|} \hline
    Fiber & Fibration Structures & Inequivalent Fibrations & Total Fibrations \\\hline\hline
$F_{1}$ &34,339,236 & 34,990,811 & 35,461,473 \\\hline
$F_{2}$ &70,340,325 & 74,432,915 & 75,476,280 \\\hline
$F_{3}$ &201,064,783 & 210,042,410 & 212,128,020 \\\hline
$F_{4}$ &316,154,156 & 323,559,047 & 325,373,736 \\\hline
$F_{5}$ &222,330,801 & 244,004,671 & 246,317,835 \\\hline
$F_{6}$ &392,614,430 & 403,991,733 & 406,116,680 \\\hline
$F_{7}$ &30,158,692 & 31,807,255 & 32,177,670 \\\hline
$F_{8}$ &249,141,445 & 255,482,339 & 256,753,239 \\\hline
$F_{9}$ &140,262,167 & 145,501,972 & 146,510,508 \\\hline
$F_{10}$ &314,373,356 & 317,175,553 & 317,945,412 \\\hline
$F_{11}$ &111,071,735 & 111,917,349 & 112,327,098 \\\hline
$F_{12}$ &53,880,209 & 55,577,013 & 55,975,484 \\\hline
$F_{13}$ &27,213,569 & 27,275,196 & 27,349,323 \\\hline
$F_{14}$ &10,750,057 & 10,798,598 & 10,859,072 \\\hline
$F_{15}$ &3,979,255 & 4,004,880 & 4,035,646 \\\hline
$F_{16}$ &182,820 & 182,915 & 184,776 \\\hline

    \end{tabular}
    \caption{Summary statistics for fibrations of each fiber type. For each fiber, we have indicated the number of fibration structures, the number of inequivalent fibrations, and the total number of fibrations  with that fiber type.}
    \label{tab:fiberstats}
\end{table}

We find that genus-one fibrations, i.e. those with fiber 
$F_1$, $F_2$, or $F_4$, 
are common: 19.3\% (436,311,489
total, 432,982,773 inequivalent) of all fibrations are genus-one,
rather than elliptic. Amongst genus-one fibrations, fiber $F_4$ is by
far the most common, representing 74.6\% (325,373,736 total,
323,559,047 inequivalent) of all genus-one fibrations.

\subsection{Bases}
\label{sec:statistics-bases}

In this section we discuss the statistics of fibrations, in particular
how fibrations are distributed as a function of the effective Hodge
number of the base $h^{1,1}_*(B)$ defined in (\ref{eq:h11s}).  We find that each of the 61,539 toric bases on the MT list appears as
the base for at least one fibration; 625 bases appear exactly
once. This is unsurprising; as explained in
\S\ref{sec:toric-fibrations}, one can construct the generic elliptic
fibration over any toric base by taking (the dual of the dual of) the
standard stacking of $F_{10}$ over that base.

\begin{figure}
    \centering
    \includegraphics[width=0.6\linewidth]{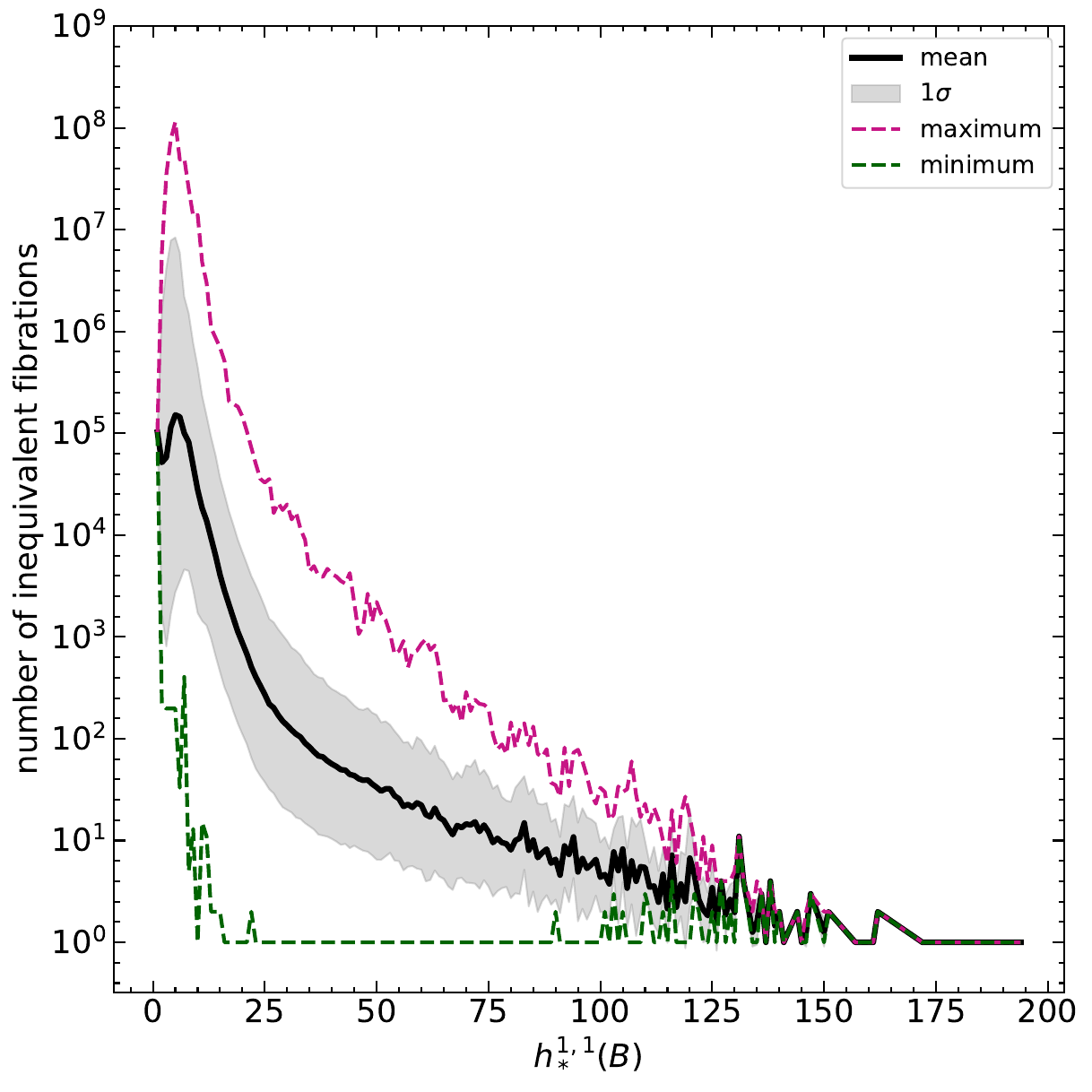}
    \caption{The mean, minimum, maximum and $1\sigma$ deviation of the
      number of inequivalent fibrations over a base $B$ with a given $h^{1,1}_*(B)$
      are plotted.
      Note that there are more structures fibered over bases with small $h^{1, 1}_*(B).$}
    \label{fig:h11B}
\end{figure}

\begin{figure}
    \centering
    \includegraphics[scale=.75]{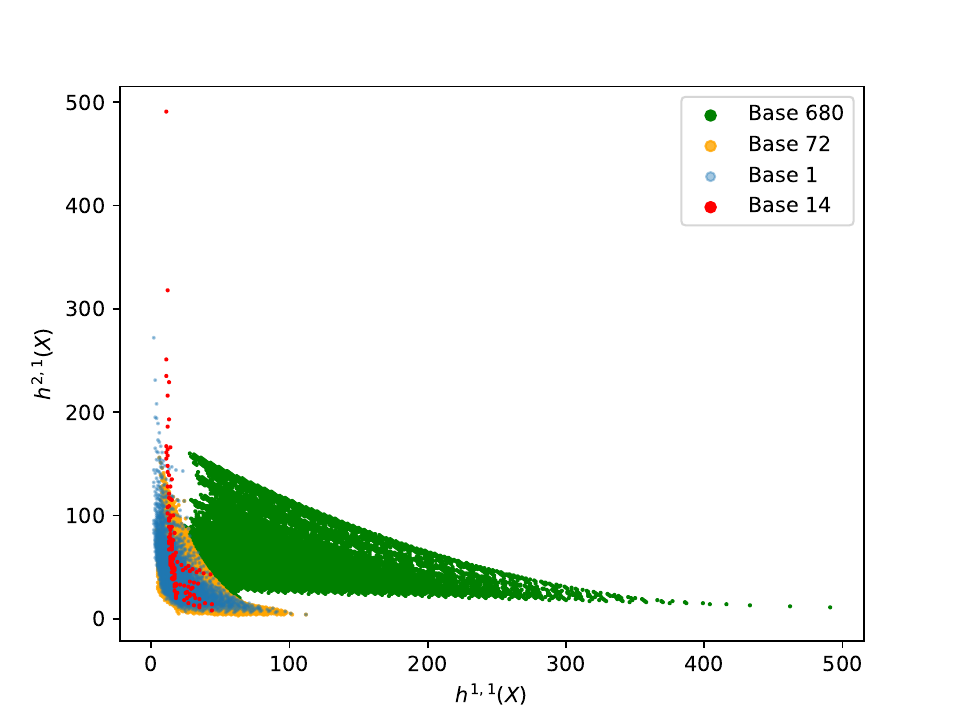}
    \caption{The Hodge numbers of fibrations over base 1, i.e. $\P^2$, discussed in \S\ref{sec:volker}--\ref{sec:non-toric-gauge}),
       base 14, Hirzebruch $\F_{12}$ which supports the elliptic CY with the largest $h^{2,1} (X) = 491$), base 72, a gdP$_4$ which is the most common base, and base 680, which appears universally in fibrations with extremely large gauge groups, as discussed in \S\ref{sec:rankincrease}.}
    \label{fig:polica}
\end{figure}

While all bases appear, there is a marked statistical preference
towards bases with small $h^{1,1}_*(B)$.
As discussed in more detail in
\S\ref{sec:typical}, this is likely because bases with larger $h^{1,1}_*
(B)$ have a large number of non-Higgsable clusters that strongly
constrain the possibility of additional tuned gauge factors.
The most common base, which is
entry 72 on the MT list, is a generalized del Pezzo surface 
$gdP_4$, with $h^{1,1}_*(B)= h^{1,1}(B)=5$, and is the base for 5.1\%
(114,961,701 total, 114,019,846 inequivalent) of all fibrations; this
is a weak Fano base, and is the toric variety defined by the $F_{12}$
polytope, as in Figure \ref{fig:16fibers}. Plotting the number of
fibrations over a base as a function of $h_*^{1, 1}(B)$,
as shown in
Figure~\ref{fig:h11B}, we find a distinct peak at small $h^{1,
  1}_*(B)$. In particular, 82.5\% (1,868,680,827 total, 1,857,694,025
inequivalent) of fibrations have $h^{1,1}_*(B)\le10$, and 99.4\%
(2,251,269,536 total, 2,237,134,428 inequivalent) have
$h^{1,1}_*(B)\le25$. We have plotted the Hodge numbers of all elliptic
fibrations over a number of specific bases with small $h^{1,1}_*(B)$ in Figure
\ref{fig:polica}.

\begin{figure}[h]
    \centering
    \includegraphics[width=0.6\linewidth]{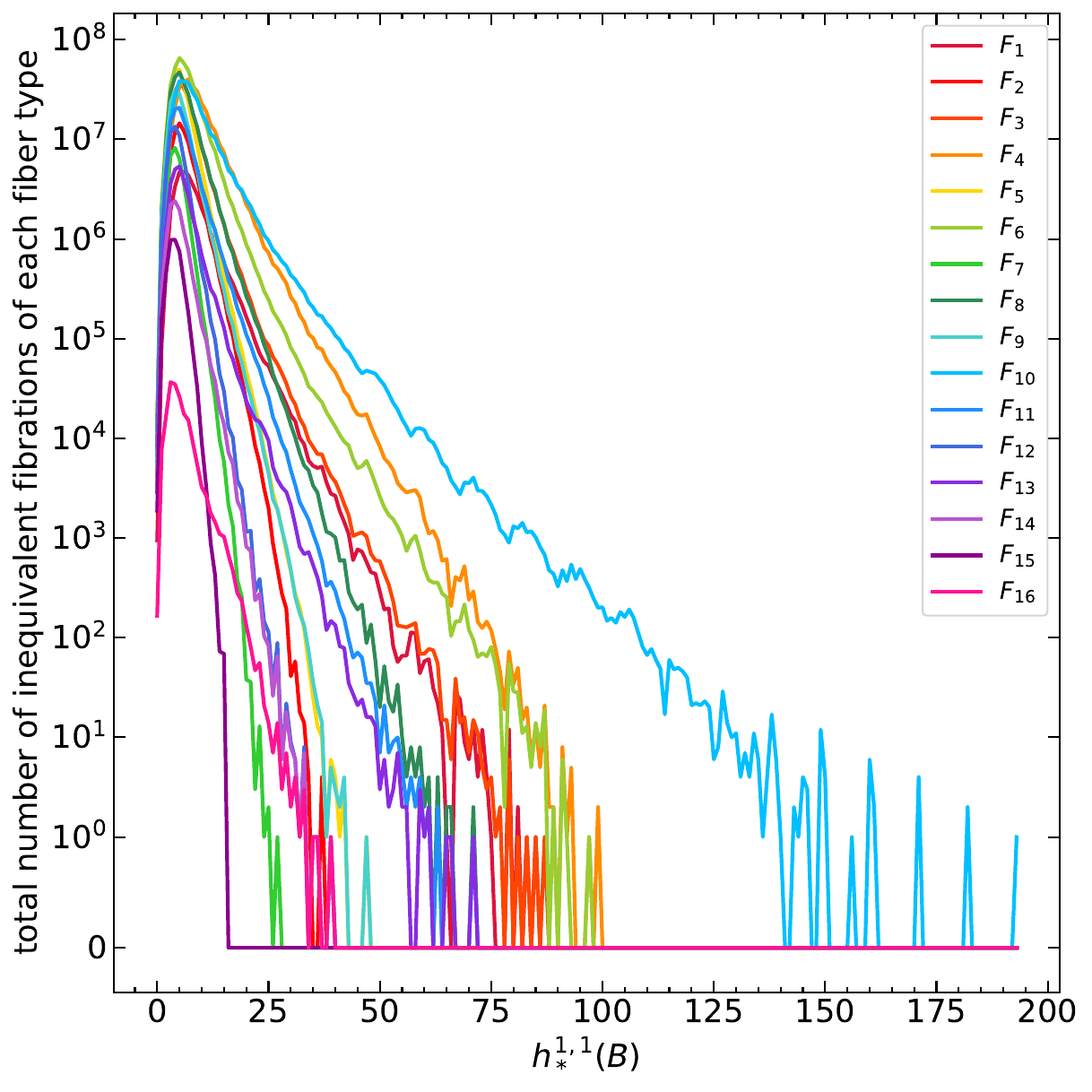}
    \caption{The number of inequivalent fibrations of each fiber type as a function of $h^{1,1}_*(B)$. Note that at large $h^{1,1}_*(B)$, only fibrations with fiber $F_{10}$ are seen.}
    \label{fig:ftype}
\end{figure}

Moreover, as we see in Fig.~\ref{fig:ftype}, the only fiber type that appears fibered over a base with $h^{1, 1}_*(B_2) > 100$ is $F_{10}$. 
This corroborates the expectation (c.f. \S\ref{sec:toric-fibrations}) that
for bases with large $h^{1,1}_*(B_2)$, the only available elliptic
fibration
is the generic elliptic fibration, associated with the
standard stacking of the fiber $F_{10}$.

\subsubsection{Weak Fano and non-weak-Fano bases}
\label{sec:fano-v-nonfano}
\begin{figure}[h]
\begin{center}
\begin{subfigure}[b]{.45\textwidth}
\begin{center}
\includegraphics[scale=.5]{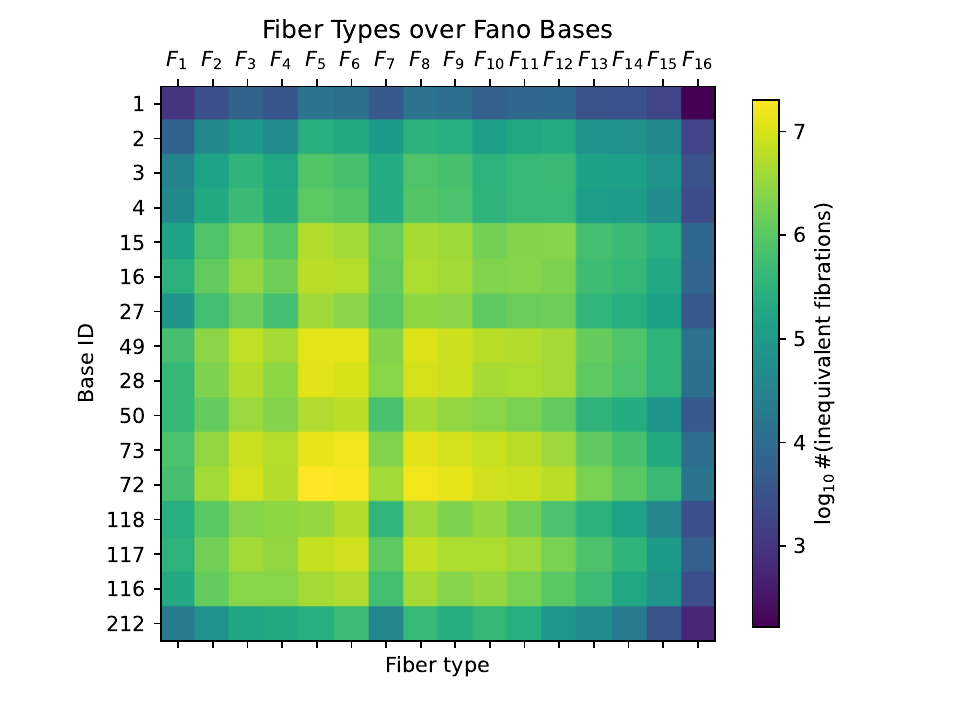}
\caption{}
\label{fig:fanofibers}
\end{center}
\end{subfigure}
~
\begin{subfigure}[b]{.45\textwidth}
\begin{center}
\includegraphics[scale=.5]{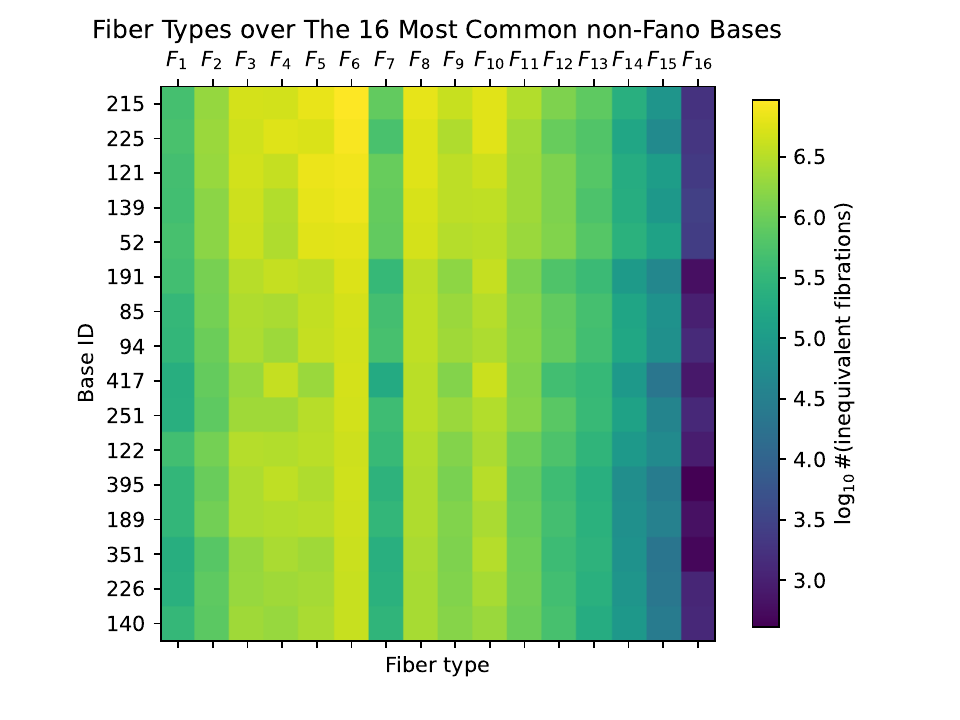}
\caption{}
\label{fig:nonfanofibers}
\end{center}
\end{subfigure}
\caption{The distribution of fiber types in irredundant fibrations over (a) the sixteen weak Fano bases and (b) the sixteen most common non-weak-Fano bases.}
\label{fig:basefibers}
\end{center}
\end{figure}

As discussed in \S\ref{sec:toric-fibrations}, of the 61,539 toric
bases, 16 are weak Fano; these exactly correspond to the 16 toric
fibers. Some summary statistics of fibrations with weak Fano bases are
given in Table \ref{tab:fanostats}. We find that fibrations with weak Fano
bases are disproportionately common, comprising 26\% (588,457,836
total, 584,970,461 inequivalent) of all fibrations. In particular, the
four most common bases are all weak Fano. Intuitively, this makes sense;
weak Fano bases lack NHCs, and so
one has
additional freedom to tune gauge groups over weak Fano bases relative to
non-weak-Fano bases. We show the distribution of fiber types in fibrations
over the sixteen weak Fano bases in Figure \ref{fig:fanofibers}.
Note that for weak Fano bases $h^{1,1} (B) = h^{1,1}_*(B)$ since there
are no toric curves of self-intersection less than -2.

\begin{table}[]
    \centering
    \small{
    \begin{tabular}{|c|c||c|c|c|c|} \hline
    Base ID & Fiber type & $h^{1,1}(B)$ & Fibration Structures & Inequivalent Fibrations & Total Fibrations \\\hline\hline
    1 & $F_{1}$ &1 & 102,563 & 102,581 & 102,600 \\\hline
2 & $F_{2}$ &2 & 1,975,645 & 2,005,188 & 2,020,060 \\\hline
3 & $F_{3}$ &2 & 5,332,362 & 5,415,109 & 5,435,366 \\\hline
4 & $F_{4}$ &2 & 6,031,392 & 6,072,869 & 6,084,817 \\\hline
15 & $F_{5}$ &3 & 28,767,691 & 29,917,327 & 30,066,698 \\\hline
16 & $F_{6}$ &3 & 34,433,015 & 34,821,744 & 34,879,206 \\\hline
27 & $F_{7}$ &4 & 19,074,969 & 20,268,222 & 20,465,927 \\\hline
28 & $F_{9}$ &4 & 63,835,026 & 67,391,516 & 67,765,284 \\\hline
49 & $F_{8}$ &4 & 74,437,389 & 77,408,588 & 77,673,426 \\\hline
50 & $F_{10}$ &4 & 32,326,443 & 32,432,358 & 32,453,682 \\\hline
72 & $F_{12}$ &5 & 98,635,610 & 114,019,846 & 114,961,701 \\\hline
73 & $F_{11}$ &5 & 82,902,723 & 87,426,934 & 87,830,312 \\\hline
116 & $F_{15}$ &6 & 26,802,481 & 29,288,912 & 29,657,220 \\\hline
117 & $F_{14}$ &6 & 45,560,240 & 48,889,834 & 49,350,175 \\\hline
118 & $F_{13}$ &6 & 25,955,132 & 26,745,791 & 26,910,622 \\\hline
212 & $F_{16}$ &7 & 2,700,198 & 2,763,642 & 2,800,740 \\\hline

    \end{tabular}
    }
    \caption{Summary statistics for fibrations with weak Fano bases. We specify the weak Fano bases both with their labels in the MT list and their labels as fiber types, as in Table \ref{tab:16fibers}. For each base, we have indicated the Hodge number $h^{1,1}(B)$ and the number of fibration structures, the number of inequivalent fibrations, and the total number of fibrations with that base.}
    \label{tab:fanostats}
\end{table}

The other 61,523 bases are non-weak Fano, and contain NHCs hosted on toric
divisors. non-weak-Fano bases comprise the remaining 74\% (1,676,534,416
total, 1,665,774,196 inequivalent) of fibrations; 96.5\% (457,234,463
total) of all polytopes admit at least one fibration with a smooth,
non-weak-Fano base, and correspondingly an NHC. The most common non-weak-Fano
base is base 215 on the MT list, which has $h^{1,1}(B)=7$ and (negative) intersection form \eqq{[2, 2, 1, 4, 1, 2, 1, 1, 1],} and thus is only non-weak-Fano because of the presence of one curve with self-intersection -4; this appears
as the base in 2.2\% (50,802,116 total, 50,472,765 inequivalent) of
fibrations. We show the distribution of fiber types in fibrations over
the sixteen most common non-weak-Fano bases in Figure
\ref{fig:nonfanofibers}.

\begin{table}[]
    \centering
    \begin{tabular}{|c||c|c|c|c|c|} \hline
    Base ID & $h^{1,1}(B)$ & Int Form & Fib. Structs. & Ineq. Fibs. & Total Fibs. \\\hline\hline
215  & 7 & [2,2,1,4,1,2,1,1,1] & 47,759,092 & 50,472,765 & 50,802,116 \\\hline
225  & 7 & [3,1,2,2,1,2,1,2,1] & 40,376,100 & 45,551,690 & 46,149,521 \\\hline
121  & 6 & [2,1,3,1,2,1,1,1] & 41,485,826 & 44,448,022 & 44,787,034 \\\hline
139 & 6 & [3,1,2,2,1,1,1,1]& 38,213,080 & 40,125,864 & 40,428,279 \\\hline
52 & 5  & [1,2,1,3,1,1,0] & 36,711,081 & 37,673,030 & 37,807,613 \\\hline
191& 7  & [2,1,3,1,4,1,2,1,0]&  28,570,661 & 29,257,553 & 29,336,318 \\\hline
85  &6 & [1,2,1,4,1,2,1,0] & 27,158,061 & 27,660,887 & 27,748,385 \\\hline
94 & 6 & [1,2,2,1,4,1,1,0] & 26,151,001 & 26,804,201 & 26,902,391 \\\hline
417 & 8 & [4,1,2,2,2,1,2,1,2,1] & 23,597,546 & 25,187,371 & 25,507,833 \\\hline
251 & 7 & [4,1,2,2,2,1,1,1,1] & 24,163,813 & 24,818,246 & 24,951,266 \\\hline
122 & 6 & [2,1,3,1,3,1,1,0] & 23,937,142 & 24,426,767 & 24,498,388 \\\hline
395 & 8 & [3,1,3,1,4,1,2,1,1,1] &23,166,420 & 23,778,385 & 23,883,798 \\\hline
189 & 7 & [2,1,3,1,3,1,2,1,1] &22,293,631 & 23,307,742 & 23,467,491 \\\hline
351 & 8 & [2,2,1,4,1,3,1,2,1,1] &20,013,295 & 20,764,018 & 20,904,914 \\\hline
226 & 7 & [3,1,2,2,2,1,2,1,1] &19,097,689 & 19,995,395 & 20,217,713 \\\hline
140 & 6 & [3,1,2,2,2,1,1,0] &19,439,118 & 19,948,414 & 20,068,453 \\\hline

    \end{tabular}
    \caption{Summary statistics for fibrations with the sixteen most
      common non-weak-Fano bases. We specify the bases by their labels
      in the MT list. For each base, we have indicated the Hodge
      number $h^{1,1}(B)= h^{1,1}_*(B)$ and the (negative of) its intersection form, as well as the number of fibration structures, the number of inequivalent fibrations, and the total number of fibrations with that base.  Note that the first five most common bases in the list differ from weak Fano by containing only a single curve of self-intersection -3 or -4.}
    \label{tab:nonfanostats}
\end{table}

\subsubsection{Singular Bases}
\label{sec:singular}
As mentioned in Section \ref{sec:algorithm},
it is possible for a smooth fibration to have a singular base. We find
that this occurs rarely, in only 0.5\% (11,184,662 total, 11,136,035
inequivalent) of fibrations. Moreover, singular bases occur only in
genus-one, rather than elliptic, fibrations. Some summary statistics
of the fiber types of fibrations with singular base are given in Table
\ref{tab:singular_fibers}.
$F_4$ remains the most common fiber type in
genus-one fibrations over singular bases, appearing in 80.8\%
(9,029,636 total, 9,002,823 inequivalent) of all such fibrations.

\begin{table}[]
    \centering
    \begin{tabular}{|c||c|c|c|c|} \hline
    Fiber & Singularity  &  Sing. Fib. Structures & Ineq. Sing. Fibs. & Total Sing. Fibs. \\\hline\hline
    $F_{1}$ &$ (\Z_3)$& 614,963 & 617,188 & 627,149 \\\hline
    $F_{2}$&$ (\Z_2)$& 1,508,437 & 1,516,024 & 1,527,877 \\\hline
    $F_{4}$& $ (\Z_2)$& 8,982,183 & 9,002,823 & 9,029,636 \\\hline

    \end{tabular}
    \caption{Summary statistics for fibrations of each fiber type over
      singular bases with a given local singularity structure. We list only the genus-one fibers; elliptic fibrations with singular bases do not occur. For each fiber, we have indicated the characteristic singularity structure, the number of fibration structures, the number of inequivalent fibrations, and the total number of fibrations  with that fiber type.}
    \label{tab:singular_fibers}
\end{table}

The singular bases have a very specific structure.  In all cases, the singular bases correspond to smooth bases in the MT list with a set of -2 curves blown down.
Blown
down -2 curves can either appear alone or in adjacent pairs, corresponding to a local $\C^2/\Z_2$ or $\C^2/\Z_3$ singularity in the base.
No other
combinations occur in our dataset; furthermore, any given singular
base has only one or the other type of singularity.  Those with a
$\Z_3$ singularity always have fiber $F_1$, while those with a $\Z_2$
singularity have fiber $F_2$ or $F_4$.  Since these singularities
corresponds precisely to the type of multisection of those genus-one
fibrations, it seems that these fibrations have nontrivial discrete
monodromy around the singularities.  We leave further investigation of
this interesting structure to further work.
  
Of the 61,539 toric bases, only 17,961 appear as the resolution of a
singular base in a genus-one fibration.
The most common base is base
417, a non-weak-Fano base with $h^{1,1}_*(B)= h^{1,1} (B)=8$ and which appears as (the
resolution of) the base in 3.3\% (372,445 total, 370,503 inequivalent)
of such fibrations. The most common overall base, MT base 72 or
equivalently the fiber $F_{12}$, appears as a singular base in only
1\% (116,300 total, 115,138 inequivalent) of such fibrations. In spite
of this shift, the statistical preference for small $h^{1,1}_*(B)$
remains; the largest singular base has $h^{1,1}_*(B)=92$, and 97.7\%
(10,929,212 total, 10,882,539) of such fibrations have
$h^{1,1}_*(B)\le25$.

\subsection{Rank increase}
\label{sec:rankincrease}
One useful way of quantifying

the structure of a fibration is through the number of divisors that
arise in the (tuned or non-Higgsable) gauge group,
which by Shioda-Tate-Wazir
is the
difference between $h^{1,1}(X)$ and $h^{1,1}(B)$. 
We define two closely
related notions of rank increase, the total rank increase
\eqq{\delta_{\text{total}} := h^{1,1}(X)- h^{1,1}(B) -1\,,} and the tuned
rank increase
\eqq{\delta_{\text{tuned}} :=
  h^{1,1}(X)-\left[h^{1,1}_*(B) +
    \operatorname{rk}\left(\text{NHC}\right)\right] -1.}
We have the
obvious bound
\eqq{\delta_{\text{tuned}}\leq\delta_{\text{total}},}
with $\delta_{\text{tuned}}=\delta_{\text{total}}$ only for fibrations
over a weak Fano base, and $\delta_{\text{tuned}}= 0$ for the generic
elliptic fibration over a given base.

While for a
weak Fano base these two notions coincide, for a non-weak-Fano base
the tuned rank increase more closely quantifies how much 
tuning
the fibration has by subtracting out a universal contribution from the
NHC (including both contributions from non-Higgsable gauge groups and
non-Higgsable SCFT's); we therefore find it to be a more useful definition. We have
computed both the total and tuned rank increase for each fibration,
but for the remainder of this section we will comment only on the
tuned rank increases; for brevity, we will refer to the tuned rank
increase simply as the rank increase in what follows.

Note that, for elliptic fibrations, the rank increase is nonnegative
by Shioda-Tate Wazir. Nevertheless, we find fibrations with negative
rank increase; this occurs only in genus-one fibrations, and is
relatively rare, occurring in a total of only 3,868,168 fibrations
(3,829,298 inequivalent). The appearance of negative rank increase
reflects the fact that the Hodge numbers of the Jacobian fibration of
a genus-one fibration are generally not equal to those of the
genus-one fibered threefold
\cite{Anderson:2023wkr,Anderson:2023tfy}. To avoid this complication, in what follows
we will consider only genuine elliptic fibrations.

\begin{figure}
    \centering
    \includegraphics[scale=.75]{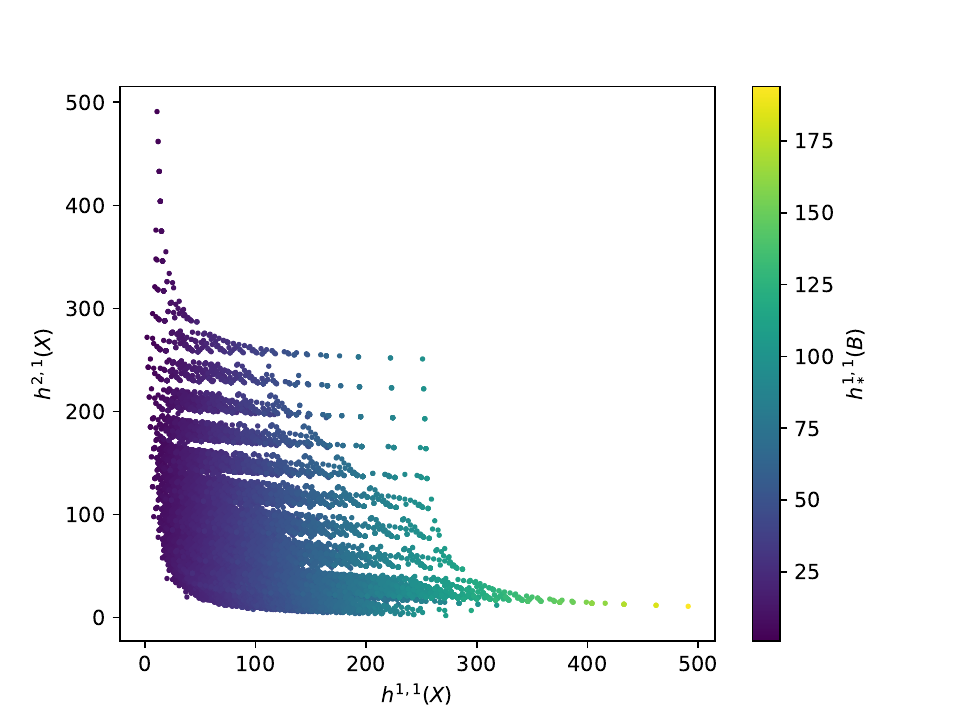}
    \caption{The Hodge numbers of the generic elliptic fibration over each base in the the MT list. For each Hodge pair, $h^{1,1}_*(B)$ is indicated by the color. 7,524 of the 30,108 unique Hodge pairs in the Kreuzer-Skarke list are realized in this way.}
    \label{fig:rank_zero_hodge}
\end{figure}

In the context of elliptic fibrations, we find a total of 630,778
(630,291 inequivalent) fibrations with rank increase zero, all of
which have fiber $F_{10}$; we find at least one such fibration over
each of the 61,539 bases in the MT list. A total of 10,558 bases
appear as the base for only a single rank-increase zero elliptic
fibration; in particular, each of the sixteen weak Fano bases appears
only once. The remaining 50,981 bases appear more than once; moreover,
all fibrations over a fixed base
with $\delta_{\rm tuned}= 0$
are found in polytopes with the same
Hodge pair, as expected from Shioda-Tate-Wazir.
For instance, the most common base for fibrations with
rank increase zero is MT base 3868, which appears as the base in 2,865
such fibrations; all of these fibrations originate in polytopes with
$h^{1,1}=h^{2,1}=39$.
\ignore{
\wati{I think the whole thing is almost by definition here, and I'm
  not sure we need to reference the standard stacking, just STW.suggest
  dropping the last sentence (commented out) and just added ``as
  expected'' above OK?}
\wati{removed arguments of... I don't think we need to repeat that
  here.  However, I'm not sure I agree with the conjecture, we could
  also have discrete gauge groups with a multisection and Jacobian
  fibration, right?  I'm not sure we have seen or ruled that out.I
  suggest modifying as below.}
In the absence of discrete gauge factors,
CY3s associated to rank increase zero fibrations
over a given base should all be equivalent
\wati{ I'm not sure what you mean here, you mean flop equivalent?
  Some of the differences might be related to different resolutions
  and/or flat fibrations.  Can you spell out a little more in detail
  what exactly this means?}
in the sense of
\cite{Gendler:2023ujl,Chandra:2023afu}, which among other things
requires that they all
have the same Hodge numbers \cite{Wall:1966rcd}.}
We have plotted the
Hodge numbers of the generic elliptic fibration over each base in the
MT list in Figure \ref{fig:rank_zero_hodge}.

\begin{figure}[h]
\begin{center}
\begin{subfigure}[b]{.45\textwidth}
\begin{center}
\includegraphics[scale=.5]{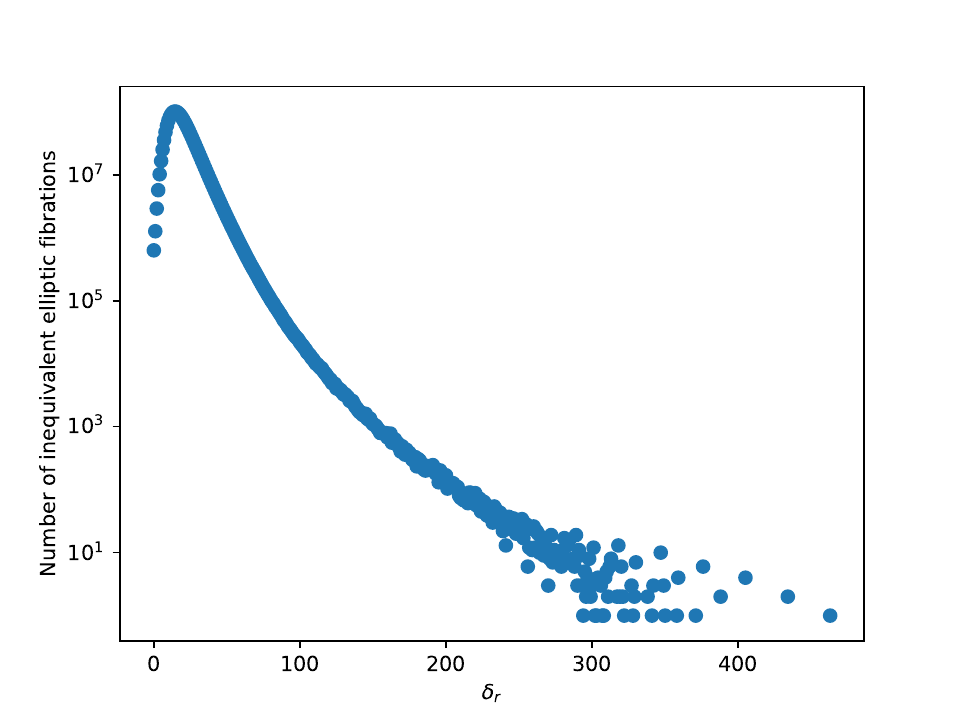}
\caption{}
\label{fig:rank_increase_elliptic_sum}
\end{center}
\end{subfigure}
~
\begin{subfigure}[b]{.45\textwidth}
\begin{center}
\includegraphics[scale=.5]{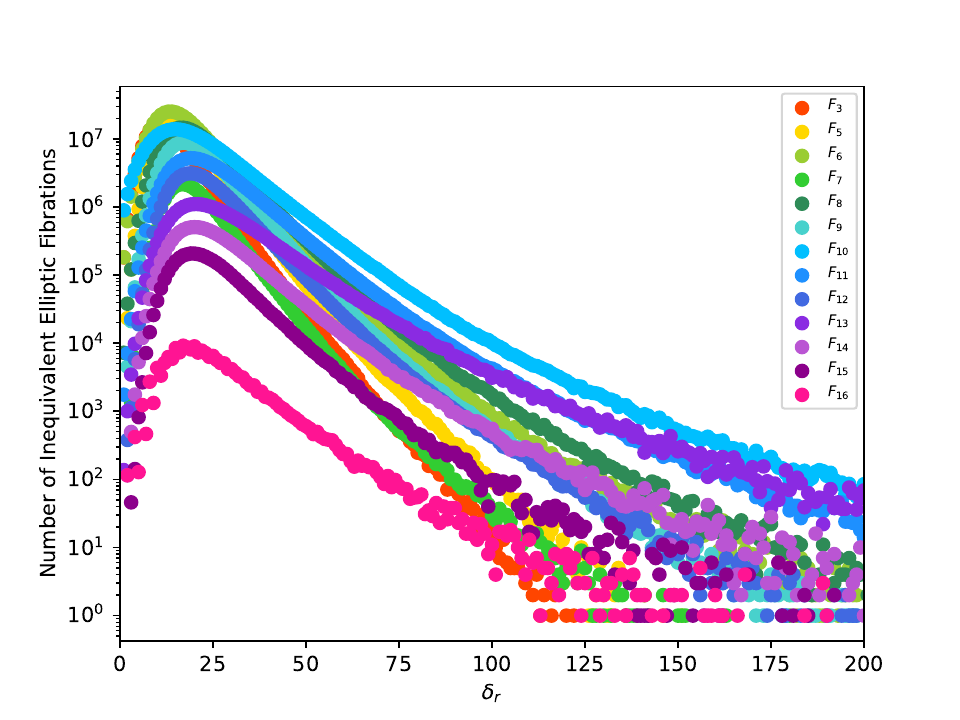}
\caption{}
\label{fig:rank_increase_elliptic_fiber}
\end{center}
\end{subfigure}
\caption{The number of inequivalent elliptic fibrations as a function of rank increase. In Figure (a), we have combined all fiber types; in Figure (b), we split the fibrations by fiber type and have zoomed in to the window $0\le\delta_r\le200$.}
\label{fig:rank_increase_elliptic}
\end{center}
\end{figure}

The remaining 1,828,049,985 elliptic fibrations (1,817,131,593 inequivalent) have positive rank increase. We have plotted the number of inequivalent elliptic fibrations of each rank increase in Figure \ref{fig:rank_increase_elliptic}. The largest bases that appear with positive rank increase are bases 61,521 and 61,523, each of which has $h^{1,1}_*(B)=150$ and appears as the base for a single fibration with $\delta_{\text{tuned}}=1$. The most common rank increase is $\delta_{\text{tuned}}=15$, which occurs in 101,205,632 elliptic fibrations (100,589,846 inequivalent). We find that large rank increases are extremely rare; only 3,933 (3,931 inequivalent) fibrations have $\delta_{\text{tuned}}\geq200$.

\begin{figure}
    \centering
    \includegraphics[scale=.75]{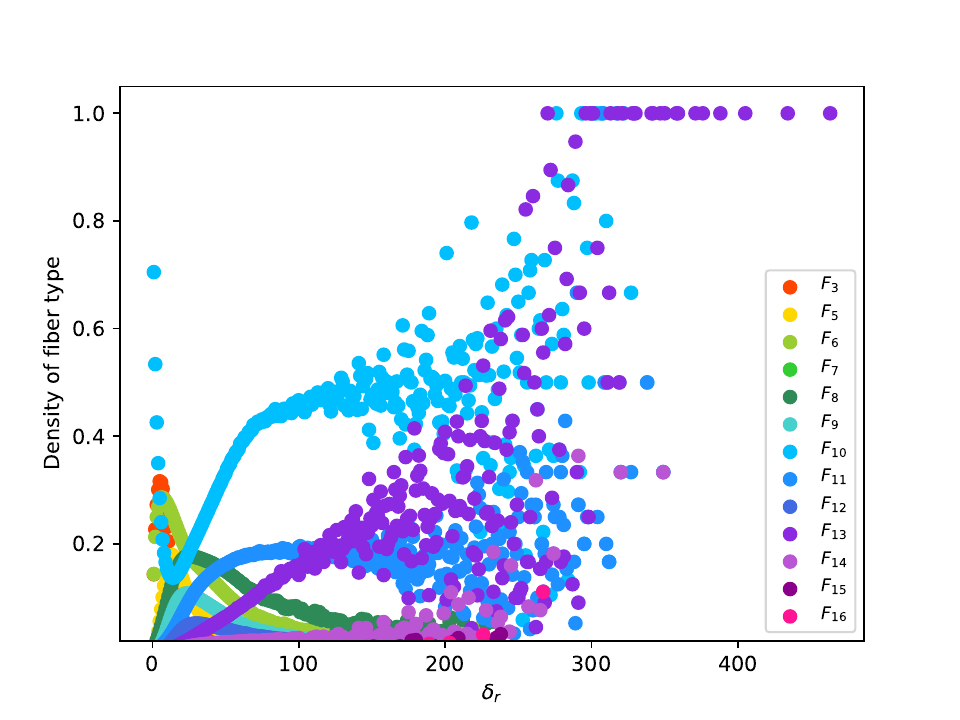}
    \caption{The relative density of each fiber type as a function of rank increase. Note that at large $\delta_r$, fibrations by fiber $F_{13}$ dominate.}
    \label{fig:rank_increase_density}
\end{figure}

We have plotted the relative density of each of the thirteen fiber
types of elliptic fibrations in Figure
\ref{fig:rank_increase_density}; from this figure, it is clear that
fibrations by fiber $F_{13}$ dominate at large $\delta_r$. The largest
rank increase is $\delta_r = 463$, found in an $F_{13}$ fibration over
base 680 in polytope v05-393, the unique polytope with
$h^{1,1}(X)=491$; we discuss this polytope and its fibrations in
\S\ref{sec:large-h11}. Indeed, this fiber-base pair is universal at
large $\delta_r$; the largest rank increase that does not come from an
$F_{13}$ fibration over base 680 is 350, found in an $F_{13}$
fibration over base 679 in a polytope v06-2907, which has
$h^{1,1}(X)=376$, $h^{2,1}(X)=10$. By Eq.~\ref{eq:272}, given an
elliptic fibration in a CY3 with Hodge numbers $h^{1,1}, h^{2,1}$, we
can sometimes trade one unit of $h^{1,1}$ for 29 units of
$h^{2,1}$. The prevalence of fibrations of $F_{13}$ over base 680 at
large $\delta_r$ can (at least in part) be explained by the mirror of
this process (which will also be described further in \cite{ot-forthcoming}): their Hodge numbers are given by $h^{1,1}(X)=491-29n$, $h^{2,1}(X)=11+n$, as expected for the images of the $F_{13}$ fibration in the $h^{1,1}=491$ polytope under the mirror of these $\Delta{T}=1$ transitions. Indeed, in Figure \ref{fig:polica} it is clear that fibrations over base 680 dominate the large $h^{1,1}(X)$ regime. 

Finally, we note a curious fact: each polytope hosting one of these large rank-increase fibrations of fiber $F_{13}$ over base 680 also possesses a second fibration, namely the generic elliptic fibration over a base with very large $h^{1,1}_*(B)$. For instance, as discussed in \S\ref{sec:large-h11}, the polytope with $h^{1,1}(X)=491$ also hosts the generic elliptic fibration over the largest toric base. Although we currently cannot explain this empirical phenomenon, we hope that the ideas of \cite{Anderson:2016cdu} can be applied to these pairs of fibrations.

\section{Examples}
\label{sec:examples}
In the previous section, we discussed properties of the entire
ensemble of fibrations. In this section, we take the opposite
approach, and analyze a small number of explicit fibrations and
their description in F-theory, focusing primarily on those that either
exhibit exotic features or connect to the existing literature. In
general, F-theory compactifications on genuine elliptic fibrations are
much better understood than those on genus-one fibrations, and so in
this section we will focus on elliptic fibrations. Our goal in this section is twofold:
to use the F-theory description of
elliptic Calabi-Yaus as a way of understanding geometric structure, and to understand what kinds of F-theory models are realized through these constructions.


\subsection{Fibrations with base $\bbP^2$}
\label{sec:volker}
In \S\ref{sec:fano-v-nonfano}, we enumerated fibrations over the
sixteen weak Fano bases. Note that one of the Fano bases, MT ID 1, is
simply $\bbP^2$. Automorphism-inequivalent toric fibrations over
$\bbP^2$ were enumerated in \cite{Braun:2011ux}; our count of 102,581
such fibrations exactly matches with the count in that
work.\footnote{We thank V. Braun for helpful discussions on this
comparison.} In the following subsections, and in particular
\ref{sec:SCFT}, we will re-analyze some specific examples discussed in
\cite{Braun:2011ux}
and provide a new perspective on their physics and geometry.

\subsection{Gauge groups over non-toric divisors; Examples: SU($N$) on $\P^2$}
\label{sec:non-toric-gauge}

One interesting feature of the fibrations found
in the KS database is that, in many cases, nonabelian groups are tuned
over non-toric divisors in a toric base.  This can occur with a variety of fiber
types, and even with the fiber $F_{10}$ with non-standard stackings.
This can occur for any fiber type when the divisor associated to one of the ``generic'' nonabelian gauge factors appearing in the appropriate line of Table~\ref{tab:16fibers}  is not simply a single toric divisor.\footnote{While these are the simplest
examples of gauge groups hosted on nontoric divisors, they are not the
only type; in the next subsection, we will encounter examples where
the nontoric divisors hosting nonabelian gauge groups are absent in
the toric description in the base, and only appear after resolving
loci of the base where the fibration is ``too singular'', i.e. after
resolving (4,6) points in the base, generally producing a new base which is itself non-toric.}
Inspecting Table \ref{tab:16fibers}, we see that the only nonabelian
gauge factors we can hope to realize in this way are
those with algebras $\agsu(2)$, $\agsu(3)$,
and $\agsu(4)$. In this subsection, we demonstrate that all three of
these types of gauge group can be realized on nontoric divisors in elliptic
fibrations over $\bbP^2$.

\subsubsection{SU(2) on non-toric divisors}

As a specific example of this kind of construction,
one can realize $SU(2)$ gauge groups on divisors of
degree up to the maximum allowed degree of twelve. Consider the family of polytopes whose vertices are given by the columns of the matrix \eqq{\left(\begin{array}{cccccc}
1&0&-2&0&0&-6+n\\
0&1&-3&0&0&-9+n\\
0&0&0&1&0&-1\\
0&0&0&0&1&-1
\end{array}\right)\label{eq:su2nhverts}}
for $1\le n\le 12$; these polytopes are summarized in Table \ref{tab:su2nhTable}.

\begin{table}[h]
\begin{centering}
\begin{tabular}{|c||c|c|c|}\hline
 ID  & $(h^{1,1},h^{2,1})$ & $n_\beta$ & $\operatorname{deg}(\beta)$ \\\hline\hline
 v07-192 & (3,231) & 3 & 1 \\\hline
 v07-190 &  (3,195) & 6 & 2 \\\hline
 v07-189 &  (3,165) & 10 & 3 \\\hline
 v07-187 &  (3,141) & 15 & 4 \\\hline
 v07-176 & (3,123) & 21 & 5 \\\hline
 v07-147 & (3,111) & 28 & 6 \\\hline
 v07-134  & (3,105) & 36 & 7 \\\hline
 v07-136 &  (3,105) & 45 & 8 \\\hline
 v05-64 & (4, 112) & 55 & 9\\\hline
 v07-748 & (4, 126) & 66 & 10\\\hline
 v07-771 & (4, 144) & 78 & 11 \\\hline
 v05-48 & (3, 165) & 91 & 12\\\hline
\end{tabular}
\caption{Polytopes with fibrations over $\bbP^2$ exhibiting an $\mathfrak{su}$(2) gauge algebra tuned on the nontoric divisor $nH$ for $1\le n\le 12$. For each polytope, we give its ID, its Hodge numbers, the number of monomials in $\beta$, and the degree of $\beta$ as a divisor in the base, so that $SU(2)$ is tuned on $\operatorname{deg}(\beta)H$.}
\label{tab:su2nhTable}
\end{centering}
\end{table}

The cases $9\le n \le 11$ are subtle, so we begin with the remaining nine cases.
These polytopes all have $h^{1,1}=3$, and $h^{2,1}$ is determined by
$n$ via \eqq{h^{2,1}(n) = 3n^2 - 45n + 273.} Each of
these polytopes has a single fibration, a nonstandard stacking of base $\bbP^2$ over fiber $F_{10}$. The total gauge group has rank 1 in each
case, and we thus expect the $SU(3)$ factors must be absent; this
implies that $\alpha$ must consist of a single monomial in each
example, and this is in fact the case. However, the $SU(2)$ factor can
be present, and
explicit analysis of the Weierstrass model shows that such a factor
indeed is present in each of these eight polytopes. In
each case, $\beta$ consists of \eqq{n_\beta = \frac{(n+1)(n+2)}{2}} monomials,
 forming a generic homogeneous polynomial of degree $n$ on $\P^2$.
(As discussed in \S~\ref{sec:Nagell}, monomials in $\beta$ correspond
to points $(2, -1; m)$ in the dual polytope, which are easily
enumerated.)
Thus the fibrations of these polytopes contain an $SU(2)$ gauge
group that has been tuned over a degree-$n$ curve in the base, or equivalently on the
nontoric divisor $nH$. 
From anomaly cancellation (see e.g.\  Table 7 in
\cite{Johnson:2016qar}), we know that  the massless matter spectrum
charged under such an SU(2) consists of $6 n^2 +16 (1 - g)$  fields in
the fundamental representation and $g$ fields in the adjoint
representation, where $g = (n -1) (n -2)/2$ is the genus of the
generic degree-$n$ curve.  Recalling that the Cartan elements of the
adjoint  do not contribute to $H_{\text{charged}}$, we see that this
matter spectrum nicely matches with  (\ref{eq:272}) and the above
formula for $h^{2,1}$.

Next, we consider the case $n=9$, for which Eq.~\ref{eq:su2nhverts}
once again furnishes a reflexive polytope, with Hodge numbers
$h^{1,1}(X)=4$, $h^{2,1}(X)= 112$. As before, this polytope has a
single fibration, a nonstandard stacking of $\bbP^2$ over of $F_{10}$, and for
$n=9$, we find that $\beta$ consists of 55 monomials, as expected for
$\mathfrak{su}$(2) hosted on a degree-nine curve.
From the above discussion, for $\mathfrak{su}$(2) on a nonic
curve we expect 28 adjoints and 54 fundamentals, which would yield
$h^{2,1}(X)=111$.
To satisfy
Eq.~\ref{eq:STW}, we require an additional gauge factor; since no
toric divisors host nonabelian gauge groups and $\alpha$ consists of a
single monomial, we conclude that the missing gauge factor is a
$U(1)$.
Furthermore, since the discrepancies in the
Hodge numbers in the model without the U(1) factor
are the same, all matter charged under this abelian
 gauge factor must also be charged under the SU(2).  These constraints
 are nicely satisfied by the theory with gauge group $(SU(2) \times
 U(1))/\Z_2$ , in which there are 27 fields in each of the
 representations $({\bf 2}, \pm{\bf 1})$.  This theory
   illustrates two general principles
that have been identified  in 6D F-theory models, and are hypothesized
to hold in general but have not yet been fully proven.  First, the
global structure of the gauge group is fixed by the  {\it massless
  charge completeness} hypothesis \cite{Morrison:2021wuv}, which suggests
that the gauge group does not contain any central elements under which
the massless spectrum is invariant; this spectrum is invariant
under the $\Z_2$ diagonal central element of $SU(2) \times U(1)$, so we select the quotient to project out this element.  Second, the  {\it automatic
  enhancement conjecture} \cite{Raghuram:2020vxm} states that the additional U(1)
factor here is a necessary consequence of the SU(2) spectrum; that is,
since the  U(1) factor cannot be broken through a Higgs mechanism
while preserving the SU(2) symmetry,
any Weierstrass model over $\P^2$ with an SU(2)
tuned over a degree 9  curve will automatically force the presence of an
additional U(1) factor of this form that enhances the Mordell-Weil
group of the associated threefold.  
Indeed, precisely this example (as well as the cases with $n = 10, 11,
12$) is discussed in
\cite{Raghuram:2020vxm}.

Next, we consider the cases $n=10,11$, for which
Eq.~\ref{eq:su2nhverts} does not furnish a reflexive
polytope. Instead, we employ the ``dual of the dual'' construction
described above to find the polytopes whose vertices are given by the
columns of the matrix \eqq{\left(\begin{array}{cccccc}
    2&0&-2&0&0&-6+n\\ 1&1&-3&0&0&-9+n\\ 0&0&0&1&0&-1\\ 0&0&0&0&1&-1
\end{array}\right).\label{eq:su2nhvertsf13}}
These are reflexive polytopes with $h^{1,1}(X)=4$ and \eqq{h^{2,1}(X) = \left\{\begin{array}{cc}
     126,& n=10 \\
     144, & n = 11
\end{array}\right. = 3n^2 - 45n + 276.}

These polytopes each have a single fiber, with vertices given by the columns of the matrix \eqq{\left(\begin{array}{ccc}
2&0&-2\\
1&1&-3\\
0&0&0\\
0&0&0
\end{array}\right).} These subpolytopes have three vertices and seven
total points each, and so we see that the effect of the ``dual of the
dual'' procedure is to replace the $F_{10}$ fiber in
Eq.~\ref{eq:su2nhverts} with an $F_{13}$ fiber. The generic gauge
group associated to $F_{13}$ fibrations was determined in
\cite{Klevers:2014bqa} to be $(SU(4)\times SU(2)^2)/\bbZ_2$; in terms
of the Weierstrass coefficients in
Eq.~\ref{eq:longWeierstrassWithAlphaBeta}, the $\mathfrak{su}(4)$ is
hosted on $\alpha$, one $\mathfrak{su}(2)$ is hosted on $\beta$ as in
$F_{10}$ fibrations, and the other $\mathfrak{su}(2)$ is hosted on
$a_4$. In these polytopes, the $\mathfrak{su}(4)$ is not turned on,
but both $\mathfrak{su}(2)$ factors are. As above, $\beta$ consists of
$(n+1)(n+2)/2$ monomials, and so for $n=10,11$, we have an
$\mathfrak{su}(2)$ tuned on divisors of degree 10 and 11,
respectively, as desired. On the other hand, $a_4$ consists of
$(13-n)(14-n)/2$ monomials, and so the other $\mathfrak{su}(2)$ is
tuned on divisors of degrees 2 and 1, respectively.
These cases exactly match the expectation from the automatic
enhancement conjecture.  Tuning SU(2) on a degree 10 or 11 curve forces a
gauge group $(SU(2) \times SU(2))/\Z_2$, where all matter charged
under the second SU(2) is also charged under the first SU(2) factor.
For degree 10, there are 36 adjoints under the first SU(2) and 20
bifundamental fields $({\bf 2}, {\bf 2})$, and in degree 11 there are
45 adjoints under the first SU(2) and 11 bifundamental fields.  Note
that in both cases the global structure of the gauge group is also
compatible with the massless charge completeness conjecture.

It is also worth noting that for the degree 12 case, the spectrum
consists of 55 adjoints, and the gauge group is actually $SO(3) =
SU(2)/\Z_2$.  This matches the massless charge completeness hypothesis
and is shown explicitly by constructing the explicit  torsion section
in this case in \cite{Morrison:2021wuv}.

Finally, it is worth noting that this set of polytopes thus includes
every single possible elliptic Calabi-Yau threefold with a $\P^2$ base
and the minimal gauge group and matter compatible with tuning only a
nonabelian SU(2) factor. For $n = 13$, the anomaly cancellation
conditions cannot be satisfied for any SU(2) matter content.
Furthermore, this set of polytopes includes all possible elliptic
Calabi-Yau threefolds with $h^{1,1} (X) = 3$ and only a nonabelian
gauge group in the F-theory picture, for similar reasons to the
above. 

\subsubsection{SU(4) on a non-toric divisor}

Using this general mechanism, we expect a variety of realizations of
SU(2), SU(3), and SU(4) gauge factors over non-toric divisors, often
in combination with U(1) and other gauge factors.  We have not
identified cases with any other gauge factors over non-toric divisors
in purely toric bases, although as mentioned above, there are gauge
groups that can be tuned over non-toric divisors in   more general
 non-toric
resolved bases.

We also provide some simple but interesting examples where
 $SU(3)$ and $SU(4)$ 
are realized in this way
 in fibrations over $\bbP^2$.
  We begin with the SU(4) case, which is conceptually slightly
  more straightforward. To realize $SU(4)$  on a non-toric divisor in $\P^2$, we consider the
polytope v05-110, whose vertices are given by the columns of the
matrix
\eqq{\left(\begin{array}{ccccc}-1&-1&1&1&1\\-2&2&0&0&0\\0&0&1&0&-1\\0&0&0&1&-1\end{array}\right).}
This polytope has $h^{1,1}=6,$ $h^{2,1}=60$, and is a stacking
of $\bbP^2$ over the vertex $(1,0)$ of the fiber $F_{13}$. The
analysis of \cite{Klevers:2014bqa} indicates that $SU(4)$ is tuned on
the vanishing locus of the  coefficient $a_{012}$ in the cubic; in this
geometry, $a_{012}$  contains 28 monomials, as expected for a
generic polynomial of degree 6, so we conclude that this
fibration has an $SU(4)$ gauge group on a sextic in $\P^2$.
  From anomaly cancellation, we can ascertain that the massless
  spectrum for such an SU(4) factor will consist of no fields in the
  fundamental, 18 fields in the antisymmetric ({\bf 6})
  representation, and 10 fields in the adjoint ({\bf 15}) representation.
 If this
 was the only gauge group and matter, we would expect to have a
 Calabi-Yau threefold with $h^{1,1} =  5, h^{2,1} = 59$.
 Thus, there is an additional hidden gauge factor.  Since there are no
 other nonabelian factors manifest in the Weierstrass model, this must again
 be  an  abelian factor.  Furthermore, since the discrepancies in the
 Hodge numbers are the same, again,
 all matter charged under this abelian
 gauge factor must also be charged under the nonabelian factor SU(4).  These constraints
 are nicely satisfied by the theory with gauge group
\begin{equation}
G =(SU(4) \times
 U(1))/\Z_2\,,
\label{eq:}
\end{equation}
 in which there are 9 fields in each of the
 representations $({\bf 6}, \pm{\bf 1})$.  This theory
 provides another nice example of both the  massless
  charge completeness hypothesis and the   automatic
  enhancement conjecture, which suggests that the additional U(1)
factor here is a necessary consequence of the SU(4) spectrum; that is,
since the  U(1) factor cannot be broken through a Higgs mechanism
while preserving the SU(4) symmetry,
it seems that any Weierstrass model over $\P^2$ with an SU(4)
tuned over a sextic  curve will automatically force the presence of an
additional U(1) factor of this form that enhances the Mordell-Weil
group of the associated threefold.  We have not checked this
explicitly but leave confirmation of this as an interesting
calculation for further work.

\subsubsection{SU(3) on a non-toric curve}
\label{sec:su3}

To realize SU$(3)$, we consider polytope v05-108, a nonstandard
stacking of a $\bbP^2$ base over fiber $F_{10}$ whose vertices are
given by the columns of the matrix \eqq{\left(\begin{array}{ccccc} 1 &
    0 & 0 & 0 & -2\\ 0 & 1 & 1 & 1 & -3\\ 0 & 1 & 0 & -1 & 0\\ 0 & 0 &
    1 & -1 & 0\end{array}\right).} This polytope has $h^{1,1}=6,$
$h^{2,1}=60$, and is a stacking of $\bbP^2$ over the vertex of
$F_{10}$ corresponding to the Weierstrass variable $y$ in
Eq.~\ref{eq:F10weierstrass}. As discussed in \ref{sec:Nagell}, we
expect the divisor $\alpha$ sitting in front of the $y^2$ term in the
generalized long Weierstrass form to host an SU(3) gauge group
whenever it consists of more than one monomial. In this polytope,
$\alpha$ consists of 28 monomials, and so we  realize an SU(3)
gauge group on a  generic sextic curve in $\P^2$.  Anomaly cancellation again
 allows us to identify the spectrum  of massless charged fields, which
 consists of 54 fields in the fundamental and 10 fields in the adjoint.  If this
 was the only gauge group and matter, we would expect to have a
 Calabi-Yau threefold with $h^{1,1} =  4, h^{2,1} = 58$.  Thus, there
 must be a hidden gauge group.  Again, in this case there is no
 manifest nonabelian factor so we must have two U(1) factors,   and
 again  all matter charged under the U(1) factors must be also charged
 under the SU(3).  A solution to this set of constraints is given by a
 theory with gauge group
\begin{equation}
G =(SU(3) \times U(1) \times U(1))/\Z_3\,,
\label{eq:3113}
\end{equation}
 containing 18 fields in each of the representations $({\bf 3},  {\bf
  1},  {\bf 1}), ({\bf 3},  {\bf
  -1},  {\bf 0}), ({\bf 3},  {\bf
  0},  {\bf -1})$.  This spectrum again manifests the principles of
 massless charge completeness and automatic enhancement; it seems that
 any tuning of SU(3) on a degree 6 curve in $\P^2$ will automatically
 have these extra U(1) factors in this form.  Note that this spectrum
 can also be understood as arising from a Higgsing of a theory with
 gauge group $(SU(3) \times SU(3))/\Z_3$, with the SU(3) factors
 respectively living on a sextic  and a cubic, where  Higgsing is
 carried out on the single adjoint representation of the second SU(3)
 factor.  This construction of the above $U(1) \times U(1)$ spectrum
 was realized in \cite{Cvetic:2015ioa}.

\subsection{SCFTs and non-toric bases; Example: Maximum rank increase over $\P^2$}
\label{sec:SCFT}

It is interesting to consider the example of the fibration over $\P^2$
with the largest $h^{1,1} (X)$.  This Calabi Yau was identified in
\cite{Braun:2011ux}, where it was noted that the apparent gauge group
is $SU(27)$, supported on the divisor $H$, and
there are an additional 84 divisors encoded in non-flat fibers, giving
a total of $h^{1,1} (X) = 112$. The polytope, whose ID is v05-65, has $h^{2,1}=4$, and the fiber type in this fibration is $F_{16}$. We can understand this geometry in terms of an interesting SCFT with a
natural geometric resolution.

Simply from the structure of the gauge group and an analysis of the
F-theory and Weierstrass models, it is known that the gauge group SU(27) cannot be
realized on the base $\P^2$ in a straightforward way.  Anomaly cancellation considerations show
that with only the gauge group SU($N$), for $N >24$ the number of fundamental
matter representations would go negative, and SU(24)
is the largest SU($N$) group that can be tuned in an F-theory
Weierstrass model without singularities outside the Kodaira
classification of Table~\ref{tab:kodaira} \cite{Morrison:2011mb}.

Indeed, as pointed out in \cite{Braun:2011ux}, in this model, there are three points
on the base where the orders of vanishing of the Weierstrass
coefficient $f, g$ reach (4, 6), signaling a non-flat fiber, which can
be interpreted as an SCFT coupled to the gravity theory \cite{HeckmanMorrisonVafa,Heckman:2015bfa}.

To analyze this geometry in more detail, we consider the explicit
Weierstrass model for this fibration.
Since the fiber is $F_{16}$, the monomials live in the dual fiber
$F_{1}$, and we can think of this as a specialized  Tate-form
Weierstrass model as in Eq.~\ref{eq:longWeierstrass}, where the only
non-vanishing $a_i$ coefficients are $a_1, a_3$.
More explicitly, we can put the polytope into (fiber; base)
coordinates where the
vertices are the columns of the matrix
\begin{equation}
  {\rm vertices} (\nabla) =
\left(\begin{array}{ccccc}
0 &  0 &  6 & -3 & -3\\
0 &  0 & -3 &  6 & -3\\
1 &  0 & -1 & -1 & -1\\
0 &  1 & -1 & -1 & -1
\end{array}\right).
\label{eq:}
\end{equation}
so the vertices of the dual polytope are
 \begin{equation}
   {\rm vertices} (\nabla^\circ) =
\left(\begin{array}{ccccc}
 1 &  0 &  0 &  0 & -1\\
 0 &  1 &  0 &  0 & -1\\
-1 & -1 &  2 & -1 & -1\\
-1 & -1 & -1 &  2 & -1
\end{array}\right)\,,
\label{eq:}
\end{equation}
or after a linear transformation
\begin{equation}
  {\rm vertices} (\nabla^\circ) =
\left(\begin{array}{ccccc}
 1 & 0 & 0 &  0 & -1\\
 0 & 1 & 0 &  0 & -1\\
 0 & 0 & 2 & -1 & -3\\
 0 & 0 & -1 &  2 & -3
\end{array}\right).
\label{eq:}
\end{equation}
This corresponds to a Tate form Weierstrass model with $a_1$ a
general cubic and $a_3$ of the form $w^9$.

The coefficients $f, g$ and the discriminant can be written in local coordinates $(x, w)$ on the
base, where $x=0$ is a (generically non-toric) root of $a_1=x
\tilde{g}(x, w) =0$, as
\begin{eqnarray}
f & =  & - (a_1^4-24 a_1 a_3)/48 \sim x (x^3 \tilde{g}(x, w)^3 + w^9)\tilde{g}(x, w)\,,\\
g & = & - (a_1^6+36 a_1^3a_3 -216 a_3^2)/864 \sim (x^6 \tilde{g}(x,
w)^6+x^3 \tilde{g}(x, w)^3w^9+ w^{18}) \,,
\end{eqnarray}
and
\begin{equation}
 \Delta = a_3^3 (a_1^3 -27 a_3) \sim w^{27} (w^{9}+ x^{3} \tilde{g}(x, w)^3) \,,
\label{eq:}
\end{equation}
where $\sim$ means that we have dropped constant coefficients.
There is a (4, 6) locus at the non-toric point $x = w = 0$.  We blow
up at this point by making the new local chart $(x, w) \rightarrow (u\hat{x},
u)$.  We then have
\begin{align}
  f & \rightarrow u^{4} [\hat{x}^{4} \tilde{g}^{4} + u^{5} \tilde{g}]\\
  g & \rightarrow u^{6} [\hat{x}^{6} \tilde{g}^{6}+
    \hat{x}^{3} \tilde{g}^{3}u^{6} + u^{15}]\\
  \Delta &\rightarrow u^{12}
         [u^{15}(u^{9}+ u^{3}x^{3}\tilde{g}^{3})] \,,
\label{eq:}
\end{align}
where $\tilde{g}$ is a nonzero constant plus higher order terms in $u, x$.
Dropping the factor of $u^{12}$ from $\Delta$ to take the proper transform, we get
\begin{equation}
 \Delta \rightarrow  u^{18} (u^{6} +  \hat{x}^{3}\tilde{g}^3) \,.
\label{eq:}
\end{equation}
We thus see an SU(18) factor on $u = 0$, which is now a -1 curve,
while $w = 0$  becomes a -2 curve after blowing up the (4, 6) point at
each of the three roots of $a_1$.
Following this branch further,
there is another (4, 6) locus at the (non-toric) point
where $\hat{x}= 0$
on $u = 0$.
Blowing that up, a similar analysis shows that
we get an SU(9) factor and one more point we must
blow up, giving a chain of curves of self-intersections -2, -2, and -1, with the -2 curves carrying SU(18) and SU(9) gauge factors, respectively.  There are three points like this on the original  SU(27)
locus, so the final geometry is as shown in Figure~\ref{f:27-SCFT}.

\begin{figure}
\begin{center}
\begin{picture}(200,90)(- 100,- 40)
{\color{blue} 
\put(-100,-30){\line(1,0){200}}
\put(0, -45){\makebox(0,0){\small SU(27) [+1 {\color{black}{$\rightarrow -2$}}]}}
\put(-73.33,-30){\circle*{4}}
\put(66.66,-30){\circle*{4}}
\put(-3.33,-30){\circle*{4}}
}
{
\put(-75,-35){\line(1,3){10}}
{\color{red}\put(-5,-35){\line(1,3){10}}}
\put(65,-35){\line(1,3){10}}
\put(-75,15){\line(1,3){10}}
{\color{red}\put(-5,15){\line(1,3){10}}}
\put(65,15){\line(1,3){10}}
\put(-55, 30){\makebox(0,0){[-1]}}
\put(-55, 5){\makebox(0,0){[-2]}}
\put(-55, -20){\makebox(0,0){[-2]}}
\put(-90, 5){\makebox(0,0){\small SU(9)}}
\put(-90, -20){\makebox(0,0){\small SU(18)}}

{\color{red}
\put(15, 30){\makebox(0,0){[-1]}}
\put(15, 5){\makebox(0,0){[-2]}}
\put(15, -20){\makebox(0,0){[-2]}}
\put(-20, 5){\makebox(0,0){\small SU(9)}}
\put(-20, -20){\makebox(0,0){\small SU(18)}}
}

\put(85, 30){\makebox(0,0){[-1]}}
\put(85, 5){\makebox(0,0){[-2]}}
\put(85, -20){\makebox(0,0){[-2]}}
\put(50, 5){\makebox(0,0){\small SU(9)}}
\put(50, -20){\makebox(0,0){\small SU(18)}}

{\color{red}\put(5,-10){\line(-1,3){10}}}
\put(75,-10){\line(-1,3){10}}
\put(-65,-10){\line(-1,3){10}}
}
\end{picture}

\end{center}
\caption[x]{\footnotesize  The elliptic fibration over $\P^2$ with
  largest $h^{1,1} (X)\ (= 112)$,  was identified by Braun, and has an
  apparently anomalous gauge group of SU(27) tuned on a +1 curve $H$.
  This is explained by the presence of three SCFT factors associated
  with (4, 6) loci on $H$.  Resolving the geometry by going out on the
tensor branch of these SCFTs gives a theory on a non-toric base with
$h^{1,1} (B)= 10$ and a gauge group SU(27) $\times$ (SU(18) $\times$
SU(9))$^{3}$.  The figure depicts the final arrangement of negative
self-intersection curves on the non-toric blown-up base.}
\label{f:27-SCFT}
\end{figure}

In the final resolved geometry, we have a non-toric (but semi-toric
\cite{MartiniTaylorSemitoric}) base with $h^{1,1} (B) = 10$.  The
gauge group is  SU(27) $\times$ (SU(18) $\times$
SU(9))$^{3}$, and all seven SU($N$) factors live on curves of
self-intersection -2.  In general, the matter content for an SU($N$)
gauge factor on a -2 curve is $2 N$  fundamental matter fields.  At
each intersection between $\SU(N) \times \SU(N')$ a bifundamental
matter field is supported.  It is easy to check that these
bifundamentals exactly saturate the anomaly cancellation condition for
each $\SU(N)$ factor.  The total number of charged matter fields is $3
\times 18 \times (9+27) = 1944.$  We have $T = 9$, and there are  $V=
1937$ vector multiplets, from which it is easy to confirm that all anomaly cancellation conditions are properly satisfied with $h^{2,1} = 4$.

This example illustrates a general feature of many fibrations in the
database: in the elliptically fibered phase, there are non-flat fibers
associated with (4, 6) fibration singularities in the Weierstrass
model, corresponding in the F-theory picture to local strongly coupled
superconformal field theory sectors.  There is a resolved (non-toric) phase for
such Calabi-Yau threefolds  in which points on the base are blown up,
giving generically a non-toric base, often with an enhanced gauge
group.
\footnote{Note that similar things can happen for different flop
phases/triangulations of a CY where both bases are toric; for example,
for a specific triangulation of the generic elliptic fibration (simple
stacking) over Hirzebruch $\F_1$, the resulting CY3 is elliptically
fibered over $\P^2$ with a non-flat fiber over the toric point that
would be blown up to form $\F_1$ \cite{rw-forthcoming}; we expect that many examples of this
that connect to a non-toric base phase similarly can be associated
with flop transitions in the extended K\"ahler cone.}
This is a much more dramatic version of the familiar presence
of (4, 6) loci on -9, -10, and -11 curves in a toric base.  
The structures of this kind that arise in the toric hypersurface
database seem to describe elliptic fibrations over a wide range of
non-toric bases.  We have not explored this in any depth, but leave
for further study the range of non-toric bases that are realized in
this fashion in the database.  It would be interesting to compare the
set of bases realized in this way with previous partial analyses of
the full set of non-toric bases \cite{MartiniTaylorSemitoric,TaylorWangNon-toric}.

\subsection{The most common base (base 72) and fiber ($F_{6}$)}
\label{sec:common}

As described in \S\ref{sec:statistics}, the base that supports the
most fibrations is base 72, and the
most common fiber is fiber $F_6$.
Thus, we may think of this base and fiber as characteristic of the
structure of a ``typical'' Calabi-Yau threefold chosen by taking a
random sample from the KS database.
We explore here the structure of fibrations with this fiber and base in more detail.

We found that the most common fiber is $F_6$, so clearly that fiber merits special attention. The generic elliptic curve in $F_6$ takes the form \begin{align}a_{300}u^3+a_{210}u^2v+a_{120}uv^2 + a_{030}v^3+a_{201}u^2w+a_{111}uvw + a_{021}v^2w + a_{102}uw^2 = 0,\end{align} which is a specialization of the generic cubic in $\bbP^2$. Passing to Weierstrass form as described in Appendix \ref{sec:Weierstrass}, we find that the Weierstrass coefficients $f,g$ are given in terms of the $a_{ijk}$ as 
\begin{subequations} \label{eq:F6weierstrass}
    \begin{align}
        f &= -\frac{1}{48} a_{111}^{4} + \frac{1}{6} a_{201} a_{111}^{2} a_{021} - \frac{1}{3} a_{201}^{2} a_{021}^{2} - \frac{1}{2} a_{030} a_{201} a_{111} a_{102} + \frac{1}{6} a_{120} a_{111}^{2} a_{102} \nonumber\\ &~~~ + \frac{1}{3} a_{120} a_{201} a_{021} a_{102}   - \frac{1}{2} a_{210} a_{111} a_{021} a_{102} + a_{300} a_{021}^{2} a_{102} - \frac{1}{3} a_{120}^{2} a_{102}^{2} + a_{210} a_{030} a_{102}^{2}, \\ 
        g &= \frac{1}{864} a_{111}^{6} - \frac{1}{72} a_{201} a_{111}^{4} a_{021} + \frac{1}{18} a_{201}^{2} a_{111}^{2} a_{021}^{2} - \frac{2}{27} a_{201}^{3} a_{021}^{3} + \frac{1}{24} a_{030} a_{201} a_{111}^{3} a_{102} - \frac{1}{72} a_{120} a_{111}^{4} a_{102} \nonumber\\ &~~~ - \frac{1}{6} a_{030} a_{201}^{2} a_{111} a_{021} a_{102} + \frac{1}{36} a_{120} a_{201} a_{111}^{2} a_{021} a_{102} + \frac{1}{24} a_{210} a_{111}^{3} a_{021} a_{102} + \frac{1}{9} a_{120} a_{201}^{2} a_{021}^{2} a_{102} \nonumber\\ &~~~ - \frac{1}{6} a_{210} a_{201} a_{111} a_{021}^{2} a_{102} - \frac{1}{12} a_{300} a_{111}^{2} a_{021}^{2} a_{102} + \frac{1}{3} a_{300} a_{201} a_{021}^{3} a_{102} + \frac{1}{4} a_{030}^{2} a_{201}^{2} a_{102}^{2} \nonumber\\ &~~~ - \frac{1}{6} a_{120} a_{030} a_{201} a_{111} a_{102}^{2} + \frac{1}{18} a_{120}^{2} a_{111}^{2} a_{102}^{2} - \frac{1}{12} a_{210} a_{030} a_{111}^{2} a_{102}^{2} + \frac{1}{9} a_{120}^{2} a_{201} a_{021} a_{102}^{2} \nonumber\\ &~~~- \frac{1}{6} a_{210} a_{030} a_{201} a_{021} a_{102}^{2} - \frac{1}{6} a_{210} a_{120} a_{111} a_{021} a_{102}^{2} + a_{300} a_{030} a_{111} a_{021} a_{102}^{2} + \frac{1}{4} a_{210}^{2} a_{021}^{2} a_{102}^{2} \nonumber\\ &~~~ - \frac{2}{3} a_{300} a_{120} a_{021}^{2} a_{102}^{2} - \frac{2}{27} a_{120}^{3} a_{102}^{3} + \frac{1}{3} a_{210} a_{120} a_{030} a_{102}^{3} - a_{300} a_{030}^{2} a_{102}^{3}.
    \end{align}
\end{subequations}
Note that these $f,g$ do not generically factor. However, $\Delta$
does: \begin{align}\Delta = -16\left(4f^3+27g^2\right) \sim
  -a_{102}^2\left(\text{ten-ic polynomial in the remaining
  }a_{ijk}\right).\end{align} Thus, whenever $a_{102}$ consists of
more than one monomial, we have a Tate-tuned $\mathfrak{su}(2)$ over
the locus $a_{102}=0$ \cite{Klevers:2014bqa}. Additionally, it was
shown in \cite{Klevers:2014bqa} that $F_6$ fibrations always have at
least a $U(1)$ abelian sector; whereas the ``generic''
$\mathfrak{su}(2)$ might or might not be realized in any particular
example, this $U(1)$ always appears, regardless of whether any of the
$a_{ijk}$ consist of a single monomial.
The presence of this extra Mordell-Weil section can also be seen
immediately from the fact that the fiber $F_6$ contains two distinct
-1 curves, as mentioned in Section~\ref{sec:genus-1}

In this subsection, we will consider fibrations by $F_6$ over base 72, which is also fiber $F_{12}$. This base has seven toric divisors, whose intersection structure is indicated in Figure \ref{fig:base72}. This is a weak Fano base, and so it has no NHCs.

\begin{figure}
    \centering
    \begin{tikzpicture}[scale=1.5]
         \draw[gray, thin] (0,-2) -- (0,2);
         \draw[gray, thin] (-2,0) -- (2,0);
         \draw[blue,thick] (0,0) -- (1,0);
         \draw[blue,thick] (0,0) -- (1,1);
         \draw[blue,thick] (0,0) -- (0,1);
         \draw[blue,thick] (0,0) -- (-1,0);
         \draw[red,thick] (0,0) -- (-1,-1);
         \draw[blue,thick] (0,0) -- (-1,-2);
         \draw[red,thick] (0,0) -- (0,-1);
         \filldraw[blue] (1,0) circle (2pt) node[anchor=north]{(1,0)};
         \filldraw[blue] (1,1) circle (2pt) node[anchor=west]{(1,1)};
         \filldraw[blue] (0,1) circle (2pt) node[anchor=south]{(0,1)};
         \filldraw[blue] (-1,0) circle (2pt) node[anchor=south]{(-1,0)};
         \filldraw[red] (-1,-1) circle (2pt) node[anchor=east]{(-1,-1)};
         \filldraw[blue] (-1,-2) circle (2pt) node[anchor=east]{(-1,-2)};
         \filldraw[red] (0,-1) circle (2pt) node[anchor=north]{(0,-1)};
    \end{tikzpicture}
    \caption{The seven toric divisors in base 72 and their intersection structure. Blue curves have self-intersection -1, and red curves have self-intersection -2.}
    \label{fig:base72}
\end{figure}

We will present three examples of (72,6) fibrations: a ``generic'' example, with Hodge numbers characteristic of the ensemble of such geometries, and two extreme examples, which have the largest $h^{1,1}$ and $h^{2,1}$ of any (72,6) fibration.

\subsubsection{A Typical (72,6) Fibration}
\label{sec:typical}

To illustrate a ``typical'' example of an $F_6$ fiber over base 72,
consider polytope v07-64854, whose vertices are given by the columns of the
matrix \begin{align} \left(\begin{array}{ccccccccc}-1& 0& 1& -3& 0& 0&
    0& 0& 1\\-1& 0& 0& -1& -1& 0& 2& 4& 4\\0& 1& 0& 2& -1& 0& -1& -1&
    -2\\2& 0& 0& 0& 0& 1& 0& -2&
    -2\end{array}\right). \label{eq:watiExample} \end{align} This
polytope has $h^{1,1}=20, h^{2,1}=26$, and contains four
two-dimensional reflexive subpolytopes: an $F_4$ fiber with base 118,
an $F_6$ fiber with base 72, an $F_8$ fiber with base 74, and an
$F_{10}$ fiber with base 392. We will analyze the $F_6$ fibration, and
accordingly  work in a basis where the subpolytope is in
standard form, with vertices given by the columns of the
matrix \begin{align} \left(\begin{array}{cccc}
    1&0&-1&-1\\0&1&1&-1\\0&0&0&0\\0&0&0&0\end{array}\right).\label{eq:f6subvertices} \end{align}
Because base 72 has $h^{1,1}(B)=5$, by STW we have a gauge group of
total rank $20-5-1=14$. Moreover, by the analysis of
\cite{Klevers:2014bqa}, the abelian sector has rank at least one, so
the total rank of the nonabelian sector of the gauge group is no
larger than 13.

In this geometry, four of the eight coefficients $a_{ijk}$, namely
$a_{030}$, $a_{111}$, $a_{021}$, and $a_{102}$, contain only a single
monomial; we thus conclude that the ``generic'' SU(2) hosted on
$a_{102}=0$ is absent. Base 72 has seven toric divisors, and on these
divisors we find three $I_0^*$ singularities, one $I_6$ singularly,
and one $III$ singularity. We find that the toric divisors host a
gauge algebra given by
\eqq{\mathfrak{g}_2\oplus\mathfrak{so}(7)\oplus\mathfrak{so}(8)\oplus\mathfrak{sp}(3)\oplus\mathfrak{su}(2),}
as summarized in Table \ref{tab:WatiExampleTate}. This algebra has
total rank $2+3+4+3+1=13$, and so we have completely identified the
gauge group of this model
(up to a possible discrete quotient)
as \begin{align} G = U(1) \times SU(2)
  \times SO(7) \times SO(8) \times Sp(3) \times
  G_2. \label{eq:watiExampleG} \end{align}

\begin{table}[h]
\begin{centering}
\begin{tabular}{|c|c||c|c|c|c|}\hline
Base point & Self-intersection & $\#$(points) & $\operatorname{ord}(f,g,\Delta)$ & Kodaira type & Gauge algebra \\\hline\hline
(0,-1) & -2 & 2 & (1,2,3) & $III$ & $\mathfrak{su}(2)$ \\\hline
(1,0) & -1 & 3 & (2,3,6) & $I_0^*$ & $\mathfrak{so}(7) $ \\\hline
(1,1) & -1 & 1 & (0,0,0) & Smooth & $\emptyset$ \\\hline
(0,1) & -1 & 3 & (2,3,6) & $I_0^*$ & $\mathfrak{so}(8)$  \\\hline
(-1,0) & -1 & 4 & (0,0,6) & $I_6$ & $\mathfrak{sp}(3)$ \\\hline
(-1,-1) & -2 & 2 & (2,3,6) & $I_0^*$ & $\mathfrak{g}_2$ \\\hline
(-1,-2) & -1 & 1 & (0,0,0) & Smooth & $\emptyset$ \\\hline
\end{tabular}
\caption{The toric gauge group of the $F_6$ fibration of the  polytope in Eq.~(\ref{eq:watiExample}). For each of the seven toric divisors in base 72, we specify its self-intersection, the number of points in four-dimensional polytope that project to the base divisor, the orders to which the Weierstrass coefficients $f$ and $g$ and the discriminant $\Delta$ vanish on the divisor, the Kodaira singularity type inferred from these vanishings, and the gauge algebra picked out by the monodromies. From this table, we identify that the total nonabelian gauge algebra is $\mathfrak{su}(2)\oplus\mathfrak{so}(7)\oplus\mathfrak{so}(8)\oplus\mathfrak{sp}(3)\oplus\mathfrak{g}_2 $, with no additional factors from nontoric divisors.}
\label{tab:WatiExampleTate}
\end{centering}
\end{table}

Now we consider the matter content of the compactification. As
mentioned above, we have $h^{1,1}(B)=5$, and hence \begin{align}T =
  h^{1,1}(B)-1 = 4\end{align} tensor multiplets. From the gauge group
  $G$ given in Eq.~(\ref{eq:watiExampleG}), we find that \begin{align}
    V = 1+3+21+28+21+14 = 88. \end{align} Accordingly, from
  Eq.~(\ref{eq:HminusV}), the spectrum contains \begin{align} H = V +
    273 - 29T = 245\end{align} hypermultiplets, of which \begin{align}
      H_{\text{neutral}} = h^{2,1}(X)+1 = 27 \end{align} are neutral
    and \eqq{H_{\text{charged}} = H - H_{neutral} = 218} are charged
    under the Cartan of the gauge group.

It is perhaps interesting to speculate on why it is natural that this
kind of base + fiber structure is in some sense ``typical''.  One
feature of base 72 is that it is a generalized del Pezzo surface,
i.e. a weak Fano base with only -1 and -2 curves, and no curves of
more negative self-intersection that would force a non-Higgsable/rigid
gauge group.  In some sense this gives the greatest flexibility for
tuning gauge groups --- for example, if a base has a curve of
self-intersection -5 or below, that curve must support a non-Higgsable
gauge factor, and no gauge groups can be supported on any curve that
intersects that curve.  Furthermore, this base has a relatively large
number (5) of curves of self-intersection -1, and a relatively large
$h^{1,1} (B)$ among weak Fano bases, so the number of places that
gauge factors can be tuned and the flexibility in tuning these
factors is relatively large.  We see that the gauge group factors in
this particular model all have relatively low rank (1--4); many of
these factors can coexist on intersecting curves, giving a large
combinatorial space for the set of possible tunings of low rank gauge
factors on all the curves in this base.  This makes it seem quite
plausible that this kind of structure is indeed potentially typical
not only for toric hypersurface Calabi-Yau threefolds, but also for
general elliptic Calabi-Yau threefolds: we may expect many
possibilities when the base is a generalized del Pezzo and the gauge
group factors are relatively small and distributed on many different divisors.

\subsubsection{The Largest $h^{2,1}$ (72,6) Fibration}
The polytope (ID v12-6536) with the largest $h^{2,1}$ of any (72,6) fibration
has vertices given by the columns of the
matrix \begin{align}\left(\begin{array}{cccccccc}1& -3& -1& -1& -1& 0&
    0& 0\\0& 3& -1& 1& 1& 0& 0& 1\\0& -1& 0& -1& 1& 0& 1& 0\\0& -1& 0&
    1& -1& 1& 0& 0\end{array}\right).\label{72,6bigh21}\end{align}
This polytope has Hodge numbers $h^{1,1}=7,h^{2,1}=97$.
Moreover, it
contains a single two-dimensional reflexive subpolytope, namely the
$F_6$ fiber; we again work in a basis where the vertices of the
subpolytope are given by Eq.~(\ref{eq:f6subvertices}).

By Shioda-Tate-Wazir, we see that the total rank of the gauge group is $r = h^{1,1}(X)-h^{1,1}(B)-1 = 1.$ On the other hand, all $F_6$ fibrations have at least one $U(1)$ factor, so in this example the entire gauge group is the universal $U(1)$ identified in \cite{Klevers:2014bqa}, with no nonabelian sector. Indeed, the Weierstrass coefficient $a_{102}$ which gives rise to the universal $SU(2)$ factor consists of only a single monomial, and hence the $SU(2)$ factor does not appear; moreover, none of the $a_{ijk}$ vanish on any toric divisor, and hence we have no toric Tate tunings, as expected.

We therefore have $V=1$. As usual in fibrations with base 72, we have
$T=4$, and thus we have $H=1+273-116=158$ hypers, of which
$H_{\text{neutral}} = h^{2,1}(X)+1 = 98$ are neutral; the remaining 60
hypers are charged under the $U(1)$ gauge group.
We can check that this matches perfectly with anomaly cancellation.
The anomaly coefficient $b$ for the U(1) factor coming from the extra
section associated with the fiber must be $- K$, with $b \cdot b =
5$.  This U(1) carries exactly 60 charged fields (which can also be
recognized as the number of fields coming from the 30 fundamentals of
an SU(2) tuned on the genus-one anticanonical divisor after Higgsing
on the single adjoint).

\subsubsection{The Largest $h^{1,1}$ (72,6) Fibration}
\label{sec:726-largest}

On the other hand, the polytope with the largest $h^{1,1}$ of any
(72,6) fibration has vertices given by the columns of
\eqq{\left(\begin{array}{cccccccccccccc}-1& -1& -1& -1& -1& -1& -1&
    -1& 0& 0& 1& 1& 2& 5\\1& 1& -5& -1& -1& 1& 1& 1& 1& 1& 0& 0& 1&
    -2\\0& 1& -1& -1& 1& -1& -1& 1& -1& 1& -1& 1& -1& -1\\1& 1& -2& 0&
    0& -2& 0& 0& 0& 0& 0& 0& -2&
    -2\end{array}\right).\label{eq:72,6bigh11}} This polytope (ID v08-22025) has
Hodge numbers $h^{1,1}(X)=68,$ $h^{2,1}(X)=8$, and has 25
two-dimensional reflexive subpolytopes; we will focus on the
unique
(72,6)
fibration, and have again worked in a basis where the vertices of the
subpolytope are given by the columns of the matrix in
Eq.~(\ref{eq:f6subvertices}). By Shioda-Tate-Wazir, and including the
universal $U(1)$ gauge group present in $F_6$ fibrations, we have
$h^{1,1}(X)-1-5-1 = 61$ total units of rank to identify in this
geometry.

We begin by considering the gauge factors that can be determined by strictly toric information. The coefficient $a_{102}$ consists of only a single monomial, so the univeral $\mathfrak{su}(2)$ is not present.  As indicated in Table \ref{tab:72,6bigh11Tate}, the toric divisors host a \eqq{\mathfrak{G}_{\text{toric}} = \mathfrak{su}(15)\oplus \mathfrak{su}(10) \oplus \mathfrak{su}(10) \oplus \mathfrak{su}(5) \oplus \mathfrak{su}(5)} gauge algebra. 

\begin{table}[h]
\begin{centering}
\begin{tabular}{|c|c||c|c|c|c|}\hline
Base point & Self-intersection & $\#$(points) & $\operatorname{ord}(f,g,\Delta)$ & Kodaira type & Gauge algebra \\\hline\hline
(0,-1) & -2 & 10 & (0,0,10) & $I_{10}$ & $\mathfrak{su}(10)$ \\\hline
(1,0) & -1 & 6 & (0,0,5) & $I_5$ & $\mathfrak{su}(5) $ \\\hline
(1,1) & -1 & 1 & (0,0,0) & Smooth & $\emptyset$ \\\hline
(0,1) & -1 & 1 & (0,0,0) & Smooth & $\emptyset$ \\\hline
(-1,0) & -1 & 6 & (0,0,5) & $I_5$ & $\mathfrak{su}(5)$ \\\hline
(-1,-1) &  -2 & 10 & (0,0,10) & $I_{10}$ & $\mathfrak{su}(10)$ \\\hline
(-1,-2) & -1 & 31 & (0,0,15) & $I_{15}$ & $\mathfrak{su}(15)$ \\\hline
\end{tabular}
\caption{The toric gauge group of the $F_6$ fibration of the  polytope in Eq.~(\ref{eq:72,6bigh11}). For each of the seven toric divisors in base 72, we specify its self-intersection, the number of points in four-dimensional polytope that project to the base divisor, the orders to which the Weierstrass coefficients $f$ and $g$ and the discriminant $\Delta$ vanish on the divisor, the Kodaira singularity type inferred from these vanishings, and the gauge algebra picked out by the monodromies. From this table, we identify that the total nonabelian gauge algebra hosted on toric divisors is $\mathfrak{su}(15)\oplus \mathfrak{su}(10) \oplus \mathfrak{su}(10) \oplus \mathfrak{su}(5) \oplus \mathfrak{su}(5)$.}
\label{tab:72,6bigh11Tate}
\end{centering}
\end{table}

On each divisor $D$ hosting a toric gauge group, the Weierstrass coefficients are given by \eqq{f = -\frac{1}{48}a_{111}^4 + \mathcal{O}(z^n),~ g = \frac{1}{846}a_{111}^6 + \mathcal{O}(z^m),} where $z=0$ is a local description of the toric divisor and $n,m>0$ depend on the particular divisor. Thus, if the divisor $D$ intersects the zero locus of $a_{111}$, we then have a $(4,6)$ point, which must be blown up. $a_{111}$ is a section of $-K$ \cite{Klevers:2014bqa}, so it does not intersect the -2 curves, each of which hosts an $\mathfrak{su}(10)$ gauge algebra. Instead, we have $(4,6)$ points on the divisors $(-1, -2), (1,0)$ and $(-1,0)$, each of which must be blown up. 

On the $\mathfrak{su}(5)$ curves $(1,0)$ and $(-1,0)$, we have \begin{subequations}
    \eqq{f &= -\frac{1}{48}a_{111}^4 + \frac{a_{102}+a_{201}}{6}a_{111}^2z^2 + \cdots\\
    g &= \frac{1}{864}a_{111}^6 - \frac{a_{102}+a_{201}}{72}a_{111}^4z^2 + \cdots\\
    \Delta &= a_{102}^2a_{111}^6z^5\left[-a_{300}a_{111}+\left(a_{300}+a_{201}+a_{300}a_{102}\right)z\right] + \cdots
    }
\end{subequations}
We thus have a $(4,6)$ point on the nontoric point $a_{111}=z=0$, which we resolve with the coordinate transformation \eqq{\left(a_{111},z\right)\to \left(\hat{a}u,u\right)\label{eq:72,6blowup}} at the cost of changing the toric divisor $z=0$ from a $-1$ curve to a $-2$ curve; this yields \begin{subequations}
    \eqq{
    f &= u^4\left(\frac{a_{102}+a_{201}}{6}\hat{a}^2 -\frac{1}{48}\hat{a}^4\right) \\
    g &= u^6\left(-\frac{a_{102}+a_{201}}{72}\hat{a}^4 +\frac{1}{864}\hat{a}^6\right) \\ 
    \Delta &= a_{102}^2\hat{a}^6u^{12}\left(a_{300}+a_{201}+a_{300}a_{102} - a_{300}\hat{a}\right)
    }
\end{subequations}
Dropping the factor $u^{12}$ as before, we find that \eqq{ \Delta &= a_{102}^2\hat{a}^6\left(a_{300}+a_{201}+a_{300}a_{102} - a_{300}\hat{a}\right),} and therefore we have no additional gauge algebra or $(4,6)$ points on the $-1$ curve $u=0$. 

Conversely, on the curve $(-1,-2)$, which hosts an $\mathfrak{su}(15)$ gauge algebra, we have
\begin{subequations}
    \eqq{
    f &= -\frac{1}{48}a_{111}^4 + \frac{a_{102}+a_{201}}{6}a_{111}^2z^6 \\
    g &= \frac{1}{864}a_{111}^4 - \frac{a_{102}+a_{201}}{72}a_{111}^4z^6\\
    \Delta &= -a_{102}^2a_{300}a_{111}^7z^{15} + a_{102}^2\left(a_{201}+a_{300}+a_{102}a_{300}\right)a_{111}^6z^{18}
    }
\end{subequations}
so that we have a $(4,6)$ point at $a_{111}=z=0$,\footnote{Note that here $z=0$ is a different divisor than above.} which we can once again resolve by means of Eq.~\ref{eq:72,6blowup} to find 
\begin{subequations}
    \eqq{
    f &= u^4\left(\frac{a_{102}+a_{201}}{6}\hat{a}^2u^4 -\frac{1}{48}\hat{a}^4\right) \\
    g &= u^6\left(-\frac{a_{102}+a_{201}}{72}\hat{a}^4u^4 +\frac{1}{864}\hat{a}^6\right) \\ 
    \Delta &= a_{102}^2\hat{a}^6u^{10}u^{12}\left[\left(a_{300}+a_{201}+a_{300}a_{102}\right)u^2 - a_{300}\hat{a}\right].
    }
\end{subequations}
Thus, on the $-1$ curve $u=0$, we have tuned an $\agsu(10)$ gauge algebra, and we find an additional $(4,6)$ point at $\hat{a}=u=0$. Resolving this $(4,6)$ point by blowing $u=0$ up to a $-2$ curve, we obtain a -1 curve with an $\mathfrak{su}(5)$ gauge algebra and yet another $(4,6)$ point; we must resolve this point as well, which introduces another $-1$ curve and no further gauge factors. Thus, in total, from blowing up $(4,6)$ points, we find an additional gauge algebra \eqq{\mathfrak{G}_{(4,6)} = \mathfrak{su}(10)\oplus \mathfrak{su}(5).}

Putting it all together, we find a compactification on a base $B'$
which is a blowup of $B$ at five points, and accordingly has
\eqq{h^{1,1}(B') = h^{1,1}(B) + 5 = 10.} The nonabelian gauge algebra
is given by \eqq{\mathfrak{G}_{\text{nonabelian}} &=
  \mathfrak{G}_{\text{toric}}\oplus \mathfrak{G}_{(4,6)} =
  \mathfrak{su}(15)\oplus \mathfrak{su}(10)^3 \oplus
  \mathfrak{su}(5)^3,\label{eq:72,6bigh11nonab}} and has rank
\eqq{\operatorname{rk}(\mathfrak{G}_{\text{nonabelian}}) =
  14+27+12=53.} Note that the intersection structure of the curves of
$B'$ and the gauge groups they host is extremely reminiscent of the
pattern in \S\ref{sec:SCFT}: a -2 curve
hosts an $\mathfrak{su}(3N)$
gauge algebra and intersects three $-2$ curves, each of which hosts an
$\mathfrak{su}(2N)$ gauge algebra and intersects an additional $-2$
curve hosting a $\mathfrak{su}(N)$ gauge group.

In addition to the nonabelian sector, there is at least one $U(1)$
gauge group. However, with the nonabelian gauge group in
Eq.~\ref{eq:72,6bigh11nonab} and an abelian sector with only a single
U(1) factor,
the fibration over base $B'$ does not satisfy STW:
\eqq{h^{1,1}(B')+1+\rk(\mathfrak{G}_{\text{nonabelian}}) + \rk(U(1)) =
  10 + 1+ 53 + 1 = 65 < h^{1,1}(X).} We conjecture that full abelian
sector is $G_{\text{abelian}}=U(1)^4$, so that the entire gauge group
is given, up to a possible discrete quotient (see below), by \begin{subequations}
  \eqq{G &= G_{\text{abelian}}\times G_{\text{nonabelian}} \\ &=
    U(1)^4 \times SU(15)\times SU(10)^3 \times
    SU(5)^3,\label{eq:72,6bigh11gauge} }
\end{subequations}
which satisfies STW. However, we emphasize that at this stage this is
only a conjecture: we have not found an explicit coordinate
representation for the missing three Mordell-Weil
sections.

We now pass to anomaly cancellation, assuming that the gauge group is
as in \ref{eq:72,6bigh11gauge}. We have \eqq{V = 224 + 297 + 72 + 4 =
  597.} We then have \eqq{H = V + 273 - 29T = 609} hypers, of which
\eqq{H_{\text{neutral}} = h^{2,1}(X)+1 = 9} are neutral under the
Cartan and \eqq{H_{\text{charged}} = H-H_{\text{neutral}}=600} are
charged.
This precisely matches the complete matter content of the SU($N$)
gauge factors: an SU($N$) gauge factor on a $-2$ curve has $2 N$
fundamentals, and intersecting divisors carrying SU($N$) gauge factors
carry bifundamental matter.  It is easy to check that all of the
matter for each factor is taken care of by the bifundamentals, and the
total number of fields is then just the number of fields charged under
the SU(10) factors: $3 \times 20 = 60$ fundamentals, for 600 total
charged fields.

This suggests that the U(1) factors must not have any
charged matter.  That is certainly true for the U(1) factor coming
from the $F_6$ factor, since it has anomaly coefficient $ b = - K$,
and $- K \cdot - K = 0$ when $T = 10$, so there is no matter charged
under this U(1).  The only way to satisfy anomaly cancellation for the
additional conjectured 3 U(1) factors would be for them to have the
same anomaly coefficient.  Given the dominant SU(5) structure in this
geometry, it is tempting to speculate that these additional U(1)
factors all are equivalent to what would arise from an SU(5) factor
tuned on $- K$, which would have only a single adjoint and no
fundamental matter, which was then broken to $U(1)^{4}$.
Note that all the charged matter in this model is invariant under a
central $\Z_5$, so that
the massless charge completeness conjecture suggests that the global
structure of the group is
\begin{equation}
 G = (U(1)^4 \times SU(15)\times SU(10)^3 \times
    SU(5)^3)/\Z_5 \,,
\label{eq:}
\end{equation}
which would fit nicely to the conjectured SU(5) structure of the
abelian factors.  This also has an interesting similarity to the
structure identified in Eq.~(\ref{eq:3113}), suggesting some interesting
as-yet unidentified structure for models with a certain global gauge
group structure and multiple U(1) factors.
We leave further investigation of this in similar
models for future work.

\subsection{Fiber $F_{11}$ and the Standard Model}
\label{sec:SM}

As indicated in Table \ref{tab:16fibers}, the generic gauge group for
fibrations with fiber $F_{11}$ is $(U(1)\times SU(2)\times SU(3))/\Z_6$,
i.e. the Standard Model gauge group; this
observation has generated significant interest in this fiber \cite{Klevers:2014bqa}, and
is the origin of, e.g., the
quadrillion Standard Models of \cite{Cvetic:2019gnh}. This arises because
$F_{11}$ can be realized by adding a vertex to $F_{10}$ so that its
defining polynomial is a specialization of
Eq.~\ref{eq:longWeierstrassWithAlphaBeta}, namely \eqq{\alpha y^2 +
  a_1xyz + a_3yz^2 = \beta x^3 + a_2x^2z^2 + a_4xz^4.} In addition to
the zero section and the $\mathfrak{su}(2)$ and $\mathfrak{su}(3)$
gauge algebras inherited from $F_{10}$, this polynomial has an
additional Mordell-Weil section $(0:0:1)$, giving a rank-one Abelian
sector and completing the Standard Model.

While we obviously do not live in six space-time dimensions, we expect that any
special feature of this fiber type would likely be shared between
Calabi-Yau threefolds and the elliptic Calabi-Yau fourfolds that are
used for compactifications to 4D, and so 
it is natural to wonder whether this structure is in some way favored
or special in the F-theory landscape.
\footnote{We would like to thank
  Damien Mayorga Pena for discussions on this issue.}
Like the other fibers, $F_{11}$ captures a particular slice of the
moduli space of Weierstrass models, with certain specific tuned
structures.  The general form of the Weierstrass model with this
tuned
gauge
group was described in \cite{Taylor:2019wnm,Raghuram:2019efb}.  From this point of view, there is no
{\it a priori} reason to expect this class of fibrations to be
  particularly favored, but it is certainly a good question to investigate.

As discussed above, any $F_{11}$ fibration where $\alpha$ and $\beta$
each contain more than one monomial contains at least the Standard
Model gauge algebra, but in general such a fibration can also have a
significantly larger gauge algebra. In this section, we classify those
fibrations whose gauge algebra is exactly that of the Standard Model.

Any such fibration must have a topological rank increase (as defined
in \S\ref{sec:rankincrease}) of 4, which is the rank of the SM gauge
group. Of the 112,327,098 total fibrations with fiber $F_{11}$, only
5,318 (5,317 inequivalent; one such polytope has two
automorphism-equivalent $F_{11}$ fibrations) meet this topological
constraint, of which 5120 (5119 inequivalent) fibrations have weak Fano
bases and 198 have non-weak-Fano bases. 1,186 fibrations (1,185
inequivalent) realize the Standard Model as described above, through
the generic features of $F_{11}$ fibrations; these are necessarily
fibrations over weak Fano bases, as fibrations over non-weak-Fano bases will
have 
other gauge factors from the NHCs.

The universal gauge group contribution is not the only way to realize
an exact SM in $F_{11}$ fibrations. Even if the nontoric divisors
$\alpha,\beta$ do not host gauge factors in a given fibration, the
missing gauge factors can be realized on toric divisors; in such
examples, the $F_{11}$ can effectively be thought of as providing the
$U(1)$ factor, which is what makes it more useful than $F_{10}$ in the
context of the SM.
Note, however that in this context the global structure of the gauge
group will generally not include the $\Z_6$ quotient, which in the 4D
context likely compromises any natural realization of Standard Model-like matter in such constructions \cite{Taylor:2019ots}.

For instance, consider $F_{11}$ fibrations over the Hirzebruch surface
$\mathbb{F}_3$, which has a single SU(3) NHC. Any such fibration has a
gauge group no smaller than $U(1)\times SU(3)$, and hence we expect
such fibrations to have $h^{1,1}(X)\geq 1 +
h^{1,1}(\mathbb{F}_3)+1+\operatorname{rk}(SU(3)) = 6$. In fact, there
is a single $F_{11}$ fibration over $\bbF_3$ with $h^{1,1}(X)=6$, with
$h^{2,1}(X)=152$; all other $F_{11}$ fibrations over $\bbF_3$ have a
larger gauge group.

There are seven $F_{11}$ fibrations over $\bbF_3$ with $h^{1,1}=7$,
which is the correct rank increase to realize the SM. Because of the
$SU(3)$ NHC, by STW we cannot activate the generic $\mathfrak{su}(3)$
on $\beta$; conversely, we can activate the generic $\mathfrak{su}(2)$
on $\alpha$, and indeed whenever that occurs we will have realized the
full SM gauge algebra. This ``hybrid'' standard model gauge algebra
(with potentially different global structure)
occurs in five of
the seven models mentioned above. One of the remaining two $h^{1,1}=7$
models is actually yet another realization of the Standard Model, with
$\mathfrak{su}(2)$ tuned on a toric divisor and $\mathfrak{su}(3)$
on the NHC; the final model has its NHC enhanced to
$\mathfrak{so}(7)$ and no additional nontoric gauge algebras.

Thus, exact Standard Model gauge groups, i.e. those with no additional gauge
factors, are rare, occurring in no more than 0.0023\% of all
fibrations.

\subsection{Example: the largest $h^{1,1} (X)$ (polytope with $h^{1,
    1}= 491$)}
\label{sec:large-h11}

It is naturally interesting to consider the polytopes with the largest
values of the Hodge numbers.  It has been known for many years that
the largest  values of the Hodge numbers $h^{1,1}, h^{2,1}$ for toric hypersurface Calabi-Yau threefolds
are threefolds with $h^{1,1} (X) = 491, h^{2,1} (X) = 11$ and $h^{2,1}
(X) = 491, h^{1,1} = 11$, respectively \cite{Candelas:1997eh, Aspinwall:1997ye}.
There is a unique polytope (ID v05-394) with $h^{2,1}(X)=491$, and it has a unique elliptic fibration, namely the standard stacking of fiber $F_{10}$ over Hirzebruch $\F_{12}$; it is easily proven that this is in fact the largest $h^{2,1}(X)$ that can be realized in any elliptic Calabi-Yau threefold, and not just the largest possible in toric hypersurfaces \cite{Taylor:2012dr}.

It is much harder to prove rigorously that $h^{1,1} (X) = 491$ is a
strict upper bound for general elliptic Calabi-Yau threefolds.
It was shown long ago
by studying dual heterotic instantons
that there are two different F-theory models
corresponding to an elliptic Calabi-Yau threefold with Hodge numbers
 $h^{2,1}
(X) = 491, h^{1,1} = 11$ \cite{Aspinwall:1997ye}.
One of these corresponds to a 6D F-theory model with a large gauge
group with many $E_8$ factors, related to the $E_8 \times E_8$ heterotic
theory, and the other has a large gauge group with $Sp(N)$ and $SO(N)$
factors (and the largest gauge group rank known for any 6D supergravity model), related to the Spin(32)$/\Z_2$ heterotic theory.  In the
intervening years these structures have been encountered in various
ways in the literature.  
The Calabi-Yau threefold with these Hodge numbers and the large
$E_8$-type gauge group
was found in
\cite{MorrisonTaylorToric} to be a generic elliptic fibration over 
the maximal toric base, with $h^{1,1}_* (B) = 194$.
From a rather different point of view,
the authors of \cite{Kim:2024eoa}
 argued (with some plausible physical assumptions) for the upper bound
 of $h^{1,1} = 491$ for an elliptic CY3, and
 argued that the two 6D $\mathcal{N}=(1,0)$ theories
associated with this Calabi-Yau are ``extreme'' in the sense that they
have the largest possible number of tensor multiplets and the highest
possible rank gauge sector,  respectively, for any 6D supergravity, not
just those coming from toric hypersurface constructions.

Explicitly, the two theories associated with this Calabi-Yau are:
\begin{enumerate}
    \item The theory with $T=193$ tensor multiplets and gauge algebra
      \eqq{\mathfrak{e}_8^{17}\oplus \mathfrak{f}_4^{16} \oplus
        \mathfrak{g}_2^{32} \oplus \mathfrak{sp}_1^{32} . \label{eq:T193gauge}}
    \item A theory with $T=9$ tensor multiplets and gauge algebra \eqq{\mathfrak{sp}(72)\oplus\mathfrak{sp}(56)\oplus\mathfrak{sp}(48)\oplus\mathfrak{so}(176)\oplus\mathfrak{sp}(24)\oplus\mathfrak{so}(128)\oplus\mathfrak{so}(80)\oplus\mathfrak{so}(64)\oplus\mathfrak{so}(32)\oplus\mathfrak{sp}(40),\label{eq:491bigValgebra}} which has rank 480.
\end{enumerate}
In this section we summarize briefly how these theories can be
understood in terms of the fiber structures analyzed in this paper.

There is a unique polytope (ID v05-393) with the Hodge numbers
$h^{2,1}
(X) = 491, h^{1,1} = 11$, the vertices of which are given by the columns of the matrix 
\begin{equation}
    \left(
    \begin{array}{ccccc}
     1 &0 &0 &21 &-63\\
     0 &1 &0 &28 &-56\\
     0 &0 &1 &36 &-48\\
     0 &0 &0 &42 &-42
    \end{array}
    \right).
\end{equation}
This polytope has two subpolytopes, an $F_{10}$ fiber and an $F_{13}$
fiber. We will consider each of these in turn.

\subsubsection{The $F_{10}$ Fiber}
We begin by analyzing the $F_{10}$ fiber; to do so, it is as usual convenient to work in a basis where the fiber lives in the first two coordinates. In one such basis, the vertices of $\nabla$ are given by the columns of 

\begin{equation}
    \left(
    \begin{array}{ccccc}
     1 &0 &-2 &-2 &-2\\
     0 &1 &-3 &-3 &-3\\
     0 &0 &42 &0 &-42\\
     0 &0 &36 &1 &-48
    \end{array}
    \right).
\end{equation}
We see that we have a ``standard stacking'' of the base over the fiber vertex $(2,3)$; in fact this is the generic elliptic fibration over the unique toric base with $h^{1,1}_*(B)=194$, constructed in \cite{MorrisonTaylorToric}. We therefore have \eqq{T = h^{1,1}_*(B)-1=193} tensor multiplets, in line with the upper bound proposed in \cite{Kim:2024eoa}.

Because this is the generic fibration over the $h^{1,1}_*(B)=194$ base,
the gauge group of the associated compactification is easy to compute:
it is exactly the NHC associated to this base, as identified in
\cite{MorrisonTaylorToric}, which 
straightforwardly reproduces the gauge algebra in
Eq.~\ref{eq:T193gauge}.\footnote{One can equally well identify this
gauge algebra by identifying appropriate tops in the polytope, or by
passing to Weierstrass form and applying the Kodaira classification;
these three approaches all give the same gauge algebra, namely the one
in \ref{eq:T193gauge}.}

\subsubsection{The $F_{13}$ Fiber}
This polytope also admits an elliptic fibration with an $F_{13}$ fiber. As usual, we work in a basis where the $F_{13}$ fiber occupies the first two coordinates; in one such basis, the vertices of the polytope are given by the columns of 
\begin{equation}
    \left(\begin{array}{ccccc}
2&2&2&0&-82\\
3&3&3&-1&-81\\
-6&1&0&0&-6\\
-14&0&1&0&-14\\
\end{array}\right).
\end{equation}
In this basis we have arranged that the fiber coordinates are the first two, and the base coordinates are the last two coordinates. The base of the fibration is smooth, and has $h^{1,1}(B)=9$. The base contains eleven curves; in our basis, they are given by \begin{align}
    (0, 1),~(1, 0),~(-5, -12),~(-4, -9),~(-3, -8),~(-3, -7),\\(-1, -4),~(-2, -5),~(-1, -3),~(-1, -2),\text{ and }(0, -1),\label{eq:491bigVtoric}
\end{align}
and all but the first two host nontrivial gauge groups. In Table \ref{tab:heecheolTate} we demonstrate that the gauge group hosted on toric divisors is given by \eqq{G_{\text{toric}} = \mathfrak{sp}(72)\oplus\mathfrak{sp}(56)\oplus\mathfrak{sp}(48)\oplus\mathfrak{so}(176)\oplus\mathfrak{sp}(24)\oplus\mathfrak{so}(128)\oplus\mathfrak{so}(80)\oplus\mathfrak{sp}(64)\oplus\mathfrak{sp}(32).}

 \begin{table}[]
     \centering
     \begin{tabular}{|c|c||c|c|c|c|}\hline
          Base point & Self-intersection  & $\#$(points) & $\operatorname{ord}(f,g,\Delta)$ & Kodaira type & Gauge algebra\\\hline
          (1,0) & 0 & 1 & (0,0,0) & $I_0$  & $\emptyset$ \\\hline
          (0,1) & 2 & 1 & (0,0,0) & $I_0$ & $\emptyset$ \\\hline
          (-1,-2) & -4 & 4 & (2,3,34) & $I_{34}^*$ & $\mathfrak{so}(64)$ \\\hline
          (-4,-9) & -1 & 57 & (0,0,112) & $I_{112}$ & $\mathfrak{sp}(56)$ \\\hline
          (-3,-7) & -3 & 45 & (2,3,90) & $I_{90}^*$& $\mathfrak{so}(176)$ \\\hline
          (-5,-12) & -1 & 73 & (0,0,144) & $I_{144}$ &$\mathfrak{sp}(72)$ \\\hline
          (-2,-5) & -4 & 4 & (2,3,66) & $I_{66}^*$ & $\mathfrak{so}(128)$ \\\hline
          (-3, -8) & -1 & 49 & (0,0,96) & $I_{96}$ & $\mathfrak{sp}(48)$ \\\hline
          (-1,-3) & -4 & 4 & (2,3,42) & $I_{42}^*$ & $\mathfrak{so}(80)$ \\\hline
          (-1,-4) & -1 & 25 & (0,0,48) & $I_{48}$ & $\mathfrak{sp}(24)$ \\\hline
          (0,-1) & -4 & 4 & (2,3,18) & $I_{18}^*$ & $\mathfrak{so}(32)$ \\\hline
     \end{tabular}
     \caption{The toric gauge group of the $F_{13}$ fibration of the  polytope with $h^{1,1}(X)=491$. For each of the eleven toric divisors in the base, we specify its self-intersection, the number of points in four-dimensional polytope that project to the base divisor, the orders to which the Weierstrass coefficients $f$ and $g$ and the discriminant $\Delta$ vanish on the divisor, the Kodaira singularity type inferred from these vanishings, and the gauge algebra picked out by the monodromies. From this table, we identify that the total nonabelian gauge group hosted on toric divisors is given by Eq.~(\ref{eq:491bigVtoric}).}
     \label{tab:heecheolTate}
 \end{table}


On all divisors hosting an $\mathfrak{so}(n)$ gauge algebra, the
generic form of $f,g$ is \eqq{f \sim a_{201}^2z^2, \quad\quad g \sim
  a_{201}^3z^3,} where $z=0$ is a local description of the
divisor. Thus, if any such divisor intersects $a_{201}$, the
intersection furnishes a $(4,6)$ point and must be blown up. This
happens for only one divisor, the $-3$ curve $(-3,-7)$ hosting an
$\mathfrak{so}(176)$ gauge algebra. We therefore blow up this $-3$ curve
to obtain a $-4$ curve and a new $-1$ curve. The discriminant along the
original $-3$ curve takes the form \eqq{\Delta \sim
  -a_{300}^2z^{90}\left[a_{210}^2 + \mathcal{O}(z)\right],} and thus
after implementing the blowup by the coordinate transformation
$(a_{210},z)\to(u,uz)$, we find that the $-1$ curve hosts an
$\mathfrak{sp}(40)$ gauge algebra, with no additional $(4,6)$ points to
resolve. We have therefore shown that the gauge algebra is exactly as
in Eq.~(\ref{eq:491bigValgebra}), and moreover the intersection
numbers of the curves hosting each summand of the algebra are exactly
as described in \cite{Kim:2024eoa}.

\section{Conclusions}
\label{sec:conclusions}

In this paper, we have carried out a systematic exploration of the
fibration structure of the toric hypersurface Calabi-Yau threefolds in
the Kreuzer-Skarke database.
The results found here follow two main themes: firstly, the elliptic
and genus-one fibered Calabi-Yau threefolds in this set provide a wide
range of examples that illustrate many aspects of the geometry and
physics of F-theory.
Secondly, the ubiquity of fibration structures and the
physical/geometric language of F-theory provide a powerful lens with
which to understand and interpret this largest class of known
Calabi-Yau threefolds.

In general, on this second theme, the analysis of this paper illustrates that the
identification of fibration structures and the geometric/physical
language of F-theory provides a powerful approach to understanding and
organizing the otherwise potentially bewildering array of possible
Calabi-Yau threefolds. Because almost all known Calabi-Yau threefolds are connected by flops to an elliptic
or genus-one fibered phase, it appears that the geometric data (and the associated physical F-theory data) characterizing these fibrations, such as the base variety and the list of divisors hosting gauge algebras in the base, is a useful way of capturing the structure of families of threefolds with a common extended K\"ahler
cone.
In particular, since elliptically fibered CY threefolds are all
connected in a single moduli space  by various (tensor,
Higgs, matter) transitions, this provides a way of placing the set
of threefolds in the KS database into this broader context.
A forthcoming work  will show how even the small fraction of
non-fibered polytopes can generally be connected into this network through a
fairly minimal set of transitions \cite{jlrw-forthcoming}.
Understanding the distribution of and complementarity between different fibration
structures for CY threefolds in the same family, i.e. related by flops
and flips in the extended K\"ahler cone,  provides another interesting
opportunity for further insight into this structure; some aspects of this will be described further in \cite{rw-forthcoming}.
Fibration structure also promises to provide new insights into mirror symmetry \cite{Huang:2018vup, ot-forthcoming}.

In terms of exploring the range of examples relevant for, e.g., 6D
F-theory, one remarkable aspect of this study is that while toric
geometry is {\it a priori} a limited framework in which to construct generic
Calabi-Yau threefolds, in many aspects, the set of geometric
structures found in the KS database is somewhat broader than might be
expected.  It is natural that toric hypersurfaces will capture
elliptic fibrations with toric bases and gauge groups tuned over toric
divisors in those bases.  We find, however, also a rich array of
examples in which gauge groups are tuned over non-toric divisors in
toric bases, and SCFT structures encode complex non-toric base
geometries for elliptic fibrations.  We have also encountered in the
course of this study a number of somewhat exotic constructions that
illustrate many ideas explored over the last decade in the 6D F-theory
literature, including SCFTs, multiple U(1) factors, automatic
enhancement, where tuning of one gauge factor forces another
complementary factor to appear, and the massive charge completeness
principle, which governs the global structure of many F-theory gauge
groups in terms of torsion Mordell-Weil sections in the geometry.  It
is worth noting that many of the  structures found in the KS
database have already been identified through independent reasoning in
the F-theory context.   We expect, however, that further exploration of
some of the more exotic structures in the database will reveal further
geometric and physical structure.
For example, the examples analyzed in \S\ref{sec:su3} and
\S\ref{sec:726-largest} seem to share an intriguing structure
involving multiple U(1) factors with a discrete quotient as would be
seen upon breaking an SU($N$) factor with a single adjoint; this hints
at some interesting structural framework that we are not yet aware of.

One interesting question to explore further is the extent to which
toric hypersurfaces provide a representative sample of the set of
possible elliptic fibrations.  As discussed in the introduction, the
finite set of topologically distinct
 elliptic Calabi-Yau threefolds can in principle more or less be
 systematically classified by first finding all bases, then
 considering the allowed rigid and tuned nonabelian gauge factors,
 then classifying abelian gauge groups and matter.  While there is not
 yet a systematic understanding of abelian U(1) or discrete gauge
 structure, the global space of bases and nonabelian gauge factors is
 reasonably well understood.  The KS hypersurface threefolds represent
 specific slices of the full available elliptic CY3 fold moduli space.
 We have found that toric hypersurfaces sample a surprisingly wide
 range of non-toric bases, through non-flat fibers and associated SCFTs.
 For the gauge groups SU(2), SU(3), SU(4), toric
hypersurfaces provide a wide range of realizations, and at least for
the base $\P^2$, all possible tunings of SU(2) on non-toric divisors
are realized.  However, we do not see tunings of any other gauge
factors on non-toric divisors in the KS database.  While many of the
examples encountered here have multiple U(1) factors, we do not yet have
a systematic understanding of which Mordell-Weil structures can and
cannot be realized in this context.

For discrete gauge factors, it is worth noting that
toric hypersurface constructions produce genus-one fibrations only
with $k$-sections for  $k = 2, 3$, since these are the only genus-one
types associated with 2D toric reflexive fibers ($F_1, F_2, F_4$).  An
important open question is  to identify an  upper bound on $k$ for
more general Calabi-Yau threefold constructions.  It is known that
complete intersection fibers can give genus-one fibrations with a
4-section \cite{Braun:2014qka}.
 While it has been proven that larger
values of
$k \geq 5$ cannot be realized through toric constructions or complete
intersections, $k = 5$ Calabi-Yau threefolds have been constructed
using more sophisticated methods \cite{Knapp:2021vkm}.
It has been conjectured \cite{Oehlmann:2016wsb} that the maximum value of $k$
possible  is $k = 6$.  
At the same time,
in \cite{Raghuram:2018hjn}, 
F-theory models were found that appear to admit a physical Higgsing to 
$\Z_{21}$,  in tension with the conjecture that $k \leq 6$.
The class of genus-one  fibered Calabi-Yau threefolds with $k > 4$
represents perhaps the least-understood corner of the space of
elliptic and genus-one fibered Calabi-Yau threefolds, and presents
important challenges for further research.

We have presented here a range of specific examples from the data
provided from the full analysis of fibration structures in the KS
database.  We expect that there is much more that can be learned from
this data.  A natural next step would be to apply some similar
analysis to fibration structures for complete intersections in toric
varieties, which would use as fibers intersections in
higher-dimensional reflexive subpolytopes, such as the intersection
fibers analyzed in \cite{Braun:2014qka}.

\acknowledgments{We would like to thank
  Lara Anderson, Volker Braun, James Gray,
Shing Yan Li, Damien Mayorga Pena, Paul Oehlmann, and Thorsten Schimannek
 for helpful discussions.
This work was supported by the
DOE under contract \#DE-SC00012567.
WT would like to thank the Santa Fe Institute, where some of this work
was carried out. RN is a Pappalardo Postdoctoral Fellow in the MIT Department of Physics. FA was supported by the Margaret MacVicar UROP Memorial Fund for this project.}

\appendix

\section{Details of Weierstrass Models}
\label{sec:Weierstrass}
In this appendix we review the Weierstrass forms associated with the
16 reflexive 2D polytope toric fibers, originally analyzed in
\cite{Klevers:2014bqa}; see also \cite{an2001jacobians}.  In principle,  the procedure is to plug
the defining polynomials for each of the 16 reflexive polytopes into
Nagell's algorithm to find their generic Weierstrass models. However,
the defining polynomials of the 16 fibers are closely related, and so
in practice this exercise would be highly redundant. For instance,
from Figure \ref{fig:16fibers} it is straightforward to see that
$F_{6}$ can be obtained from $F_1$ by adding a vertex, namely the
point $(-1,1)$. Adding a point to a polytope removes points from its
dual, and hence the defining polynomial of the Calabi-Yau hypersurface
defined by $F_6$ can be obtained by removing monomials from the
polynomial of $F_1$, i.e. by setting their coefficients to be
zero. Actually, $F_6$ can be obtained in a second way, by adding the
vertex $(0,1)$ to $F_4$; this provides an alternative but equivalent
way of deriving the Weierstrass form of the defining polynomial of
$F_6$.

In this way, the Weierstrass forms of each of the sixteen 2D reflexive polytopes can be obtained from specializations of one of four\footnote{In fact, this is still a slight overcounting. $F_{10}$ itself, and therefore $F_{13}$, can be obtained from a generic cubic. Nevertheless, we find it instructive to treat $F_{10}$, and its close analogy to the long Weierstrass model in Eq.~(\ref{eq:longWeierstrass}), by hand.}  base cases. These are:

\paragraph{The cubic in $\bbP^2$}. Applying Eq.~(\ref{eq:definingPolynomial}) to $F_1$ and its dual polytope $F_{16}$, we see that the defining polynomial takes the form of a generic cubic in $\bbP^2$: \eqq{a_{300}u^3+a_{210}u^2v+a_{120}uv^2+a_{030}v^3+a_{201}u^2w+a_{111}uvw+a_{021}v^2w+a_{102}uw^2+a_{012}vw^2+a_{003}w^3 = 0.}
Note that, for generic $a_{ijk}$, this equation has no nonzero solutions, and hence the defining polynomial does not have a point in $\bbP^2$. Thus the Calabi-Yau hypersurface defined by $F_1$ is a genus-one curve, and correspondingly fibrations with $F_1$ fiber are genus-one fibrations. Nevertheless, we can apply Nagell's algorithm to find a Weierstrass model for the Jacobian of this curve; the Weierstrass coefficients $f,g$ are given in terms of the $a_{ijk}$ as  
\begin{subequations}
\label{eq:F1weierstrass}
{\footnotesize
\begin{align}
    f = &-\frac{1}{48} a_{111}^{4} + \frac{1}{6} a_{201} a_{111}^{2} a_{021} - \frac{1}{3} a_{201}^{2} a_{021}^{2} - \frac{1}{2} a_{030} a_{201} a_{111} a_{102} + \frac{1}{6} a_{120} a_{111}^{2} a_{102}  \nonumber\\ &+\frac{1}{3} a_{120} a_{201} a_{021} a_{102} - \frac{1}{2} a_{210} a_{111} a_{021} a_{102} + a_{300} a_{021}^{2} a_{102} - \frac{1}{3} a_{120}^{2} a_{102}^{2} + a_{210} a_{030} a_{102}^{2} \nonumber \\&+ a_{120}^{2} a_{201} a_{003} - 3 a_{210} a_{030} a_{201} a_{003} - \frac{1}{2} a_{210} a_{120} a_{111} a_{003} + \frac{9}{2} a_{300} a_{030} a_{111} a_{003} + a_{210}^{2} a_{021} a_{003} \nonumber\\&- 3 a_{300} a_{120} a_{021} a_{003} + a_{030} a_{201}^{2} a_{012} - \frac{1}{2} a_{120} a_{201} a_{111} a_{012} + \frac{1}{6} a_{210} a_{111}^{2} a_{012} + \frac{1}{3} a_{210} a_{201} a_{021} a_{012} \nonumber\\&- \frac{1}{2} a_{300} a_{111} a_{021} a_{012} + \frac{1}{3} a_{210} a_{120} a_{102} a_{012} - 3 a_{300} a_{030} a_{102} a_{012} - \frac{1}{3} a_{210}^{2} a_{012}^{2} + a_{300} a_{120} a_{012}^{2} \nonumber\\
    g &= 1/864 (a_{111}^6 - 12 a_{111}^4 (a_{102} a_{120} + a_{021} a_{201} + a_{012} a_{210}) + 
   36 a_{111}^3 (a_{030} a_{102} a_{201} + a_{012} a_{120} a_{201} + a_{021} a_{102} a_{210} \nonumber\\&+ 
      a_{003} a_{120} a_{210} + a_{012} a_{021} a_{300} + 15 a_{003} a_{030} a_{300}) + 
   24 a_{111}^2 (2 a_{021}^2 a_{201}^2 - 3 a_{012} a_{030} a_{201}^2 + 
      a_{012} a_{021} a_{201} a_{210} \nonumber\\&+ 2 a_{012}^2 a_{210}^2 + 
      a_{102} a_{120} (a_{021} a_{201} + a_{012} a_{210}) + 
      a_{102}^2 (2 a_{120}^2 - 3 a_{030} a_{210}) - 
      3 (a_{021}^2 + 9 a_{012} a_{030}) a_{102} a_{300} \nonumber\\&- 3 a_{012}^2 a_{120} a_{300} - 
      3 a_{003} (a_{120}^2 a_{201} + 9 a_{030} a_{201} a_{210} + a_{021} a_{210}^2 + 
         9 a_{021} a_{120} a_{300})) - 
   144 a_{111} (a_{012}^2 a_{210} (a_{120} a_{201} + a_{021} a_{300}) \nonumber\\&+ 
      a_{210} (a_{003} a_{102} a_{120}^2 + a_{021} a_{120} (a_{102}^2 - 5 a_{003} a_{201}) + 
         a_{021}^2 (a_{102} a_{201} - 6 a_{003} a_{300})) + 
      a_{012} (a_{021} a_{120} a_{201}^2 + a_{021}^2 a_{201} a_{300} \nonumber\\&+ 
         a_{003} a_{120} (a_{210}^2 - 6 a_{120} a_{300}) + 
         a_{102} (a_{120}^2 a_{201} + a_{021} a_{210}^2 - 5 a_{021} a_{120} a_{300})) + 
      a_{030} (-6 a_{012}^2 a_{201} a_{300} \nonumber\\&+ a_{102}^2 (a_{120} a_{201} - 6 a_{021} a_{300}) + 
         a_{102} (a_{021} a_{201}^2 - 5 a_{012} a_{201} a_{210} - 6 a_{003} a_{210}^2 + 
            9 a_{003} a_{120} a_{300}) \nonumber\\&+ 
         a_{003} (-6 a_{120} a_{201}^2 + 9 a_{021} a_{201} a_{300} + 
            9 a_{012} a_{210} a_{300}))) + 
   8 (-72 a_{003} a_{021} a_{120}^2 a_{201}^2 - 8 a_{021}^3 a_{201}^3 - 
      108 a_{003} a_{030}^2 a_{201}^3 \nonumber\\&+ 108 a_{003} a_{021} a_{030} a_{201}^2 a_{210} + 
      27 a_{003}^2 a_{120}^2 a_{210}^2 - 72 a_{003} a_{021}^2 a_{201} a_{210}^2 - 
      108 a_{003}^2 a_{030} a_{210}^3 + 
      54 a_{003} a_{120} (-2 a_{003} a_{120}^2 \nonumber\\&+ 2 a_{021}^2 a_{201} + 
         9 a_{003} a_{030} a_{210}) a_{300} - 
      27 a_{003} (4 a_{021}^3 + 27 a_{003} a_{030}^2) a_{300}^2 - 
      4 a_{102}^3 (2 a_{120}^3 - 9 a_{030} a_{120} a_{210} + 27 a_{030}^2 a_{300}) \nonumber\\&+ 
      3 a_{012}^2 (9 a_{120}^2 a_{201}^2 + 
         4 a_{201} a_{210} (-6 a_{030} a_{201} + a_{021} a_{210}) - 
         6 a_{021} a_{120} a_{201} a_{300} + 9 a_{021}^2 a_{300}^2) - 
      4 a_{012}^3 (2 a_{210}^3 - 9 a_{120} a_{210} a_{300} \nonumber\\&+ 27 a_{030} a_{300}^2) + 
      3 a_{102}^2 (4 a_{012} a_{120}^2 a_{210} + 
         a_{021} (4 a_{120}^2 a_{201} - 6 a_{030} a_{201} a_{210}) + 
         3 a_{021}^2 (3 a_{210}^2 \nonumber\\&- 8 a_{120} a_{300}) + 
         3 a_{030} (3 a_{030} a_{201}^2 - 8 a_{012} a_{210}^2 + 
            12 a_{012} a_{120} a_{300})) + 
      6 a_{102} (a_{120} ((2 a_{021}^2 - 3 a_{012} a_{030}) a_{201}^2 + 
            a_{012} a_{021} a_{201} a_{210} \nonumber\\&+ 2 a_{012}^2 a_{210}^2) + 
         6 a_{021}^3 a_{201} a_{300} - 
         3 a_{012} (4 a_{012} a_{120}^2 + 9 a_{021} a_{030} a_{201} + a_{021}^2 a_{210} - 
            6 a_{012} a_{030} a_{210}) a_{300} + 
         3 a_{003} (2 a_{120}^3 a_{201} \nonumber\\&- 
            a_{120} a_{210} (9 a_{030} a_{201} + a_{021} a_{210}) + 
            6 a_{021} a_{120}^2 a_{300} + 
            9 a_{030} (3 a_{030} a_{201} - a_{021} a_{210}) a_{300})) + 
      6 a_{012} (2 a_{021}^2 a_{201}^2 a_{210} \nonumber\\&- 
         3 a_{003} a_{201} (a_{210} (a_{120}^2 - 6 a_{030} a_{210}) + 
            9 a_{030} a_{120} a_{300}) + 
         3 a_{021} (2 a_{030} a_{201}^3 + 2 a_{003} a_{210}^3 - 
            9 a_{003} a_{120} a_{210} a_{300} + 27 a_{003} a_{030} a_{300}^2)))).
\end{align}}
\end{subequations}

While $F_1$ is only a genus-one curve, by specializing some of the coefficients $a_{ijk}$ to vanish it is straightforward to obtain a defining polynomial with solutions, and therefore whose zero-locus is an elliptic curve. For instance, upon setting $a_{003}=0$ we find that the point $(0:0:1)$ is a solution. The Jacobian of an elliptic curve is simply the curve itself, so Eq.~(\ref{eq:F1weierstrass}) (with the appropriate coefficients set to zero) still gives the Weierstrass coefficients with specialized coefficients. In fact, the specialization $a_{003}=0$ yields the defining polynomial of the polytope $F_3$. 

\paragraph{The biquadric in $\bbP^1\times\bbP^1$}
The polytope $F_2$ describes $\bbP^1\times\bbP^1$, and its defining polynomial is a general biquadric in four variables: \eqq{&\phantom{+} a_{1111}tuvw + a_{2200} t^2u^2 + a_{2020}t^2v^2 + a_{0202}u^2w^2 + a_{0022}v^2w^2 \nonumber\\&+ a_{2110}t^2uv + a_{1201}tu^2w + a_{1021}tv^2w + a_{0112}uvw^2 = 0.}

Applying Nagell's algorithm to the Jacobian of this curve, we find the Weierstrass model
\begin{subequations}

{\tiny
    \label{eq:F2weierstrass}
    \begin{align}
        f = &-\frac{1}{48} a_{1111}^{4} + \frac{1}{6} a_{1111}^{2} a_{2020} a_{0202} - \frac{1}{3} a_{2020}^{2} a_{0202}^{2} + \frac{1}{6} a_{1111}^{2} a_{2200} a_{0022} - \frac{14}{3} a_{2200} a_{2020} a_{0202} a_{0022} \nonumber\\&- \frac{1}{3} a_{2200}^{2} a_{0022}^{2} + a_{0202} a_{0022} a_{2110}^{2} - \frac{1}{2} a_{1111} a_{0022} a_{2110} a_{1201} + a_{2020} a_{0022} a_{1201}^{2} - \frac{1}{2} a_{1111} a_{0202} a_{2110} a_{1021} \nonumber\\&+ \frac{1}{6} a_{1111}^{2} a_{1201} a_{1021} + \frac{1}{3} a_{2020} a_{0202} a_{1201} a_{1021} + \frac{1}{3} a_{2200} a_{0022} a_{1201} a_{1021} + a_{2200} a_{0202} a_{1021}^{2} - \frac{1}{3} a_{1201}^{2} a_{1021}^{2} \nonumber\\&+ \frac{1}{6} a_{1111}^{2} a_{2110} a_{0112} + \frac{1}{3} a_{2020} a_{0202} a_{2110} a_{0112} + \frac{1}{3} a_{2200} a_{0022} a_{2110} a_{0112} - \frac{1}{2} a_{1111} a_{2020} a_{1201} a_{0112} \nonumber\\&- \frac{1}{2} a_{1111} a_{2200} a_{1021} a_{0112} + \frac{1}{3} a_{2110} a_{1201} a_{1021} a_{0112} + a_{2200} a_{2020} a_{0112}^{2} - \frac{1}{3} a_{2110}^{2} a_{0112}^{2} \\\nonumber
        g &= 1/864 (a_{1111}^6 - 
   12 a_{1111}^4 (a_{1021} a_{1201} + a_{0202} a_{2020} + a_{0112} a_{2110} + 
      a_{0022} a_{2200}) + 
   36 a_{1111}^3 (a_{0112} a_{1201} a_{2020} + a_{0202} a_{1021} a_{2110} \nonumber\\&+ 
      a_{0022} a_{1201} a_{2110} + a_{0112} a_{1021} a_{2200}) + 
   24 a_{1111}^2 (2 a_{0202}^2 a_{2020}^2 + a_{0112} a_{0202} a_{2020} a_{2110} + 
      2 a_{0112}^2 a_{2110}^2 - 3 a_{0112}^2 a_{2020} a_{2200} \nonumber\\&+ 2 a_{0022}^2 a_{2200}^2 + 
      a_{1021} a_{1201} (a_{0202} a_{2020} + a_{0112} a_{2110} + a_{0022} a_{2200}) + 
      a_{1021}^2 (2 a_{1201}^2 - 3 a_{0202} a_{2200}) + 
      a_{0022} (-3 a_{1201}^2 a_{2020} \nonumber\\&- 3 a_{0202} a_{2110}^2 - 
         20 a_{0202} a_{2020} a_{2200} + a_{0112} a_{2110} a_{2200})) + 
   8 (-72 a_{0022} a_{0202} a_{1201}^2 a_{2020}^2 - 8 a_{0202}^3 a_{2020}^3 + 
      27 a_{0022}^2 a_{1201}^2 a_{2110}^2 \nonumber\\&- 72 a_{0022} a_{0202}^2 a_{2020} a_{2110}^2 - 
      24 a_{0022} (-11 a_{0202}^2 a_{2020}^2 + 
         3 a_{0022} (a_{1201}^2 a_{2020} + a_{0202} a_{2110}^2)) a_{2200} + 
      264 a_{0022}^2 a_{0202} a_{2020} a_{2200}^2 \nonumber\\&- 8 a_{0022}^3 a_{2200}^3 + 
      a_{1021}^3 (-8 a_{1201}^3 + 36 a_{0202} a_{1201} a_{2200}) + 
      a_{0112}^3 (-8 a_{2110}^3 + 36 a_{2020} a_{2110} a_{2200}) + 
      3 a_{0112}^2 (9 a_{1201}^2 a_{2020}^2 - 
         4 (a_{0202} a_{2020} \nonumber\\&+ a_{0022} a_{2200}) (-a_{2110}^2 + 6 a_{2020} a_{2200})) - 
      3 a_{1021}^2 (-4 a_{0202} a_{1201}^2 a_{2020} - 4 a_{0112} a_{1201}^2 a_{2110} - 
         4 a_{0022} a_{1201}^2 a_{2200} - 9 a_{0112}^2 a_{2200}^2 \nonumber\\&+ 
         6 a_{0202} a_{2200} (a_{0112} a_{2110} + 4 a_{0022} a_{2200}) + 
         a_{0202}^2 (-9 a_{2110}^2 + 24 a_{2020} a_{2200})) + 
      6 a_{0112} a_{2110} (2 a_{0202}^2 a_{2020}^2 + 2 a_{0022}^2 a_{2200}^2 \nonumber\\&+ 
         a_{0022} (-3 a_{1201}^2 a_{2020} + 6 a_{0202} a_{2110}^2 - 
            20 a_{0202} a_{2020} a_{2200})) + 
      6 a_{1021} a_{1201} (2 a_{0202}^2 a_{2020}^2 + a_{0112} a_{0202} a_{2020} a_{2110} + 
         2 a_{0022}^2 a_{2200}^2 \nonumber\\&+ a_{0112}^2 (2 a_{2110}^2 - 3 a_{2020} a_{2200}) + 
         a_{0022} (6 a_{1201}^2 a_{2020} - 3 a_{0202} a_{2110}^2 - 
            20 a_{0202} a_{2020} a_{2200} + a_{0112} a_{2110} a_{2200}))) \nonumber\\&- 
   144 a_{1111} (a_{0112}^2 a_{2110} (a_{1201} a_{2020} + a_{1021} a_{2200}) + 
      a_{2110} (a_{0202}^2 a_{1021} a_{2020} + 
         a_{0022} a_{1201} (a_{1021} a_{1201} + a_{0022} a_{2200}) + 
         a_{0202} (a_{1021}^2 a_{1201} \nonumber\\&- 5 a_{0022} a_{1201} a_{2020} - 
            5 a_{0022} a_{1021} a_{2200})) + 
      a_{0112} (a_{0202} a_{1201} a_{2020}^2 + a_{1021}^2 a_{1201} a_{2200} + 
         a_{0022} a_{1201} (a_{2110}^2 - 5 a_{2020} a_{2200}) \nonumber\\&+ 
         a_{1021} (a_{1201}^2 a_{2020} + a_{0022} a_{2200}^2 + 
            a_{0202} (a_{2110}^2 - 5 a_{2020} a_{2200}))))) 
    \end{align}
}
\end{subequations}

\paragraph{The quartic in $\bbP^{1,1,2}$}
The polytope $F_4$ describes the weighted projective space $\bbP^{1,1,2}$, and so its defining polynomial is given by \eqq{a_{400}u^4 + a_{310}u^3v + a_{220}u^2v + a_{201}u^2w + a_{130}uv^3 + a_{111}uvw + a_{040}v^4 + a_{021} v^2w + a_{002}w^2 = 0.} This polynomial does not have a nonzero solution for generic $a_{ijk}$, and so we apply Nagell's algorithm to the Jacobian of this genus-one curve to find the Weierstrass model

\begin{subequations} \label{eq:F4weierstrass}
    \begin{align}
        f = &-\frac{1}{48} a_{111}^{4} + \frac{1}{6} a_{201} a_{111}^{2} a_{021} - \frac{1}{3} a_{201}^{2} a_{021}^{2} - \frac{1}{2} a_{201} a_{130} a_{111} a_{002} + \frac{1}{6} a_{220} a_{111}^{2} a_{002} + a_{201}^{2} a_{040} a_{002} \nonumber\\&+ \frac{1}{3} a_{220} a_{201} a_{021} a_{002} - \frac{1}{2} a_{310} a_{111} a_{021} a_{002} + a_{400} a_{021}^{2} a_{002} - \frac{1}{3} a_{220}^{2} a_{002}^{2} + a_{310} a_{130} a_{002}^{2} - 4 a_{400} a_{040} a_{002}^{2} \\
        g = &\phantom{+}\frac{1}{864} a_{111}^{6} - \frac{1}{72} a_{201} a_{111}^{4} a_{021} + \frac{1}{18} a_{201}^{2} a_{111}^{2} a_{021}^{2} - \frac{2}{27} a_{201}^{3} a_{021}^{3} + \frac{1}{24} a_{201} a_{130} a_{111}^{3} a_{002} \nonumber\\&- \frac{1}{72} a_{220} a_{111}^{4} a_{002} - \frac{1}{12} a_{201}^{2} a_{111}^{2} a_{040} a_{002} - \frac{1}{6} a_{201}^{2} a_{130} a_{111} a_{021} a_{002} + \frac{1}{36} a_{220} a_{201} a_{111}^{2} a_{021} a_{002} \nonumber\\&+ \frac{1}{24} a_{310} a_{111}^{3} a_{021} a_{002} + \frac{1}{3} a_{201}^{3} a_{040} a_{021} a_{002} + \frac{1}{9} a_{220} a_{201}^{2} a_{021}^{2} a_{002} - \frac{1}{6} a_{310} a_{201} a_{111} a_{021}^{2} a_{002} \nonumber\\&- \frac{1}{12} a_{400} a_{111}^{2} a_{021}^{2} a_{002} + \frac{1}{3} a_{400} a_{201} a_{021}^{3} a_{002} + \frac{1}{4} a_{201}^{2} a_{130}^{2} a_{002}^{2} - \frac{1}{6} a_{220} a_{201} a_{130} a_{111} a_{002}^{2} \nonumber\\&+ \frac{1}{18} a_{220}^{2} a_{111}^{2} a_{002}^{2} - \frac{1}{12} a_{310} a_{130} a_{111}^{2} a_{002}^{2} - \frac{2}{3} a_{220} a_{201}^{2} a_{040} a_{002}^{2} + a_{310} a_{201} a_{111} a_{040} a_{002}^{2} \nonumber\\&- \frac{2}{3} a_{400} a_{111}^{2} a_{040} a_{002}^{2} + \frac{1}{9} a_{220}^{2} a_{201} a_{021} a_{002}^{2} - \frac{1}{6} a_{310} a_{201} a_{130} a_{021} a_{002}^{2} - \frac{1}{6} a_{310} a_{220} a_{111} a_{021} a_{002}^{2} \nonumber\\&+ a_{400} a_{130} a_{111} a_{021} a_{002}^{2} - \frac{4}{3} a_{400} a_{201} a_{040} a_{021} a_{002}^{2} + \frac{1}{4} a_{310}^{2} a_{021}^{2} a_{002}^{2} - \frac{2}{3} a_{400} a_{220} a_{021}^{2} a_{002}^{2} \nonumber\\&- \frac{2}{27} a_{220}^{3} a_{002}^{3} + \frac{1}{3} a_{310} a_{220} a_{130} a_{002}^{3} - a_{400} a_{130}^{2} a_{002}^{3} - a_{310}^{2} a_{040} a_{002}^{3} + \frac{8}{3} a_{400} a_{220} a_{040} a_{002}^{3}.
    \end{align}
\end{subequations}

\paragraph{The generalized Weierstrass model in $\bbP^{1,2,3}$}
The polytope $F_{10}$ describes the weighted projective space $\bbP^{1,2,3}$, and so its defining polynomial is the most general sextic polynomial:
\begin{equation}
\alpha y^2 + a_1 x y z + a_3 yz^3 = \beta x^3 + a_2 x^2z^2+ a_4 xz^4 + a_6z^6 \,.
\end{equation}
We have suggestively chosen variable names to emphasize the close relationship between this polynomial and the long Weierstrass model in Eq.~(\ref{eq:longWeierstrass}); while this has the same functional form as the long Weierstrass model in Eq.~(\ref{eq:longWeierstrass}), because of the presence of $\alpha,\beta$ the short Weierstrass form of this model is no longer given by Eq.~(\ref{eq:shortWeierstrassInTermsOfTate}). However, Nagell's algorithm is still applicable in this more general context. We find that the short Weierstrass coefficients are given by
\begin{subequations}
    \label{eq:F10weierstrass}

    \eqq{f = &-\frac{1}{48} a_{1}^{4} - \frac{1}{6} a_{1}^{2} a_{2} \alpha - \frac{1}{3} a_{2}^{2} \alpha^{2} + \frac{1}{2} a_{1} a_{3} \alpha \beta + a_{4} \alpha^{2} \beta \\ 
    g = &\phantom{-}\frac{1}{864} a_{1}^{6} + \frac{1}{72} a_{1}^{4} a_{2} \alpha + \frac{1}{18} a_{1}^{2} a_{2}^{2} \alpha^{2} + \frac{2}{27} a_{2}^{3} \alpha^{3} - \frac{1}{24} a_{1}^{3} a_{3} \alpha \beta \nonumber \\&- \frac{1}{6} a_{1} a_{2} a_{3} \alpha^{2} \beta - \frac{1}{12} a_{1}^{2} a_{4} \alpha^{2} \beta - \frac{1}{3} a_{2} a_{4} \alpha^{3} \beta + \frac{1}{4} a_{3}^{2} \alpha^{2} \beta^{2} + a_{6} \alpha^{3} \beta^{2},}
\end{subequations}

Each of the sixteen reflexive polytopes can be realized in at least one way by adding points to one of $F_1$, $F_2$, $F_4$, and $F_{10}$, and therefore specializations of Eqs. \ref{eq:F1weierstrass}, \ref{eq:F2weierstrass}, \ref{eq:F4weierstrass}, and \ref{eq:F10weierstrass} suffice to specify the defining polynomials of all 16 fiber types. The appropriate specializations have been worked out in e.g. \cite{Klevers:2014bqa}, and are summarized for in Table \ref{tab:16fibers}. From the specializations, one can readily work out the generic gauge groups of \cite{Klevers:2014bqa} by plugging the forms of $f,g$ into the formula for the discriminant $\Delta$ and factoring it, as described in the main text.

\section{Statistics of Polytope Automorphisms}
\label{sec:autostats}
In the course of our work, we enumerated the automorphisms of every polytope in the KS list, using the built-in capabilities of CYTools \cite{Demirtas:2022hqf}; to the best of our knowledge, this is the first time that automorphisms have been exhaustively and systematically studied. While it is not the primary purpose of our work, we exhibit some statistics of those automorphisms in this Appendix.

Of the 473,800,776 polytopes, only 1.3\% (6,095,530 total) have at least one nontrivial automorphism. The polytope with the largest automorphism group is the 24-cell, studied in e.g. \cite{Braun:2011hd,Braun:2015jdy}. This polytope, which in our convention is labeled as v24-506398, is a self-dual polytope with Hodge numbers $h^{1,1}(X)= h^{2,1}(X) = 20 $ and has an automorphism group of order 1,152. The symmetry group is the Weyl group of the $F_4$ root system, and can be given explicitly as \eqq{\left(\left(\left(\bbZ_2^3\rtimes \bbZ_2^2\right)\rtimes\bbZ_3^2\right)\rtimes\bbZ_2\right)\rtimes\bbZ_2.}

\begin{figure}
    \centering
    \includegraphics[width=0.75\linewidth]{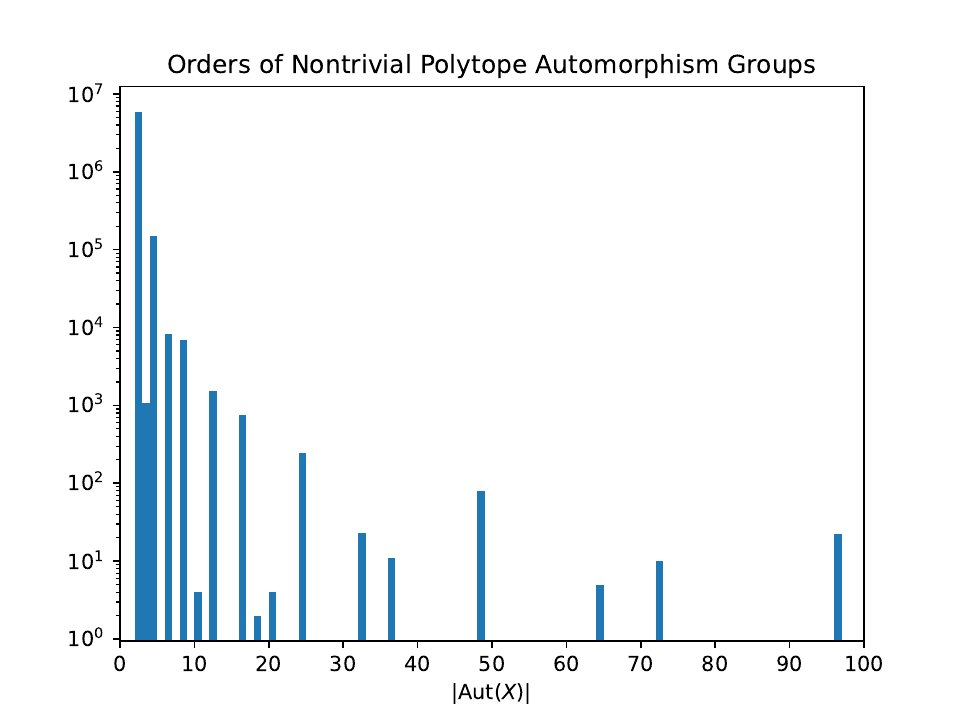}
    \caption{The orders of the automorphism groups of the 6,095,511 polytopes with $1<|\operatorname{Aut(X)|<100}$.}
    \label{fig:autostats}
\end{figure}

Large automorphism groups are the exception, rather than the rule. Of the polytopes with nontrivial automorphisms, 97.2\% (5,925,190 total) have automorphism group $\bbZ_2$, and only 19 of the remaining 170,340 polytopes have $|\operatorname{Aut(X)}|\geq100$. We have shown a histogram of the orders of the automorphism groups of the 6,095,511 polytopes with $1<|\operatorname{Aut(X)|<100}$ in Figure \ref{fig:autostats}. A total of 35 finite groups appear as the automorphism group of a 4D reflexive polytope; we have listed these groups, and the number of times they appear, in Table \ref{tab:autostats_groups}.

\begin{table}[]
    \centering
    \begin{tabular}{|c||c|c|c|} \hline
        Group & GAP ID \cite{GAP4} & Order & \#(Polytopes) \\\hline\hline
        $\bbZ_2$  & [2,1] &   2 & 5,925,190 \\\hline
        $\bbZ_3$  & [3,1] & 3 & 1,080 \\\hline
        $\bbZ_4$ & [4, 1] & 4 & 135 \\\hline
        $\bbZ_2^2$ & [4, 2] & 4 & 151,281 \\\hline
        $S_3$ & [6, 1] & 6 & 8,161 \\\hline
$\bbZ_6$ & [6, 2] & 6 & 57 \\\hline
$D_8$ & [8, 3] & 8 & 3,387 \\\hline
$Z_2^4$ & [8, 5] & 8 & 3,548 \\\hline
$D_{10}$ & [10, 1] & 10 & 4 \\\hline
$D_{12}$ & [12, 4] & 12 & 1,509 \\\hline
$D_{16}$ & [16, 7] & 16 & 2 \\\hline
$\bbZ_2\times D_{8}$ & [16, 11] & 16 & 735 \\\hline
$\bbZ_2^4$ & [16, 14] & 16 & 19 \\\hline
$\bbZ_3\times S_3$ & [18, 3] & 18 & 2 \\\hline
$\bbZ_5\rtimes\bbZ_4$ & [20, 3] & 20 & 4 \\\hline
$S_4$ & [24, 12] & 24 & 106 \\\hline
$\bbZ_2^2\times S_3$ & [24, 14] & 24 & 141 \\\hline
$\bbZ_2^4\rtimes\bbZ_2$ & [32, 27] & 32 & 5 \\\hline
$\bbZ_2^2\times D_8$ & [32, 46] & 32 & 18 \\\hline
$S_3^2$ & [36, 10] & 36 & 11 \\\hline
$D_8 \times S_3$ & [48, 38] & 48 & 8 \\\hline
$\bbZ_2 \times S_4$ & [48, 48] & 48 & 68 \\\hline
$\bbZ_2^3 \times S_3$ & [48, 51] & 48 & 3 \\\hline
$D_8^2$ & [64, 226] & 64 & 5 \\\hline
$S_3^2\rtimes \bbZ_2$ & [72, 40] & 72 & 6 \\\hline
$\bbZ_2\times S_3^2$ & [72, 46] & 72 & 4 \\\hline
$\bbZ_2 \times D_8 \times S_3$ & [96, 209] & 96 & 4 \\\hline
$\bbZ_2^2\times S_4$ & [96, 226] & 96 & 18 \\\hline
$S_5$ & [120, 34] & 120 & 2 \\\hline
$D_8^2 \rtimes \bbZ_2$ & [128, 928] & 128 & 2 \\\hline
$\bbZ_2 \times (S_3^2 \rtimes \bbZ_2)$ & [144, 186] & 144 & 2 \\\hline
$\bbZ_2 \times S_5$ & [240, 189] & 240 & 4 \\\hline
$\bbZ_3^2\rtimes\left(\bbZ_2^4\rtimes\bbZ_2\right)$ & [288, 889] & 288 & 2 \\\hline
$\left(\left(\left(\bbZ_2^3\rtimes \bbZ_2^2\right)\rtimes \bbZ_3\right)\rtimes \bbZ_2\right)\rtimes \bbZ_2$ & [384, 5602] & 384 & 6 \\\hline
         $\left(\left(\left(\bbZ_2^3\rtimes \bbZ_2^2\right)\rtimes\bbZ_3^2\right)\rtimes\bbZ_2\right)\rtimes\bbZ_2$ & [1152, 157478] & 1,152 & 1\\\hline
    \end{tabular}
    \caption{The 35 nontrivial groups that appear as the automorphism group of a 4D reflexive polytope. For each group, we list its order and the number of polytopes for which it is the automorphism group. ``Gap ID'' refers to the entry of the group in the online library of small groups, accessed through the GAP system \cite{GAP4}. }
    \label{tab:autostats_groups}
\end{table}

\begin{figure}[h]
\begin{center}
\begin{subfigure}[b]{.45\textwidth}
\begin{center}
\includegraphics[scale=.5]{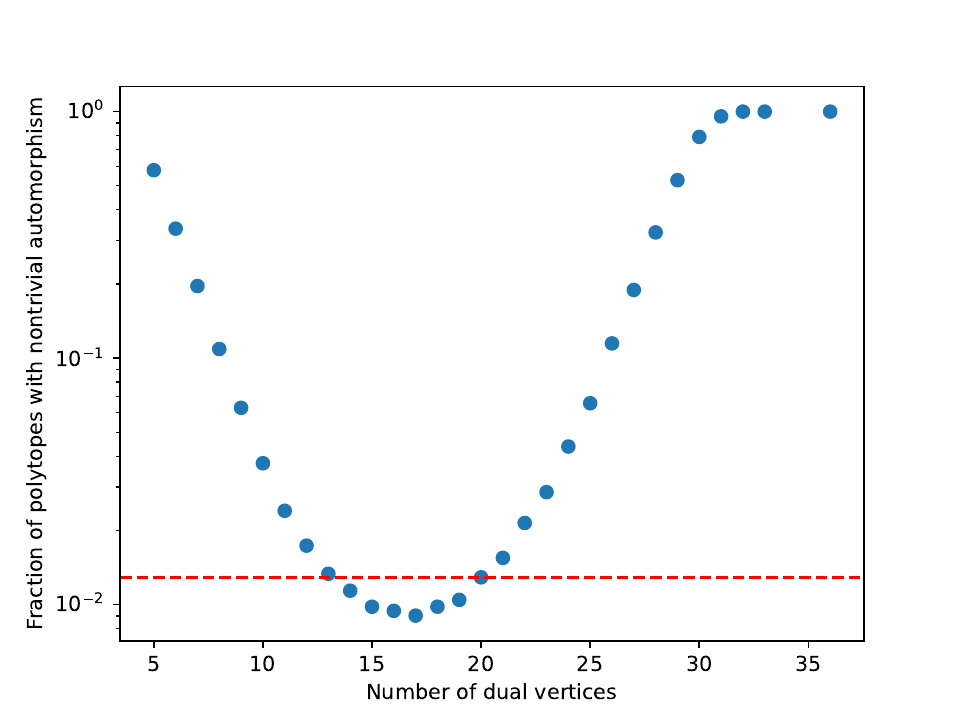}
\caption{}
\label{fig:autostats_fraction}
\end{center}
\end{subfigure}
~
\begin{subfigure}[b]{.45\textwidth}
\begin{center}
\includegraphics[scale=.5]{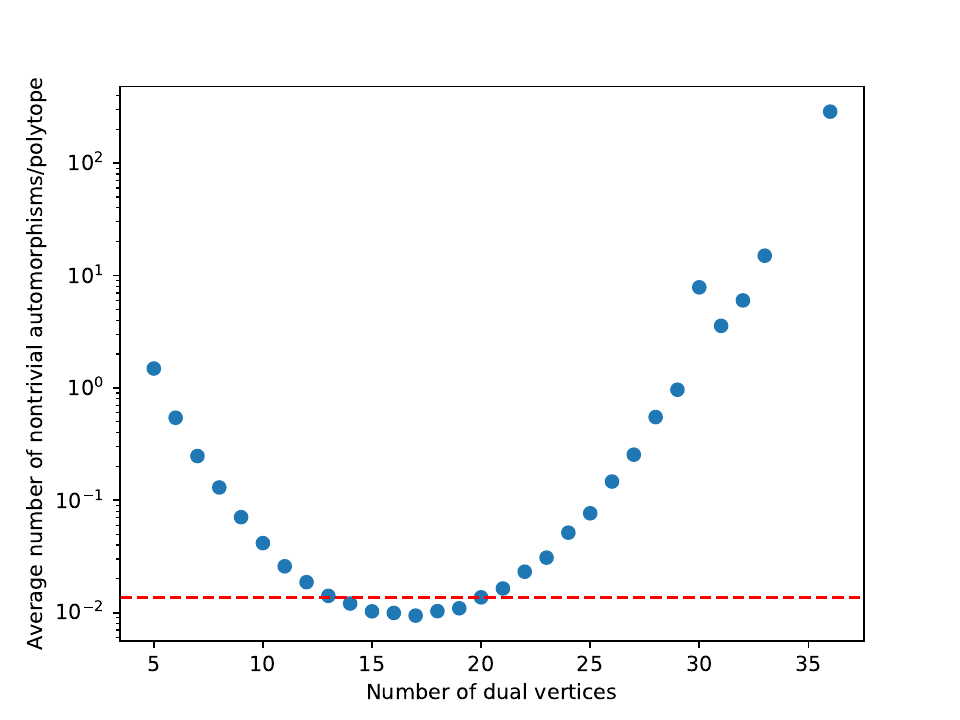}
\caption{}
\label{fig:autostats_average}
\end{center}
\end{subfigure}
\caption{(a) The fraction of polytopes with nontrivial automorphisms and (b) the average number of nontrivial automorphisms per polytope as a function of the number of dual vertices. The averages of both quantities are shown in red. We see in both figures that polytopes with either very many or very few dual vertices enjoy more automorphisms than generic polytopes.}
\label{fig:autostats_subfig}
\end{center}
\end{figure}

Next we study the distribution of automorphism groups as a function of the number of vertices of the dual polytope. We observe, but have not attempted to explain, a curious trend: polytopes whose dual polytopes have either very many or very few vertices tend to have more automorphisms than polytopes whose dual polytopes are generic.\footnote{The average 4D reflexive polytope has 15.8 dual vertices.} This can be seen in two ways, by considering either the fraction of polytopes admitting a nontrivial automorphism or by considering the average number of nontrivial automorphisms per polytope; these quantities are plotted as a function of the number of dual vertices in Figures \ref{fig:autostats_fraction} and \ref{fig:autostats_average}, respectively. We leave explaining this trend to future work. 

\bibliography{references}
\bibliographystyle{JHEP}

\end{document}

%% file: fibers_tikz.tex
\begin{figure}[p!]
\begin{center}
\begin{subfigure}[b]{.2\textwidth}
\begin{center}
\begin{tikzpicture}[scale=1]
         \draw[black, thick] (1,0) -- (-1,-1);
         \draw[black, thick] (-1,-1) -- (0,1);
         \draw[black, thick] (0,1) -- (1,0);
         \draw (0,0) node {$\times$};
         \filldraw[black] (1,0) circle (2pt);
         \filldraw[black] (0,1) circle (2pt);
         \filldraw[black] (-1,-1) circle (2pt);
    \end{tikzpicture}
\caption{$F_1$}
\end{center}
\end{subfigure}
\hspace{.5cm}
\begin{subfigure}[b]{.2\textwidth}
\begin{center}
\begin{tikzpicture}[scale=1]
         \draw[black, thick] (1,0) -- (0,1);
         \draw[black, thick] (0,1) -- (-1,0);
         \draw[black, thick] (-1,0) -- (0,-1);
         \draw[black, thick] (0,-1) -- (1,0);
         \draw (0,0) node {$\times$};
         \filldraw[black] (1,0) circle (2pt);
         \filldraw[black] (0,1) circle (2pt);
         \filldraw[black] (-1,0) circle (2pt);
         \filldraw[black] (0,-1) circle (2pt);
    \end{tikzpicture}
\caption{$F_2$}
\end{center}
\end{subfigure}
\hspace{.5cm}
\begin{subfigure}[b]{.2\textwidth}
\begin{center}
\begin{tikzpicture}[scale=1]
         \draw[black, thick] (1,0) -- (-1,1);
         \draw[black, thick] (-1,1) -- (-1,0);
         \draw[black, thick] (-1,0) -- (0,-1);
         \draw[black, thick] (0,-1) -- (1,0);

         \draw (0,0) node {$\times$};
         \filldraw[black] (1,0) circle (2pt);
         \filldraw[black] (-1,1) circle (2pt);
         \filldraw[black] (-1,0) circle (2pt);
         \filldraw[black] (0,-1) circle (2pt);
    \end{tikzpicture}
\caption{$F_3$}
\end{center}
\end{subfigure}
\hspace{.5cm}
\begin{subfigure}[b]{.2\textwidth}
\begin{center}
\begin{tikzpicture}[scale=1]
         \draw[black, thick] (1,0) -- (-1,1);
         \draw[black, thick] (-1,1) -- (-1,-1);
         \draw[black, thick] (-1,-1) -- (1,0);
         \draw (0,0) node {$\times$};
         \filldraw[black] (-1,0) circle (2pt);
         \filldraw[black] (1,0) circle (2pt);
         \filldraw[black] (-1,1) circle (2pt);
         \filldraw[black] (-1,-1) circle (2pt);
    \end{tikzpicture}
\caption{$F_4$}
\end{center}
\end{subfigure}
\\
~ 
\\
\begin{subfigure}[b]{.2\textwidth}
\begin{center}
\begin{tikzpicture}[scale=1]
\draw[black,thick] (1,0) -- (-1,1);
\draw[black,thick] (-1,1) -- (-1,0);
\draw[black,thick] (-1,0) -- (0,-1);
\draw[black, thick] (0,-1) -- (1,-1);
\draw[black, thick] (1,-1) -- (1,0);
\draw (0,0) node {$\times$};
\filldraw[black] (1,0) circle (2pt);
\filldraw[black] (-1,1) circle (2pt);
\filldraw[black] (-1,0) circle (2pt);
\filldraw[black] (0,-1) circle (2pt);
\filldraw[black] (1,-1) circle (2pt);
\end{tikzpicture}
\caption{$F_5$}
\end{center}
\end{subfigure}
\hspace{.5cm}
\begin{subfigure}[b]{.2\textwidth}
\begin{center}
\begin{tikzpicture}[scale=1]
\draw[black,thick] (1,0) -- (0,1);
\draw[black,thick] (0,1) -- (-1,1);
\draw[black,thick] (-1,1) -- (-1,-1);
\draw[black,thick] (-1,-1) -- (1,0);
\draw (0,0) node {$\times$};
\filldraw[black] (1,0) circle (2pt);
\filldraw[black] (0,1) circle (2pt);
\filldraw[black] (-1,1) circle (2pt);
\filldraw[black] (-1,0) circle (2pt);
\filldraw[black] (-1,-1) circle (2pt);
\end{tikzpicture}
\caption{$F_{6}$}
\end{center}
\end{subfigure}
\hspace{.5cm}
\begin{subfigure}[b]{.2\textwidth}
\begin{center}
\begin{tikzpicture}[scale=1]
\draw[black,thick] (1,0) -- (0,1);
\draw[black,thick] (0,1) -- (-1,1);
\draw[black,thick] (-1,1) -- (-1,0);
\draw[black,thick] (-1,0) -- (0,-1);
\draw[black,thick] (0,-1) -- (1,-1);
\draw[black,thick] (1,-1) -- (1,0);
\filldraw[black] (1,0) circle (2pt);
\filldraw[black] (0,1) circle (2pt);
\filldraw[black] (-1,1) circle (2pt);
\filldraw[black] (-1,0) circle (2pt);
\filldraw[black] (0,-1) circle (2pt);
\filldraw[black] (1,-1) circle (2pt);
\draw (0,0) node {$\times$};
\end{tikzpicture}
\caption{$F_7$}
\end{center}
\end{subfigure}
\hspace{.5cm}
\begin{subfigure}[b]{.2\textwidth}
\begin{center}
\begin{tikzpicture}
\draw[thick] (1,0) -- (1,1);
\draw[thick] (1,1) -- (0,1);
\draw[thick] (0,1) -- (-1,1);
\draw[thick] (-1,1) -- (-1,-1);
\draw[thick] (-1,-1) -- (1,0);
\draw (0,0) node {$\times$};
\filldraw[black] (1,0) circle (2pt);
\filldraw[black] (1,1) circle (2pt);
\filldraw[black] (0,1) circle (2pt);
\filldraw[black] (-1,1) circle (2pt);
\filldraw[black] (-1,0) circle (2pt);
\filldraw[black] (-1,-1) circle (2pt);
\end{tikzpicture}
\caption{$F_8$}
\end{center}
\end{subfigure}
\\
~ 
\\
\begin{subfigure}[b]{.2\textwidth}
    \begin{center}
        \begin{tikzpicture}
            \draw[thick] (1,0) -- (0,1);
            \draw[thick] (0,1) -- (-1,1);
            \draw[thick] (-1,1) -- (-1,-1);
            \draw[thick] (-1,-1) -- (0,-1);
            \draw[thick] (0,-1) -- (1,0);
            \draw (0,0) node {$\times$};
            \filldraw[black] (1,0) circle (2pt);
            \filldraw[black] (-1,0) circle (2pt);
            \filldraw[black] (0,1) circle (2pt);
            \filldraw[black] (-1,1) circle (2pt);
            \filldraw[black] (-1,-1) circle (2pt);
            \filldraw[black] (0,-1) circle (2pt);
        \end{tikzpicture}
        \caption{$F_9$}
    \end{center}
\end{subfigure}
\hspace{.5cm}
\begin{subfigure}[b]{.2\textwidth}
    \begin{center}
        \begin{tikzpicture}
            \draw[thick] (1,0) -- (0,1);
            \draw[thick] (0,1) -- (-2,-3);
            \draw[thick] (-2,-3) -- (1,0);
            \draw (0,0) node {$\times$};
            \filldraw[black] (1,0) circle (2pt);
            \filldraw[black] (0,1) circle (2pt);
            \filldraw[black] (0,-1) circle (2pt);
            \filldraw[black] (-1,-1) circle (2pt);
            \filldraw[black] (-1,-2) circle (2pt);
            \filldraw[black] (-2,-3) circle (2pt);
        \end{tikzpicture}
        \caption{$F_{10}$}
    \end{center}
\end{subfigure}
\hspace{.5cm}
\begin{subfigure}[b]{.2\textwidth}
    \begin{center}
        \begin{tikzpicture}
            \draw[thick] (1,0) -- (-1,2);
            \draw[thick] (-1,2) -- (-1,-1);
            \draw[thick] (-1,-1) -- (0,-1);
            \draw[thick] (0,-1) -- (1,0);
            \draw (0,0) node {$\times$};
            \filldraw[black] (1,0) circle (2pt);
            \filldraw[black] (-1,2) circle (2pt);
            \filldraw[black] (-1,-1) circle (2pt);
            \filldraw[black] (0,-1) circle (2pt);
            \filldraw[black] (-1,0) circle (2pt);
            \filldraw[black] (0,1) circle (2pt);
            \filldraw[black] (-1,1) circle (2pt);
        \end{tikzpicture}
        \caption{$F_{11}$}
    \end{center}
\end{subfigure}
\begin{subfigure}[b]{.2\textwidth}
    \begin{center}
        \begin{tikzpicture}
            \draw[thick] (1,0) -- (1,1);
            \draw[thick] (1,1) -- (0,1);
            \draw[thick] (0,1) -- (-1,0);
            \draw[thick] (-1,0) -- (-1,-2);
            \draw[thick] (-1,-2) -- (1,0);
            \draw (0,0) node {$\times$};
            \filldraw[black] (-1,-2) circle (2pt);
            \filldraw[black] (-1,0) circle (2pt);
            \filldraw[black] (0,1) circle (2pt);
            \filldraw[black] (1,0) circle (2pt);
            \filldraw[black] (1,1) circle (2pt);
            \filldraw[black] (-1,-1) circle (2pt);
            \filldraw[black] (0,-1) circle (2pt);
        \end{tikzpicture}
        \caption{$F_{12}$}
    \end{center}
\end{subfigure}
\\
~ 
\\
\begin{subfigure}[b]{.2\textwidth}
    \begin{center}
        \begin{tikzpicture}
            \draw[thick] (1,0) -- (-1,2);
            \draw[thick] (-1,2) -- (-1,-2);
            \draw[thick] (-1,-2) -- (1,0);
            \draw (0,0) node {$\times$};
            \filldraw[black] (1,0) circle (2pt);
            \filldraw[black] (0,1) circle (2pt);
            \filldraw[black] (-1,-2) circle (2pt);
            \filldraw[black] (-1,1) circle (2pt);
            \filldraw[black] (-1,0) circle (2pt);
            \filldraw[black] (-1,-1) circle (2pt);
            \filldraw[black] (-1,-2) circle (2pt);
            \filldraw[black] (0,-1) circle (2pt);
            \filldraw[black] (-1,2) circle (2pt);
        \end{tikzpicture}
        \caption{$F_{13}$}
    \end{center}
\end{subfigure}
\hspace{.5cm}
\begin{subfigure}[b]{.2\textwidth}
    \begin{center}
        \begin{tikzpicture}
            \draw[thick](0,-1) -- (2,1);
            \draw[thick] (2,1) -- (-1,1);
            \draw[thick] (-1,1) -- (-1,-1);
            \draw[thick] (-1,-1) -- (0,-1);
            \draw (0,0) node {$\times$};
            \filldraw[black] (1,0) circle (2pt);
            \filldraw[black] (-1,0) circle (2pt);
            \filldraw[black] (-1,-1) circle (2pt);
            \filldraw[black] (0,-1) circle (2pt);
            \filldraw[black] (-1,1) circle (2pt);
            \filldraw[black] (-1,0) circle (2pt);
            \filldraw[black] (-1,1) circle (2pt);
            \filldraw[black] (2,1) circle (2pt);
            \filldraw[black] (0,1) circle (2pt);
            \filldraw[black] (1,1) circle (2pt);
            \filldraw[white] (0,-2) circle (2pt);
        \end{tikzpicture}
        \caption{$F_{14}$}
    \end{center}
\end{subfigure}
\hspace{.5cm} 
\begin{subfigure}[b]{.2\textwidth}
    \begin{center}
        \begin{tikzpicture}
\draw[thick] (1,-1) -- (1,1);
\draw[thick] (1,1) -- (-1,1);
\draw[thick] (-1,1) -- (-1,-1);
\draw[thick] (-1,-1) -- (1,-1);
\draw (0,0) node {$\times$};
\filldraw[black] (-1,0) circle (2pt);
\filldraw[black] (1,0) circle (2pt);
\filldraw[black] (-1,-1) circle (2pt);
\filldraw[black] (0,-1) circle (2pt);
\filldraw[black] (1,-1) circle (2pt);
\filldraw[black] (-1,1) circle (2pt);
\filldraw[black] (0,1) circle (2pt);
\filldraw[black] (1,1) circle (2pt);
            \filldraw[white] (0,-2) circle (2pt);
        \end{tikzpicture}
        \caption{$F_{15}$}
    \end{center}
\end{subfigure}
\hspace{.5cm}
\begin{subfigure}[b]{.2\textwidth}
    \begin{center}
        \begin{tikzpicture}
            \draw[thick] (2,-1) -- (-1,-1);
            \draw[thick] (-1,-1) -- (-1,2);
            \draw[thick] (-1,2) -- (2,-1);
            \filldraw[black] (-1,-1) circle (2pt);
            \filldraw[black] (0,-1) circle (2pt);
            \filldraw[black] (1,-1) circle (2pt);
            \filldraw[black] (2,-1) circle (2pt);
            \filldraw[black] (-1,0) circle (2pt);
            \draw (0,0) node {$\times$};
            \filldraw[black] (1,0) circle (2pt);
            \filldraw[black] (-1,1) circle (2pt);
            \filldraw[black] (0,1) circle (2pt);
            \filldraw[black] (-1,2) circle (2pt);
            \filldraw[white] (-1,-2) circle (2pt);
        \end{tikzpicture}
        \caption{$F_{16}$}
    \end{center}
\end{subfigure}
\caption{The sixteen two-dimensional reflexive polytopes. Vertices are connected by black lines, the origin is indicated by a cross, and all other points are indicated by a black dot. }
\label{fig:16fibers}
\end{center}
\end{figure}